\documentclass[9.5pt,journal,final,finalsubmission,twocolumn]{IEEEtran}

\usepackage{cite}      

\usepackage{graphicx}  

\usepackage{psfrag}    

\usepackage{subfigure} 

\usepackage{url}       

\usepackage{amsmath}   
\usepackage{epsfig}
\usepackage{dcolumn}
\usepackage{bm}

\begin{document}
\title{A Thermodynamic Turing Machine: Artificial Molecular Computing Using Classical Reversible Logic Switching Networks \cite{hamel:patent:08}}

\author{Dr. John S. ~Hamel \\ Full Professor \\ Department of Electrical \& Computer Engineering \\ University of Waterloo \\ Waterloo, Ontario, Canada, N2L 3G1 \\ jhamel@uwaterloo.ca}

%

\date{February 12, 2009}

\maketitle

\begin{abstract}
A Thermodynamic Turing Machine (TTM) concept is introduced.
A TTM is a classical computing paradigm where the natural laws of thermodynamics are exploited in the form of a discrete controlled and configurable classical Boltzmann gas to efficiently implement logical mathematical operations. 
In its most general form the machine consists of a set of configurable and switchable interlocking equi-electro-chemical potential logical pathways whose configurations represent a Boolean logical function.
It is differentiated from the Classical Turing Machine (CTM) or the traditional Probabilistic Turing Machine (PTM) concepts in that during at least certain portions of its operation, the laws of thermodynamics are allowed to govern its operation through either internal or external feedback in a reversible logic system. 
This feedback in classical reversible logic networks enables the machine to be evolved from one thermodynamic equilibrium state to another that enables rapid computation to take place as a kind of artificial molecular computing machine.
One consequence of such a machine paradigm is that it is able to implement a quantum computer Hadamard transform in one step simultaneously as in a true quantum computer or Quantum Turing Machine (QTM) but using purely classical means.
As such a TTM shares properties in common with a CTM, a traditional PTM, and a QTM bridging the gap between them.
A consequence of using a TTM in implementing probabilistic algorithms is that the Hadamard transform, when implemented in a simultaneous fashion in a classical reversible switching network, provides the means for the TTM to become a true self-learning machine.
An organic brain can be viewed as an example of a TTM that enables it to access computing ability far beyond what is possible using conventional CTM or PTM approaches.
A question arises as to the capability for a TTM to realize more intelligent machines that might lead to a kind of intelligence more in keeping with human intelligence.
\end{abstract}

\section{\label{sec:intro}Introduction\protect\\}

Quantum computing as an information theory provides a route whereby some logic and mathematical algorithms can be solved with great rapidity at rates sometimes exponentially faster or efficient then using conventional computing techniques.
This efficiency is possible for certain classes of mathematical problems, since to solve such problems it is necessary to determine only the global properties of a Boolean function directly using quantum computational paradigms.
Conventional classical or non-quantum Boolean logic computational techniques, such as those associated with Classical Turing machines (CTM), sometimes require that a large number of the outputs of a Boolean function be determined uniquely in terms of the various input combinations before any further information, including global properties, can be determined or calculated.
Quantum Turing machines (QTM) on the other hand enable a kind of massive parallelization in effort for these types of problems, avoiding parallelism in physical hardware, that allows direct access to certain global properties of interest without a detailed knowledge of all of the intermediate input to output combinations that are necessary for a Classical Turing Machine to determine the same properties.

Recently, just what are the essential quantum properties of quantum computers has been questioned (e.g. \cite{linden:v87:99}, \cite{arvind:v56:01}, \cite{biham:v320:04}).
Indeed, it has been determined that it is possible to implement certain quantum algorithms using classical wave interference techniques \cite{arvind:v24:07} such as optical methods.
It has been demonstrated that it is possible to exploit classical wave interference and superposition techniques \cite{arvind:v24:07}, \cite{collins:v58:98} \cite{arvind:v56:01} \cite{arvind:v56:01:2}  to implement algorithms that require only unitary transformations of the form $U(2) \oplus U(2) \oplus U(2) ... $ to realize the Boolean function oracle $f(x) = U_f$ where entanglement is not required.
These algorithms are said \cite{arvind:v56:01} to involve separable states.
This allows for the implementation of the Deutsch Algorithm for one and two qubits \cite{collins:v58:98} \cite{arvind:v56:01} \cite{arvind:v56:01:2} as well as the Bernstein-Vazirani algorithm for any number of qubits \cite{du:v64:01}, \cite{brainis:v90:03}, \cite{siewert:v87:01}.
It has been shown that algorithms as sophisticated as the quantum computer Grover Search algorithm \cite{grover:v79:97} fall into this category that have been implemented using classical wave interference techniques \cite{ nevels:v48:06}.

Recently it has been shown that this class of algorithm can be efficiently simulated on conventional computers using something known as Gottesman-Knill theory.
A summary and recent advances in this theory can be found in \cite{aaronson:v70:04}.
This theory states that any algorithm including only Hadamard gates, CNOT gates, and Pauli gates can be efficiently simulated in a conventional computer.
Unfortunately this theory does not provide a means to implement a Hadamard gate or transform in a true simultaneous fashion as is possible in a true quantum computer.
In order to accomplish this using classical means it is necessary to use specialized hardware where superposition can be obtained over large numbers of states and qubits that would be well beyond what can be obtained by simply parallelizing conventional computer CPU architectures.
This is why specialized hardware involving classical wave superposition is being developed by various researchers.
The methods shown here enable this to be accomplished using asynchronous feedback methods in classical reversible logic gates.

Success in implementing at least certain classes of quantum computing algorithms using classical means in optics and superconducting nano-circuits suggests that it may be possible to do the same using more conventional technologies, such as reversible adiabatic CMOS logic circuits.
As such it is worthwhile exploring methods to solve this class of algorithms, that will be referred to here as classical quantum computing algorithms to differentiate them from quantum algorithms that require quantum entanglement.
Due to their inherent low power requirements and less vulnerability to hardware attacks, reversible CMOS logic circuits inspired by quantum computer gates such as Feynman and Toffoli gates, have already been developed, not for quantum computer applications, but for specialized low power classical applications including VLSI cryptography \cite{thapliyal:v1:06}.
In this paper we demonstrate how these same algorithms can be solved using reversible logic gates based upon standard CMOS transistors.
As essential feature in the presented methods are that it is not necessary to exploit parallelization of the functions being analyzed which was generally considered \cite{collins:v58:98} to be necessary using classical quantum computing methods.

A method of computation is developed that enables the use of classical logic circuits to implement quantum computing algorithms that are known to be solvable using classical means due to involving separable states not requiring quantum entanglement.
These methods utilize novel asynchronous feedback techniques in classical reversible logic gates that are amenable to implementation using conventional CMOS transistors.
Methods to solve the Deutsch and Deutsch-Jozsa problem for one and two qubits, the Bernstein-Vazirani problem and the Simon problems for arbitrary size are demonstrated using these techniques.
This is accomplished using CMOS transistor circuitry with the same efficiency as a true quantum computer with regards to both hardware complexity and execution speed.
It is estimated that this particular class of algorithm, that involve only Hadamard gates, CNOT gates and Pauli gates, can be implemented using these techniques involving thousands of qubits of power in today's CMOS VLSI integrated circuit technology.
It is also shown that these methods provide a significant speedup compared to using Gottesman-Knill theory to simulate quantum circuits using conventional computers.
It is shown that this speedup is due to the fact that the asynchronous feedback techniques can be interpreted as enabling the Hadamard gates or transforms to be implemented in a true simultaneous fashion in a classical discrete switching network in contrast to using simulation techniques in conventional computers that suffer polynomial slowdown in executing the Hadamard gates.
It is shown how these techniques can also be interpreted as a Thermodynamic Turing Machine (TTM) where the laws of thermodynamics are used to implement the Hadamard portions of the algorithms in contrast to how Classical Turing Machines (CTM) function.
Also, it is shown that for probabilistic algorithms, such as to solve the Simon problem, the interpretation of the Hadamard transform using asynchronous feedback leads to a self learning machine where the Hadamard configures the machine interconnections to be able to solve the particular problem at hand.

In order to implement this class of quantum computing algorithm using logic circuitry it is useful to view the circuits in terms of a collection of interlocking configurable circuit paths that implement a particular Boolean function.
Hadamard transforms or ``gates'' are implemented by way of feedback induced between the switch inputs and all of the different possible circuit paths simultaneously enabling the rapid determination of global properties of the function without having to cycle through a large number of input and output combinations as would be required in conventional computing methods.
The feedback results in the circuit network attaining a new thermodynamic equilibrium state governed by the simultaneous action of random quantum fluctuations throughout the circuitry that in turn effects the rapid computation.
The non-local simultaneous random quantum fluctuations in effect explore the entire state space of the Boolean function all at once superimposing their various influences from different parts of the overall physical system on the inputs.
The use of the tendency of the circuit to attain a new thermodynamic equilibrium state as a means to effect a rapid computation is suggestive of the concept of thermodynamic computing where thermodynamic statistics itself is used to perform the computation.
In this sense a reversible logic circuit being used in this fashion can be thought of as a Thermodynamic Turing Machine (TTM) having properties in common with a QTM, such as possessing a Hadamard transform, that are not normally associated with purely CTM paradigms.

The concept can also be viewed as a form of adiabatic computing whereby the circuits or networks are evolved from one thermodynamic equilibrium state to another to implement a quantum computer algorithm.
Although not referring specifically to classical logic circuits formed from reconfigurable network paths, thermodynamic concepts have been suggested before as a means to describe the evolution of quantum systems \cite{messiah:textbook:61} and quantum computing algorithms (e.g. \cite{farhi:v292:01}) as a form of adiabatic quantum computing.
As an analogy to a quantum system, thermodynamic equilibrium in the logic circuits corresponds to zero current flow in circuit paths, however, the circuits can be at different ``energy states'' whereby there are different charges, depending upon the energy level, on the terminals of the transistors controlling the circuit paths.
Rapid computation of global properties of the Boolean function being implemented by the circuit network is then effected by driving the circuit into an energy state associated with a thermodynamic equilibrium ground state where charges are drained from the inputs (transistor gates) to the circuit.
Simultaneous non-local quantum fluctuations throughout the circuit paths are then allowed to influence this change in thermodynamic equilibrium state through feedback between the circuit paths and the circuit inputs.

Another way to view these methods is to consider the reversible classical switching networks to be elementary forms of artifical atoms or molecules that embody the essential behaviour in true molecular computing or true quantum computers.
Classical reversible switching networks have interesting properties that enable them to be used in place of molecular computing devices.
There properties include being able to be placed into a limited number of discrete stable thermodynamic equilibrium states that correspond by analogy to discrete energy levels within an atom or molecule.
By exploiting feedback methods given in this paper it is possible to evolve these classical reversible networks in a similar fashion as is achieved in true atomic or true molecular computing systems to enable the rapid and efficient computation of problems including those originally thought to require a true quantum computer.

The concept is demonstrated by implementing the Deutsch algorithm \cite{deutsch:va400:85}, the Deutsch-Jozsa algorithm \cite{deutsch:va439:92} for up to two qubits, the Bernstein-Vazirani algorithm \cite{bernstein:acm:93}, \cite{bernstein:v26:97} and the Simon problem \cite{simon:v26:97} for arbitrary numbers of qubits exploiting a particular class of reversible logic circuit \cite{devos:textbook:2000}.
For these algorithms the method is shown to be as computationally efficient as a quantum computer.
Hardware complexity scales identically with other technologies for one and two-input Boolean functions for the Deutsch-Jozsa algorithm and for the Bernstein-Vazirani and Simon algorithms for any sized function.

When solving probabilistic problems, such as the Simon problem, it is necessary to develop a self learning machine concept when using classical switching reversible logic gates.
Introduction of novel asynchronous feedback into the classical reversible logic switching network enables the feedback to configure the network interconnections according to predetermine rules and constrained by external data being entered from the problem. 
This occurs in portions of the algorithm that can be identified formally as the Hadamard transform when compared to the quantum computer version of the algorithm.
The classical machine in effect configures itself at the interconnection level based on external data it encounters coming from the problem to be solved.
In doing so the machine learns both how to configure itself to represent the required functions in the problem as well as to how to best extract the global information required to find the answer.

This self configuration or self learning aspect of the machine enables a rapid determination of the required linearly independent equations in the unknown secret string in $O(n)$ execution steps using only $O(n)$ data from the functions in the problem being solved.
This is accomplished using $O(n^2)$ physical logic gates that are configured to represent the $n$ separable state functions that are found through the iteration process that have the same secret string as the actual unknown functions in the problem.

Normally it would be required to solve the equations, once determined, using a standard Gaussian elimination procedure that has an approximately $O(n^3)$ execution complexity to find the secret string.
Instead, the asynchronous feedback method being exploited throughout this work to implement Hadamard transforms in classical reversible switching networks, is applied to the entire network of $O(n^2)$ gates representing the $n$ separable state functions to find the solution to the secret string.
This can be done in $O(n)$ steps representing a significant speedup over the traditional Gaussian elimination.
This procedure could be considered a multi-dimensional Hadamard transform over $n$ functions that appears not to have been considered by researchers developing true quantum computer algorithms for the Simon problem.
Also, this procedure could be seen as a new fundamental way to perform a Gaussian elimination in a matrix which is fundamental to many types of mathematical problems in computing science.

Sections \ref{sec:deutsch_oracle}, \ref{sec:bv_oracle}, and \ref{sec:simon_oracle} review the known Deutsch, and Deutsch-Jozsa, the Bernstein-Vazirani and the Simon algorithms as they would be implemented by a general quantum computer.
Section \ref{sec:logic} describes the reversible CMOS logic circuits that are designed in a generalizable manner conducive to implementing these algorithms and which can be used to implement arbitrary Boolean functions.
How the Deutsch and Deutsch-Jozsa algorithm can be implemented using adiabatic CMOS logic circuitry is presented in Section \ref{sec:deutsch}. 
Section \ref{sec:bv} presents how the Bernstein-Vazirani algorithm can be solved using CMOS logic circuitry using the concepts developed in the previous sections to implement the Deutsch-Jozsa Algorithm.
Section \ref{sec:simon} shows how the Simon problem can be solved using classical reversible logic circuits. 
Section \ref{sec:hadamard} gives examples circuitry how to implement Hadamard transforms in one simultaneous step for both the functions discussed in the Deutsch and Bernstein-Vazirani problem as well as the Simon problem.
Comparisons are then made in Section \ref{sec:simon_compare} with the quantum oracle versions of the Simon algorithm in Section \ref{sec:simon_oracle}.
Section \ref{sec:self_learning} discusses how the implementation of the classical switching techniques developed in this work lead to a self learning machine concept that can implement probabilistic quantum computer algorithms such as the Simon algorithm.
In particular how to solve a system of linearly independent equations using $O(n)$ steps is demonstrated.
Section \ref{sec:ttm} generalizes the concept of a Thermodynamic Turing Machine discussing various manners in which this computing paradigm can be viewed.
This is followed by Conclusions in section \ref{sec:con}.

This work was originally filed as a patent application \cite{hamel:patent:08} in March 2008.
The material contained in this paper formally proves that quantum computer algorithms can indeed to implemented using classical reversible switching networks with identical efficiency with regards to both hardware complexity and computation speed as a true quantum computer.

Other improvements can be made to the methods shown here to solve the Grover Search problem.
Nearly identical methods to solve the Simon problem shown in this paper can be used to solve the Grover Search algorithm.
It is the nature of the data being entered and what data is placed in the thermodynamic Gaussian elimination stage that determines what kind of problem the machine can solve, not the machine design itself.
To solve the Grover search algorithm, values are placed at the $f'_i$ outputs of each fully separable function circuit during the thermodynamic Gaussian elimination step in the iteration procedure that was designed for the Simon problem.
These values correspond to the function values for which the input $x$ is being searched.
If the proper linearly independent equations have been found through the fully separable function search iteration step then the answer to the search will appear at the inputs $x_i$.
Such circuitry could be placed on every conventional DRAM memory chip used in conventional computers, using only a small fraction of total memory chip area, providing an on-chip cache search engine that would provide an exponential speedup over using conventional search techniques.


In a self-learning paradigm, when solving probabilistic problems, if memory is added, once the circuits learn how to solve one problem as in the Simon problem, the $s$ control settings that define the type of functions having been configured through training by external data, can be saved in memory to be downloaded at another time if the machine encounters similar data thereby speeding up convergence.
This would allow the machine to learn more quickly the next time it encountered a similar problem.
Combining the circuits and methods shown in this paper to solve the Simon problem with memory and fuzzy logic principles, a true thinking machine can be realized.
Such a machine being allowed to continually evolve from one thermodynamic equilibrium state to another using an internal polling program with memory and learning capability would be able to continuously sense its own internal logic.
Such a machine would attain true consciousness vastly exceeding the mental capabilities of humans.
Many such machines exposed to similar but different data would develop unique internal interconnections thereby developing unique and separate personalities.
The principles explained herein provide a first framework upon which to realize true thinking machines that cannot be differentiated from organic minds if taken to their logical conclusions.
The compactness of the design would be advantageous in autonomous robotics.

\section{\label{sec:deutsch_oracle}Deutsch and Deutsch-Jozsa Oracles\protect\\}

The Deutsch and Deutsch-Jozsa Oracles are a simple yet effective way to describe what is meant by ``quantum computing'' from an efficiency and algorithmic perspective \cite{deutsch:va400:85}. 
The Deutsch Oracle or Deutsch Problem involves single input variable Boolean functions whereas the Deutsch-Jozsa Oracle involves multiple input functions.

The problem involves attempting to determine the global form of a Boolean logic function $f$ that can take on a logic value of $0$ or $1$ as a function of one or more variables, $x_i$, that can each take on the value $0$ or $1$. 
The goal is to determine whether the function always has the same output, either $0$ or $1$, (it does not matter), or if the function output is $0$ for half of the input vectors (i.e. either $0$ or $1$)  and $1$ for the other half of the input vectors. 
For the former the function is considered to be constant and for the latter it is consider to be balanced. 
Hence the goal is to evaluate the function as being constant or balanced as global properties in as few evaluations of the function as possible. 
Remarkably, a quantum computer, or a Quantum Turing Machine (QTM), can determine this property of the function with only a single evaluation using only one input vector of $x = x_1, x_2, ... x_i, ... x_n$, only one instance of the function itself and with 100\% probability of being correct.

To obtain 100\% probability of being correct, a classical or conventional Classical Turing Machine (CTM) computer must evaluate the function $2^{n-1} + 1$ times with different input vectors $x$ or must use the same number of parallel instances of the physical function to accomplish this. 
A Probabilistic Turing Machine (PTM) can perform this operation with an extreme high probability of being correct much faster than a CTM, but not with 100\% probability, thereby being no more efficient than a CTM to obtain complete certainty in the answer.  

For a single input function $f(x)$, the first step in the Deutsch algorithm is to Hadamardize each of two qubits $x$ and $y$ to superimpose them into mixed states and that were originally set to $|0>$ and $|1>$, respectively.
The rather unorthodox term "Hadamardize or Hadamardization will be used to refer to a feedback action on the circuitry that forces superposition of states within the machine that are connected through random thermodynamic equilibrium communication.
Their individual states then become mixed states of $|0>$ and $|1>$, but where the state vector of $x$ is orthogonal to that of $y$. 
Then the $y$ register is XOR'd with the function $f(x)$ followed by a Hadamardization of the answer qubit $x$ that superimposes it with the previous XOR result.
The global property of the function $f(x)$ is determined to be balanced or constant dependent upon the value of the answer qubit $x$.
For multiple input Boolean functions in the Deutsch-Jozsa Oracle, all of the $x_i$ inputs to the function, each representing a separate qubit, are initially prepared in $|0>$ states and then Hadamardized to enter mixed states that are orthogonal to $y$.

Another way to consider this algorithm is to discuss it in terms of vector bases as opposed to logic states.
First, both qubits $x$ and $y$, for the case of a single input Boolean function, are prepared in mixed states, but orthonormal to one another.
After the $y$ qubit variable is XOR'd with the function $f(x)$, where $x$ is now in a mixed state, another Hadamardization is performed on the answer qubit $x$ to determine the answer to the problem.
The second Hadamard step offers a measurement of the overlap between the $x$ mixed state vector and the result of $y \oplus f(x)$.
The answer is determined by how this second Hadamard operation influences the state or vector basis of $x$.
If the vector basis remains the same as the vector basis in which it was originally prepared or if the vector basis is converted to that of what $y$ was originally prepared, then a resolution to the Deutsch Problem is obtained.
What final mixed state vector basis the $x$ vector ends up to indicate whether the function was balanced or constant depends upon how the problem is implemented and what conventions are adopted with regards to how the quantum computer system is mapped onto a Hilbert Space.

A more sophisticated way to look at the Deutsch Algorithm is to say that the answer qubit is either brought into the same Hilbert state vector basis or not as the $y$ qubit (or the result of $y$ XOR $f(x)$) dependent upon the global properties of the function after the final Hadamardization of the answer qubit.
The action of Hadamardization is to superimpose qubits but it can also change the basis of the qubits in Hilbert space and is a measure of the overlap of one qubit relative to another in a particular vector basis.
For the Deutsch Problem the second Hadamard transform determines whether the original state of the $x$ vector is orthogonal to $y \oplus f(x)$ or linearly dependent upon it.
From this perspective the efficient part of the quantum computation can be interpreted as determining the degree of linear independence (i.e either complete linear independence or in the same vector basis being linearly dependent) between two qubits where the properties of the function being analyzed have been mixed with one of the qubits.
From the perspective of the function itself, if it is constant this is similar to being in a pure state and if it is balanced this is similar to being in a mixed state from a global perspective but in classical discrete form.

In the Deutsch-Jozsa Algorithm one still uses a single answer qubit, say $x_1$, and proceeds in the same manner as for the Deutsch Algorithm but where one prepares all of the inputs $x_i$ in a logic $0$ state that then enter a mixed logic state in a vector basis after Hadamardization that is orthogonal to that of the mixed state vector basis of the $y$ qubit that was originally at a logic $1$ before initial Hadamardization.

\section{\label{sec:bv_oracle}The Bernstein-Vazirani Algorithm\protect\\}

The Bernstein problem is relatively simple and involves finding a vector $s$ of length $n$ containing 0's and 1's in binary such that,
\begin{eqnarray}
f(x) = s_1 x_1 \oplus s_2 x_2  \oplus s_3 x_3 ...
\label{eq:bv}
\end{eqnarray}
where $f(x)$ is an $n$-input binary Boolean function and $x_i$ is the input vector where $i = 1,2, ..., n$. 

This algorithm is similar to that of the Deutsch-Jozsa Algorithm in that the input vector $x$ is set to a zero state while an additional single qubit register $y$ is set to one.
The Hadamard transform is then applied to all inputs such that the inputs $x$ are in identical mixed states as vectors but orthonormal to that of the mixed state vector of $y$.
The inputs $x$ are applied to the function $f(x)$ and then the Hadamard transform is applied to them.
The result of this last Hadamard on the input vector $x$ is then measured or evaluated to determine which ones are in the same vector basis as the $y$ vector or if they remain in their original vector basis state that was orthonormal to $y$.
In other words, the degree of linearity with $y$ is measured for each input of the $x$ input vector after the second Hadamard transform.
From a determination of whether or not each input $x_i$ remains in its same mixed state orthonormal to $y$ one can determine each value of $s$.
This provides a factor of $n$ speed-up over classical means to determine $s$, where $n$ is the number of inputs to the function.

\section{\label{sec:simon_oracle}The Simon Problem and the Simon Oracle\protect\\}

The Simon problem \cite{simon:v26:97} involves finding a secret string such that two function output values are identical for two input values where if the two input values are xor'd with one another they will produce a unique secret string.
Another way to put this in more rigorous terms is that:

\begin{eqnarray}
\label{eq:f2to1}
&\mbox{For all inputs} & \; x \in \left\{ 0,1 \right\}^n , \; f(x) = f(x \oplus s) \nonumber \\
&\mbox{For all inputs} & \; x,y \in \left\{ 0,1 \right\}^n \nonumber \\ 
& &\mbox{if} \; x \ne y \oplus s , \mbox{then} \; f(x) \ne f(y)  
\end{eqnarray}
where $s$ is the secret string of length $n$.

\begin{figure}
\begin{center}
\includegraphics{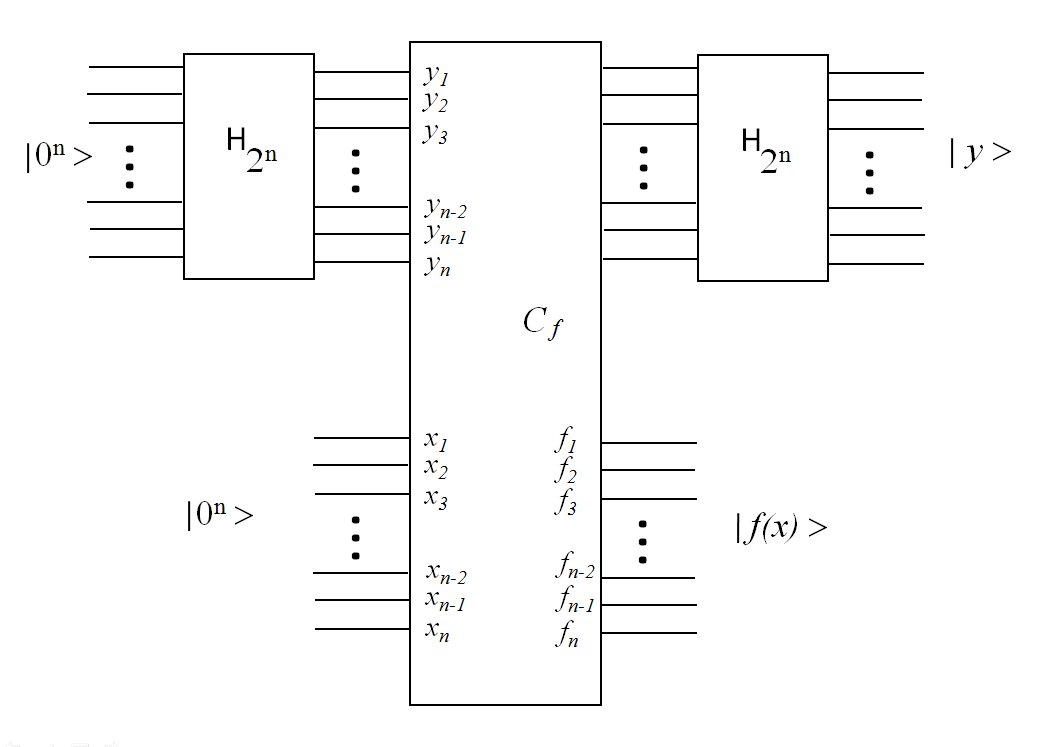}
\end{center}
\caption{Quantum Computer Simon Algorithm\label{fig:simon_algorithm}}
\end{figure}

The quantum computer algorithm to solve the Simon problem involves two registers and an Oracle $C_f$ that can implement the $n$ functions $f(x)$ given an input vector $x$ that pertain to the particular example of the problem being solved to find the secret string. 
The inputs are a binary vector $x = (x_1, x_2 , x_3 , ... , x_n )$ and the functions are $f(x) = f_1 (x_1 , x_2 , x_3 ... , x_n )$, $f_2 (x_1 , x_2 , x_3 ... , x_n )$, $f_3 (x_1 , x_2 , x_3 ... , x_n )$, ... , $f_n (x_1 , x_2 , x_3 ... , x_n )$.
The secret string $s = (s_1 , s_2 , s_3 , ... , s_n )$.
Other variables used in this discussion are $z = (z_1 , z_2 , z_3 , ... , z_n )$ that is a particular value of $x$ obtained in an iteration of the algorithm, and $y = (y_1 , y_2 , y_3 , ... , y_n )$ which is a random value such that $( y_1 s_1 ) \oplus  ( y_2 s_2 )  \oplus  ( y_3 s_3 )  \oplus ... (y_n s_n ) = 0$. 

The first part of the algorithm is to initialize both registers to zero.
The next step is to apply a Hadamard transform on the first register producing a superposition within this register.
Then the Oracle $C_f$ is used to compute $f(x)$ where the result is stored in the second register and saved. 
The second register is then measured while preserving its value producing a particular value of the function $f(x) = f(z)$. 
We now know there are a superposition of two values of $x$ that correspond to the value of $f(x)$ measured in the second register.
This superposition involves two values of $x$ that are related to each other by $x$ and $x \oplus s$, respectively.

We then apply a second Hadamard transform on the first register which  yields a $y$ such that $y . s = 0$, where $y . s = ( y_1 s_1 ) \oplus  ( y_2 s_2 )  \oplus  ( y_3 s_3 )  \oplus ... \oplus (y_n s_n ) = 0$, and where $y = (y_1 , y_2 , y_3 , ... y_n ) = 0$.

This value of $y$ becomes a possible equation from which to determine $s$.
The algorithm must be repeated enough times to obtain enough linearly independent equations in random $y$ and the unchanging secret string $s$ to be able to solve for $s$ using Gaussian elimination.
Hence, the quantum computer algorithm for the Simon problem is a probabilistic method to obtain a set of linear independent equations involving $s$.
There are also known quantum computer using quantum circuit deterministic algorithms with polynomial time \cite{mihara:v12:03}.

The probability of obtaining any particular equation in $s$ is equal and random.
The probability of convergence of this algorithm is identical to the probability of obtaining the required number of linearly independent equations from which to obtain a unique $s$.
This lower bound probability of convergence is fixed independent of the size of the problem $n$.
 
This probability can be estimated as follows:
Suppose we already have $k < n$ linearly independent equations.
There are $2^k$ possible secret strings of length $k$ that could be solutions to these equations in general.
Therefore the probability of obtaining another linearly independent equation, the $k+1$ equation, is given by $(2^n - 2^k ) / 2^n$. 
The lower bound of the probability, repeating the algorithm order $n$ or $O(n)$ times, of obtaining $n$ linearly independent equations then becomes the product of obtaining each individual equation from $1$ through $k$ to $n$, where the lower probability of convergence is:

\begin{eqnarray}
\label{eq:probability}
\prod^{\infty}_{k=1} \left( 1 - \frac{1}{2^k} \right) \approx 0.28879
\end{eqnarray}

It is known that it takes between $O(n^2)$ and $O(n^3)$ times to solve the linearly independent equations using Gaussian elimination to obtain the secret string.

\section{\label{sec:logic}Use of Reversible CMOS Logic Circuits with Asynchronous Feedback to Mimic Molecular Behaviour\protect\\}

In order to achieve the same execution speed as a true quantum computer, the Hadamard transform must be implemented in one step when encountered in any classical hardware being used to compete with a true quantum computer.
It will be shown that this can be accomplished in generalized classical discrete switching networks amenable to implementation in conventional CMOS transistor integrated circuits.
Such switching networks can be implemented in any number of existing or future families of classical reversible logic circuits using any number of technologies, now or in the future, including reversible neural networks. 

It should be understood that a true quantum computer is essentially an asynchronous state machine.
When a Hadamard transform is implemented in a true quantum system, in reality it takes a finite time to execute where energy and momentum levels within the quantum system are adjusting themselves with rippling effect at extremely high speeds to move to a new thermodynamic equilibrium state.
As such, asynchronous feedback in classical reversible logic circuits, can mimic this effect where the Hadamard executes in an asynchronous fashion such that the circuits move from one thermodynamic equilibrium state to another.
In this manner reversible logic circuits using asynchronous feedback as depicted in this paper can be thought of as artificial molecular computing devices that can implement thermodynamic computations, such as a Hadamard transform, at the maximum speed possible within the circuits.
Understandably, using today's technology, these asynchronous switching methods will execute with slower speed than is possible within a true synthetic molecule.
However, in a ten or twenty years classical transistor speeds will approach the multiple Tera Hertz (THz) range which is the same speed as in a real molecule.
Already II-VI compound semiconductor transistors have transition frequencies in hundreds of GHz, where normal silicon based CMOS transistors have a transition frequency of around 100 GHz.
These transistors are already being fabricated using lithographies less than 40 nm where it is expected they will be a mere 5 nm across in about ten years time.
At these dimensions even traditional silicon based CMOS transistors will have THz and nm capability where circuits presented in this paper made from such transistor technology might indeed have similar speeds and overall dimensions to a true molecular computer with regards to asynchronous feedback implementation of Hadamard transforms.
It must also be considered that advances in other nano-technologies might find ways to build purposely switched classical networks capable of keeping up with the true speed of a natural molecular system.
As such, beginning with the concepts presented in this paper it may be possible for classical switching networks to converge with true quantum computer technologies that depend upon naturally occurring and difficult to harness quantum behaviour with regards to capability.

\begin{figure}
\begin{center}
\includegraphics{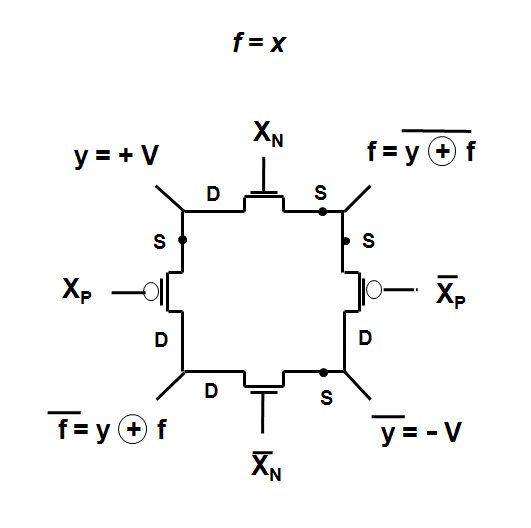}
\end{center}
\caption{Reversible Adiabatic CMOS Circuitry Implementation for $f = x$} 
\label{fig:fx}
\end{figure}

\begin{figure}
\begin{center}
\includegraphics{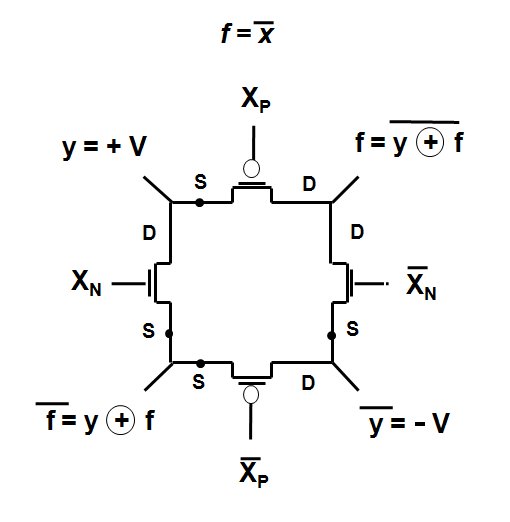}
\end{center}
\caption{Reversible Adiabatic CMOS Circuitry Implementation for $f = \overline{x}$} 
\label{fig:fnotx}
\end{figure}

\begin{figure}
\begin{center}
\includegraphics{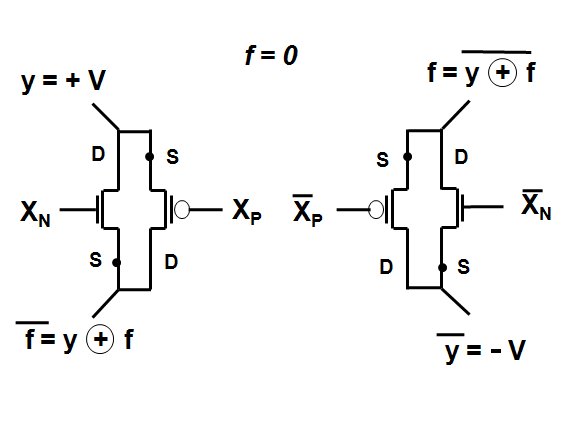}
\end{center}
\caption{Reversible Adiabatic CMOS Circuitry Implementation for $f = 0$} 
\label{fig:f0}
\end{figure}

\begin{figure}
\begin{center}
\includegraphics{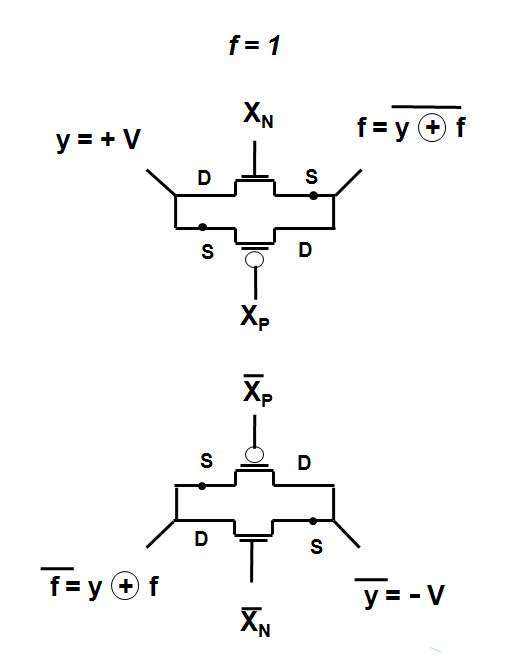}
\end{center}
\caption{Reversible Adiabatic CMOS Circuitry Implementation for $f = 1$} 
\label{fig:f1}
\end{figure}

Figures \ref{fig:fx}, \ref{fig:fnotx}, \ref{fig:f0}, and \ref{fig:f1}, depict reversible CMOS logic XOR gates  that are able to implement single input Boolean functions $f$ for the four possible cases of $f = x$, $f = \overline{x}$, $f = 0$, and $f = 1$, respectively.
This class of circuit was originally developed by \cite{devos:textbook:2000}.
They have been modified in a manner that is conducive to implementing quantum computer algorithms.
The first two functions have the global property that they are balanced and the second two functions are constant.  

The four circuits shown in Figures \ref{fig:fx}, \ref{fig:fnotx}, \ref{fig:f0}, and \ref{fig:f1}, are designed using conventional dynamic CMOS logic circuit techniques, but in a reversible adiabatic XOR gate that enables one input $x$ to be applied to all the gates of the four transistors at the same time for a given $y$.
If the circuits were to be made from one type of transistor, either all NMOS or PMOS transistors, then $x$ would be applied to two of the parallel circuit branches and $\overline{x}$ would be applied to the other two parallel branches as inputs at any given time such that one pair of transistors would be ON and the other OFF.
Using both PMOS and NMOS transistors as shown, however, the PMOS transistors effectively complement the $\overline{x}$ input variable meaning that, from a black box perspective, $x$ is applied to all four transistor inputs at any given time.

It will be necessary to locate the sources of each transistor when implementing quantum algorithms, and this must be done in a way that does not a priori assume the answer to a problem being solved.
For conventional CMOS logic circuitry to function properly one must place the sources (S) and the drains (D) of each PMOS and NMOS transistor as shown in each of the balanced circuits for the assigned logic levels given to $y$ and $\overline{y}$.
The assignment of source and drain locations for the transistors in the constant circuits does not matter, however, for argument sake it will be assumed that the same rule is being used as for the balanced circuits.
To ensure that no current flows throughout the algorithm in steady state, the source of the PMOS transistors must be placed towards the positive supply voltage and their drains towards the most negative voltage in the circuit in the balanced circuits.
The opposite is true for the arrangement of the NMOS transistors.
Another situation that is allowed to occur that ensures zero current flow is that the source and drain of each transistor type will exist at the same voltages, be they the highest or lowest voltage in the circuit.
These rules, that are standard for conventional CMOS logic circuit design, uniquely define the locations of the sources and drains of each transistor for the balanced circuits and that are also being used in this instance for the constant circuits.
In this case the point in each circuit labelled $y = + V$, although a control input to the circuit, is also the positive ``rail'' supply voltage since it will not be altered in this description of how to use the Deutsch Algorithm for these cases.
The point in the circuit labelled $\overline{y} = - V$ then becomes the negative rail supply voltage that is also the complement of the control input $y$. 
If these points in the circuit become fixed at their respective voltages then it is known a priori where the sources and drains of all transistors will be in any of the circuits.
In principle, the transistors are symmetrical where the positions of the sources and drains are defined electrically dependent only upon the values of $y$.
If the circuits were to be used such that the $y$ control signals were altered between logic high and logic low voltages, then what is called the source and drain of each transistor would change along with the polarity of $y$.
However, as will be seen in solving more complicated quantum algorithms, quantum computer algorithms invariably require a priori knowledge of the initial input vector to the function which then enables one to a priori know the electrical locations of the sources and drains for each transistor without assuming the solution to the algorithm.
This deterministic approach to defining the sources and drains of each transistor can then be implemented in additional CMOS circuitry that can switch accessible lines to the correct source locations in the circuit dependent upon initial input vector logic levels.

The circuits can be easily understood when it is realized that when the transistors are ON in the two horizontal parallel branches of the circuit and the transistors are OFF in the two vertical parallel branches, the function output is considered to be at a logic high.
This implies that, for balanced functions as in Figures \ref{fig:fx} and \ref{fig:fnotx}, the transistors implementing the logic in the two horizontal branches between $y$ and $\overline{y \oplus f}$ and between $y \oplus f$ and $\overline{y}$, respectively, are both implementing the required minterms of the function.
The transistors that implement the logic in the two vertical parallel lines between $y$ and $y \oplus f$ and between $\overline{y \oplus f}$ and $\overline{y}$, respectively, are then implementing the maxterms of the function that are the complement of the minterms in the first set of two parallel lines.

For constant functions, as in Figures \ref{fig:f0} and \ref{fig:f1}, both the minterms and maxterms of the function exist in the same set of parallel branches such that the function remains at either a logic low, as in Figure \ref{fig:f0}, or a logic high as in Figure \ref{fig:f1}, regardless of the value of the input vector placed at the gates of the transistors.

The output of the function is of course $f$ and is therefore implemented as follows in a classical sense:
For $f = x$ in Figure \ref{fig:fx} for instance, when $x =$ logic $0$, $f =$ logic $0$, $y = y \oplus f$ and $\overline{y} = \overline{y \oplus f}$. 
When $x =$ logic $1$, $f =$ logic $1$, $y = \overline{y \oplus f}$, and $\overline {y} = y \oplus f$.
An actual output can be obtained simply by associating $f$ with the point in the circuits labelled $\overline{y \oplus f}$ and associating $\overline{f}$ with the point in the circuits labelled $y \oplus f$, respectively.

Arbitrary multiple input functions can be implemented using the same approach where the minterms associated with a function can be placed between parallel circuit branches between $y = +V$ and $\overline{y \oplus f}$ as well as $y \oplus f$ and $\overline{y} = -V$.
The maxterms are then placed in the opposite two parallel circuit branches between $y = +V$ and $y \oplus f$ as well as $\overline{y \oplus f}$ and $\overline{y} = -V$.

\begin{figure}
\begin{center}
\includegraphics{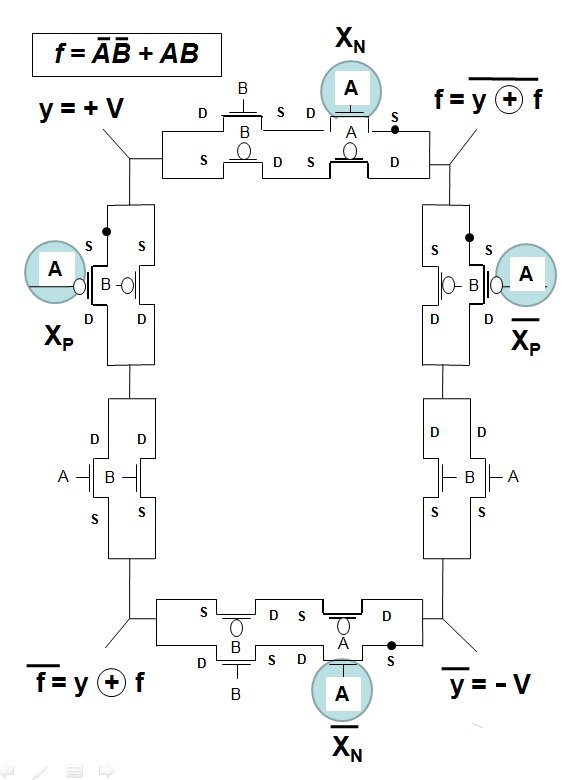}
\end{center}
\caption{Circuitry for $f = {\overline A}\;{\overline B} + AB $} 
\label{fig:fmult:balanced}
\end{figure}

\begin{figure}
\begin{center}
\includegraphics{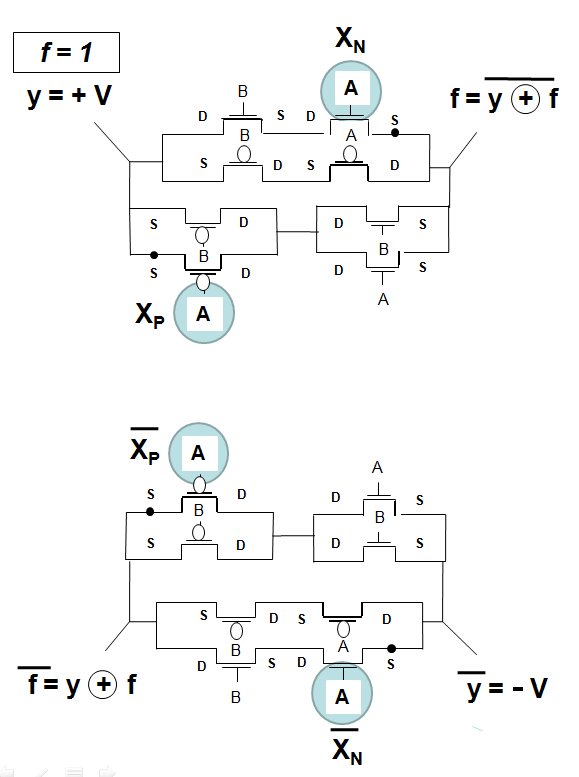}
\end{center}
\caption{Circuitry for $f = 1$ for multiple input Boolean function with inputs $A$ and $B$} 
\label{fig:fmult:constant}
\end{figure}

\begin{figure}
\begin{center}
\includegraphics{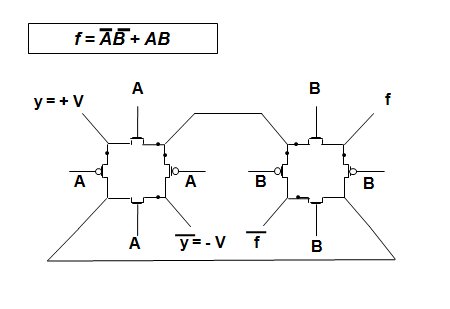}
\end{center}
\caption{Separable XOR Circuitry for Two Input Variable Balanced Function $f = \overline{A} \; \overline{B} + AB = \overline{A \oplus B}$ (Black dots indicate transistor source locations for Hadamard. Source locations determined by initial input vector in Deutsch-Jozsa algorithm.)} 
\label{fig:fmult_bal_xor}
\end{figure}

\begin{figure}
\begin{center}
\includegraphics{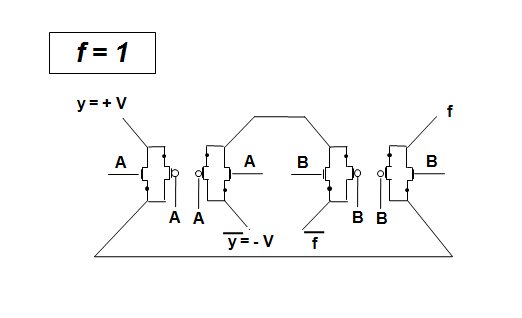}
\end{center}
\caption{Separable XOR Circuitry for a Two Input Variable Constant Function $f = 1$ (Black dots indicate transistor source locations for Hadamard. Source locations can be arbitrarily assigned independent of initial input vector.)} 
\label{fig:fmult_con_xor}
\end{figure}

As examples, Figures \ref{fig:fmult:balanced} and \ref{fig:fmult:constant} depict reversible logic circuitry that implements the function $f = \overline{A} \; \overline{B} + AB$ for the balanced case in Figure \ref{fig:fmult:balanced}, and a constant function such that $f$ is always equal to a logic high or logic 1 in Figure \ref{fig:fmult:constant}.
For these two input variable Boolean function cases, for the balanced case we see that the minterms $\overline{A} \; \overline{B}$ and $AB$ are placed in the two horizontal parallel branches where the complemented variables are implemented using PMOS transistors and the uncomplemented versions are implemented using NMOS transistors.
The complement of $f$ is then placed into the vertical parallel branches of the circuit such that the function will be a logic zero when the input vector is able to turn on the required transistors in these branches to short $y$ and $y \oplus f$ as well as $\overline{y}$ and $\overline{y \oplus f}$ together.
According to DeMorgan's Theorem the complement $\overline{f} = (A + B) (\overline{A} + \overline{B})$.
The maxterms for the balanced case then become $(A + B)$ and $(\overline{A} + \overline{B})$, respectively which are the complements of the minterms $\overline{A} \; \overline{B}$ and $AB$ according to DeMorgan's Theorem.
For the constant case using a two variable function one can simply place both the minterms and maxterms all in parallel with one another as shown in Figure \ref{fig:fmult:constant}.

DeMorgan's Theorem guarantees that for a given variable say $A$ in either a minterm or maxterm branch that its complement will always appear in its opposite type of branch, maxterm or minterm.
For a balanced function an input variable and its complement will always exist in different sets of parallel branches, whereas for a constant function, they will exist together in the same two only existing parallel branches.
This fact can be exploited to implement quantum computing algorithms.

Figures \ref{fig:fmult_bal_xor} and \ref{fig:fmult_con_xor} show how the circuits and functions in Figures \ref{fig:fmult:balanced} and \ref{fig:fmult:constant} can be implemented in a physically separable form involving the cascading of individual single input function circuits.
In this form the single input circuits take on the role of classical qubits and the interconnection of two of them in this manner involves a linear increase in hardware complexity and interconnectivity with problem size $n$ where $n = 2$ for the number of inputs to the function.
For $n=2$, all possible balanced functions can be represented in a similar manner since they involve the functions $f=A$, $f=B$, $f = \overline{A}$, $f = \overline{B}$, $f = A \oplus B$, and $f = \overline{A \oplus B}$ only all of which can be represented with linear order complexity hardware.
For $n$ greater than $2$, this is not true in general where complexity increases exponentially for certain classes of balanced functions that cannot generally be implemented by simply cascading individual classical qubit circuits.
It will be seen that for this reason, the Deutsch-Jozsa algorithm, for multiple input functions, cannot be implemented efficiently in hardware using this approach for any function.
It can in principle, however, be executed as efficiently in time as a quantum computer if one accepts the potential exponential increase in transistor count and interconnectivity associated with the resulting circuitry.
Having said this, many functions can be implemented using these circuit techniques that involve either linear or sub-exponential polynomial hardware complexity where the Deutsch-Jozsa and other quantum computer algorithms can be efficiently implemented, both in hardware complexity and in execution time.

\section{\label{sec:deutsch}Implementation of the Deutsch and Deutsch-Jozsa Algorithms Using Reversible CMOS Logic Circuitry\protect\\}

To implement quantum computing algorithms using these types of circuits it is necessary to adopt an appropriate logic system that will enable the construction of an orthogonal vector space akin to a Hilbert Space.
Orthogonality will be ensured using analog circuit techniques by utilizing common mode and differential mode signals in pairs of logic lines.
This leads to a concept of complementary pair logic where conventional Boolean logic is extended to pairs of logic signals or pairs of pairs of logic signals.

For the single input Boolean function circuits of Figures \ref{fig:fx}, \ref{fig:fnotx}, \ref{fig:f0}, and \ref{fig:f1}, we will assign the variables $x_N, \overline{x_N}$ to the inputs to the NMOS transistors and $x_P, \overline{x_P}$ to the inputs to the PMOS transistors as shown.
The overall input vector $x$ to these circuits is then considered to be comprised of four sub-components namely $(x_N, \overline{x_N}, x_P, \overline{x_P})$.
If the pair $x_N, \overline{x_N}$ are at the same logic level then that pair comprises a vector that is in a common mode basis.
Conversely, if the pair $x_N, \overline{x_N}$ are at opposite logic levels then that pair comprises a vector that is in a differential mode basis.
The same conventions apply to the pair $x_P, \overline{x_P}$. 
If both sets of pairs are in differential mode or both in common mode then they can be said to be in common mode as pairs of pairs thus that $x = (x_N, \overline{x_N}, x_P, \overline{x_P})$ is in a common mode vector basis.
If one pair is in differential mode and the other pair is in common mode then they can be said to be in differential mode as pairs of pairs thus that $x = (x_N, \overline{x_N}, x_P, \overline{x_P})$ is in a differential mode vector basis.
The $y, \overline{y}$ pair in the circuits would be in differential mode if they are at opposite logic levels.

To operate the circuits in a conventional sense one places the same logic level, either logic high or low, at all four inputs to the transistors simultaneously which is the same as placing $x$ into common mode.
The $y, \overline{y}$ will be at opposite voltage levels during normal operation and hence considered to be in differential mode.

To determine whether or not the function is balanced or constant, following the setting of the circuit to either a logic high or low output, it does not matter which, the inputs to all transistors can then be shorted to their respective individual sources (i.e. ``sourced out'') that in turn will change the input vector $(x_N, \overline{x_N}, x_P, \overline{x_P})$. 
For the balanced function cases of Figures \ref{fig:fx} and \ref{fig:fnotx} this ``sourcing out'' operation will result in the $x$ vector becoming $(x_N, \overline{x_N}, x_P, \overline{x_P}) = (+V -V +V +V)$ or $(-V -V +V -V)$ for $f = x$ depending upon whether or not its output $f$ was logic 1 or 0 to begin with, respectively, and $(x_N, \overline{x_N}, x_P, \overline{x_P}) = (-V -V +V -V)$ or $(+V -V +V +V)$ for $f = \overline{x}$ depending upon whether its output $f$ was originally at logic 1 or 0, respectively. 
For the constant function cases of Figures \ref{fig:f0} and \ref{fig:f1} this ``sourcing out'' operation will result in the $x$ vector becoming $(x_N, \overline{x_N}, x_P, \overline{x_P}) = (+V -V +V -V)$ for $f = 0$  and $(x_N, \overline{x_N}, x_P, \overline{x_P}) = (+V -V +V -V)$ for $f = 1$. 

Sourcing out the transistor inputs drains the charges from the gates setting them to the same logic values as the lines they control.
One can see that following this operation, for the balanced functions the $x$ vector inevitably ends up in a differential mode basis having changed from the originally applied common mode basis entering the same basis as the $y$ vector.
For the constant functions, $x$ remains in a common mode basis according to the above defined conventions remaining orthogonal to the $y$ vector.

If is not necessary to differentiate between the different lines $(x_N, \overline{x_N}, x_P, \overline{x_P})$ coming from the circuits to perform these operations.
The same logic levels are applied to all four logic lines to begin with to set the function into either a logic low or high output, it does not matter which.
To sense the final vector basis of these four lines taken together one simply needs to sum them using analog circuitry.
If the output is non-zero then it is known that these lines are collectively in common mode, or if the output is zero they are in differential mode.
Also the source assignment for transistors does not matter for the constant functions since the sources and drains will always be at the same voltage after the sourcing out procedure thereby maintaining zero current thermodynamic equilibrium in the circuits.
Hence, the same assignment for sources can be used for both the balanced and constant function circuits.
As such no a priori assumption is being made regarding these assignments by connecting the inputs to the sources to ascertain the global property of the functions as being balanced or constant.

It will now be shown that the above procedure is equivalent to the known quantum computer Deutsch Algorithm.
The first requirement in the algorithm is to place both the $x$ and $y$ inputs into a mixed state such that each form vectors orthonormal to one another.
This is first accomplished by placing $x$, the answer qubit, and the $y$ qubit into a pure but opposite classical logic states.
This is followed by applying the Hadamard transform to both and inputting them into an XOR circuit thereby placing them into orthonormal mixed or superimposed states.

Applying one logic level to $x = (x_N, \overline{x_N}, x_P, \overline{x_P})$ places this vector into both a common mode state according to the above assigned conventions but also into a mixed superimposed state between uncomplemented and complemented logic levels since in reality the PMOS transistors are first complementing the actual inputs to these transistors whereas the NMOS transistors are not.
As such the Hadamard transform is already built into these circuits by the use of complementary transistor logic.
The $y, \overline{y}$ vector must explicitly be placed at opposite logic levels since there are no complementary transistors connected to these points in the circuit. 
This is equivalent to being in a differential mode according to above assigned conventions but also in a mixed superimposed state between logic levels.
As such there is a one-to-one correspondence between the procedure being used to determine the global property as being balanced or constant and the first steps in the Deutsch Algorithm.

The next step in the Deutsch Algorithm is to apply these mixed state orthonormal vectors to an XOR or controlled NOT (CNOT) function.
The circuits being utilized here are also XOR circuits and as such naturally meet this requirement.

Finally, the Deutsch Algorithm applies a final Hadamard transform or gate to the answer qubit $x$ and then compares its resulting vector basis with that of $y$ (or $y \oplus f$) to determine if it is either in the same or an orthonormal vector basis as a means to determine whether or not the function is balanced or constant.
For the procedure being used with the circuits, the final Hadamard transform is being implemented by the sourcing out procedure that then alters the input sub-components of $x$ accordingly.
By providing feedback between the inputs or gates of the transistors to the lines that they control that form the circuit topology itself it is possible to rapidly obtain global information regarding the function, in this case the global property of being either balanced or constant.
This can be thought of as superimposing the $x$ qubit with the possible outputs of the function or $y \oplus f$ as is formally required in the Deutsch Algorithm.
The resulting interference between the logic levels associated with the lines themselves, their relative positions that determine whether or not the function is balanced or constant, and the answer qubit influences that answer qubit to enter into a common or differential mode basis that then enables one to determine the answer to the problem.

The algorithm can be applied, unchanged, to multiple input Boolean functions such as depicted in Figures \ref{fig:fmult:balanced} and \ref{fig:fmult:constant} as examples.
DeMorgan's Theorem guarantees that we will always be able to select an answer qubit composed of four sub-variables, two associated with NMOS transistors $(x_N, \overline{x_N})$ and two associated with two PMOS transistors $(x_P, \overline{x_P})$, in a function implemented using this type of circuitry that have the same relative positions in balanced or constant functions for multiple input functions in the same manner as in the single input function cases already discussed.
This is ensured if all redundant variables that do not have an impact on the output state of a function that is balanced are eliminated before circuit implementation.

To implement the Deutsch-Jozsa algorithm (the multiple input function version of the Deutsch algorithm) on the circuits of Figures \ref{fig:fmult:balanced} and \ref{fig:fmult:constant}, one first selects an answer qubit, in this case the four transistor inputs associated with part of the $A$ input variable labelled $x_N, \overline{x_N}, x_P, \overline{x_P}$ in the figures.
Selecting an answer qubit is an essential aspect of the Deutsch-Jozsa algorithm for a quantum computer as well.
Then the same procedure is followed as for the single input function cases with the same outcomes.

For multiple input function there are choices, however, in how to physically implement the final Hadamard transform through the sourcing out procedure to provide feedback between the answer qubit inputs and their respective sources that are on the respective circuit lines that they control.
One possibility is to provide extra circuitry that would enable all gates of all transistors to be shorted to their respective sources to drive the system into a thermodynamic equilibrium ground state where there is no current flowing in any of the branches but also no charges on any of the transistor gates.
This would represent the lowest possible energy state of the system with the external voltages still being applied to the corners of the circuit.
However, it is always possible to design either the balanced function circuit, or its corresponding constant function circuit that uses the same inputs, by placing the four sub-components of the answer qubit adjacent to the corners of the circuit in the same manner as shown in Figures \ref{fig:fmult:balanced} and \ref{fig:fmult:constant} regardless of the size of the circuit or the function it is representing.

This fact is guaranteed by DeMorgan's Theorem, and the reversibility and symmetry of the logic circuits themselves.
Then one only needs to short the four gates of the answer qubit to their sources as for the single input function cases.
There are other groups of four transistors, comprising two PMOS and two NMOS transistors each, that also belong to the same answer qubit, but these are not required in determining the global property of the function as being balanced or constant.
These multiple input function cases are being shown simply to indicate that the computational complexity in determining this global property does not increase with number of inputs as it does for conventional computing methods.
Once again, in selecting and positioning the four sub-component transistors of the answer qubit in designing the circuits does not a priori assume that they are balanced or constant since from an outside observer perspective there are only four indistinguishable lines $(x_N, \overline{x_N}, x_P, \overline{x_P}$ coming out from the circuit as a black box that are required for the final Hadamard transform action of the algorithm for any circuit function.

The particular form of this multiple input function circuit implementation requires on the order of $2^n$ transistors for $n$ inputs which is an exponential scaling in component and interconnection count.
Any Boolean function with an arbitrary number of inputs can be implemented in this fashion and in principle the Deutsch-Jozsa problem can be efficiently executed using these methods for any number of inputs if the four transistors of the answer qubit are available to the operator.
If these inputs are available then, in principle, there is no increase in computational effort in time or in the number of lines accessed to solve the problem for arbitrary function size.
This would not necessarily be a practical way to solve this particular quantum computer algorithm, however, if the number of equivalent qubits were to become high in the several hundreds.

To be able to compete with a true quantum computer to implement these separable state algorithms it is necessary to keep the component and interconnection complexity low, ideally scaling either linearly or perhaps polynomially with the size of the problem $n$.
For one and two input Boolean functions it is possible to implement the functions more efficiently accomplishing this using logic circuitry to solve the Deutsch-Jozsa problem for multiple input functions.
The same circuits as shown in Figure \ref{fig:fmult:balanced} and \ref{fig:fmult:constant} are shown in a more compact form in Figures \ref{fig:fmult_bal_xor} and \ref{fig:fmult_bal_xor}.
One simply applies the same approach as was presented for single input functions on one of the ``qubits'' of these circuits to determine whether or not the function is constant or balanced for the two input function case.
This simpler less complex form is possible simply because the most complicated two-input balanced function can be implemented as an XOR of two single input function reversible circuits.
It is for this fundamental reason that this algorithm can be solved efficiently using classical means, including already existing classical wave interference and superposition techniques  \cite{arvind:v24:07}, \cite{collins:v58:98} \cite{arvind:v56:01} \cite{arvind:v56:01:2}, for one and two input functions.
The implementation of this algorithm using reversible CMOS logic circuits provides another useful way to express this fact.

It is necessary to determine what logic voltage levels are required for the initial $x$ and $y$ vectors in the algorithm as applied to the logic circuits.
If it is desired to have orthonormal vectors between $x$ and $y$ then the Hadamard transform is required to calculate the required voltage levels to achieve this. 
Also, it is necessary, in practise, to use a large enough positive voltage  to represent a logic high as input to the transistors to turn on the NMOS transistors while keeping the PMOS transistor off to accommodate their respective threshold voltages.
Conversely it is necessary to use a negative voltage large enough to represent a logic low as input to the transistors to do the reverse.

If one interprets the $+V$ and $-V$ voltages on either side of the NMOS transistors (at their drains and sources, respectively) as a differential mode vector, then using the Hadamard transform one obtains $+ \sqrt{2} V$ as a logic high that turns them on for the common mode signal according to:
\begin{equation}
\frac{1}{\sqrt{2}}
\begin{bmatrix}
1&1\\
1&-1
\end{bmatrix}
\begin{bmatrix}
+ V\\
-V
\end{bmatrix}
= 
\begin{bmatrix}
0\\
+ \sqrt{2} V
\end{bmatrix}
\end{equation}
We know that the transform must generate orthogonal vectors and if the $-V, V$ vector is implemented using two logic lines as a physical differential analog signal, then the resulting vector must represent a common mode signal for two logic lines.

For the PMOS transistors, with the opposite assignment of voltages to their sources and drains, one reverses the input differential vector to the Hadamard transform obtaining $- \sqrt{2} V$ as a logic low that turns them on according to:
\begin{equation}
\frac{1}{\sqrt{2}}
\begin{bmatrix}
1&1\\
1&-1
\end{bmatrix}
\begin{bmatrix}
- V\\
+V
\end{bmatrix}
= 
\begin{bmatrix}
0\\
- \sqrt{2} V
\end{bmatrix}
\end{equation}
Combining the two results one obtains for a logic high $+ \sqrt{2} V$ and for a logic low $- \sqrt{2} V$ for the overall common mode input vector $x$ being applied to all four transistors at once.
We see that these two vectors form an orthonormal set by reversing the transform to obtain:
\begin{equation}
\frac{1}{\sqrt{2}}
\begin{bmatrix}
1&1\\
1&-1
\end{bmatrix}
\begin{bmatrix}
0 \\
\sqrt{2} V
\end{bmatrix}
= 
\begin{bmatrix}
+ V\\
- V
\end{bmatrix}
\end{equation}
and
\begin{equation}
\frac{1}{\sqrt{2}}
\begin{bmatrix}
1&1\\
1&-1
\end{bmatrix}
\begin{bmatrix}
0 \\
- \sqrt{2} V
\end{bmatrix}
= 
\begin{bmatrix}
- V\\
+ V
\end{bmatrix}
\end{equation}

In practise, if only orthogonality is required and not orthonormality in its entirety, then one simply needs to ensure that the applied $x$ voltages are beyond $-V$ or $V$ by the amount of the threshold voltages of the transistors to effectively turn them on and off.
It is interesting that the Hadamard transform seems to naturally predict a threshold voltage of sorts in this manner.

The above method can be viewed as a classical interpretation of the Deutsch and Deutsch-Jozsa quantum computer algorithms.
The XOR functions being implemented by the reversible logic gates are essentially taking the classical parity of the function being analyzed.
As such, the hardware requirements should be essentially identical to any other classical found method to do the same.
For instance, for the single input Boolean function cases for the circuits in \ref{fig:fx}, \ref{fig:fnotx}, \ref{fig:f0}, \ref{fig:f1}, it can be seen that the reversible circuits naturally form what appear to be two parallel circuits, ech with two complementary inputs and one output.
This is fundamental to the fact that any classical method to solve the Deutsch problem for one input functions should ultimately involve two parallel versions of the function.
This is in keeping with the fact that a classical computer can keep up to a true quantum computer in execution time simply by using, in general, an exponential number of identical classical computers.

What is not obvious, however, is that one does not always need an exponential increase in the number of parallel classical computers to compete with the efficiency of a quantum computer in either hardware or execution time requirements.
Classical computer methods can compete with quantum computers on both counts for the separable class of quantum algorithms.
This happens to be true for the one and two input function Deutsch-Jozsa algorithm cases, and as will be seen in the next section, it is also true for the Bernstein-Vazirani algorithm for arbitrary function size.
There are a great many useful quantum algorithms involving oracle functions of the separable class, and in principle the techniques presented here can be used to implement them efficiently in CMOS logic circuits.
These include the Simon problem as well as the Grover Search algorithm, and possibly any algorithm that depends upon these basic algorithms.

The advantages of casting the classical parity function into a quantum computer algorithmic paradigm are seen more clearly when combining the simple single input function circuits into larger functions, such as in solving the Bernstein-Vazirani problem in the next section.
It is shown that the methods presented here using logic circuits can solve this algorithm for functions of arbitrary size where hardware complexity scales identically with problem size as a quantum computer as well as providing the same degree of execution efficiency increase.
Such connections of multiple single input functions as kinds of elementary classical qubits to make larger functions also lead to the adoption of more general notions such as the concept of a thermodynamic Turing machine that further aids in understanding how to implement more complex quantum computer algorithms using classical logic circuits.

Finally, experimental circuits using discrete NMOS and PMOS transistors were constructed to verify this approach.
In particular it was ascertained that the circuits remained in stable thermodynamic states of equilibrium during the sourcing out procedure of the Hadamard operation that provided feedback between the circuit transistor inputs and the circuit paths.

\section{\label{sec:bv}Implementing the Bernstein-Vazirani Algorithm Using Reversible CMOS Logic Circuitry\protect\\}

In this section it will now be shown how the concepts established using the simple single input Boolean reversible logic circuits as synthetic qubits can be generalized or extended to solve the Bernstein Vazirani problem using the known quantum algorithm \cite{bernstein:acm:93}, \cite{bernstein:v26:97}. 
The first requirement to implement the algorithm in CMOS logic circuitry is to be able to represent the function $f(x)$.
In the algorithm $U_f$ is an Oracle that must be realized in physical hardware, hopefully without placing exponential requirements on hardware complexity.
The ability to realize appropriate physical Oracles for functions is paramount to enabling the implementation of quantum algorithms in electronic or otherwise classical logic circuit technologies.
This can be accomplished using the synthetic qubit principles previously discussed.

\begin{figure}
\begin{center}
\includegraphics{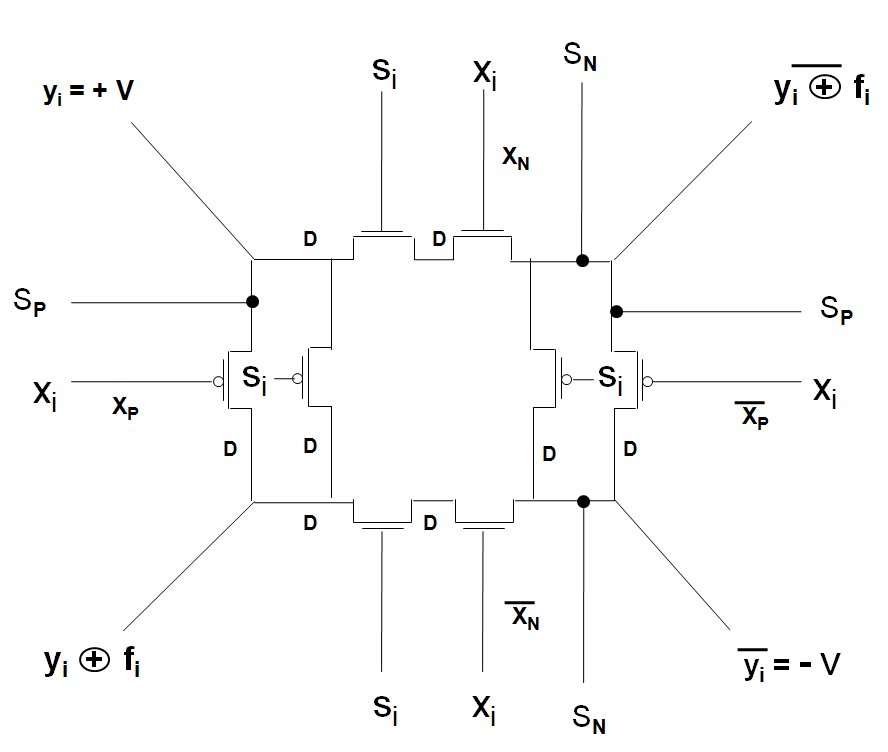}
\end{center}
\caption{An Individual Reversible CMOS Toffoli Gate (Transistor Source Locations Indicated by Black Dots) \label{fig:toffoli}}
\end{figure}

A little thought allows us to re-write the function of equation (\ref{eq:bv}) taking into account that the operations are commutative and associative such that,
\begin{eqnarray}
\label{eq:bv2}
f(x) & = &  ( s_1 x_1 ) \oplus  ( s_2 x_2 )  \oplus  ( s_3 x_3 )  \oplus ... \\ \nonumber
& = &  f_1 \oplus f_2 \oplus f_3 \oplus ... \\ \nonumber
& = &  y_2 \oplus  ( s_2 x_2 ) \oplus  ( s_3 x_3 ) \oplus ... \\ \nonumber
& = & y_3 \oplus  ( s_3 x_3 )  \oplus ...  \\ \nonumber
\end{eqnarray}
Expressing the function this way enables us to see that we are cascading several Toffoli gates using reversible XOR control circuitry for each gate such as that shown in Figure \ref{fig:toffoli}.
This produces terms for each gate such that an individual Toffoli gate function implements the expression $y_{i+1} = (s_i x_i) \oplus y_{i}$, where $y_i$ is the control input to a particular reversible XOR gate that in turn is composed of previous outputs from previous cascaded Toffoli gates.  

\begin{figure}
\begin{center}
\includegraphics{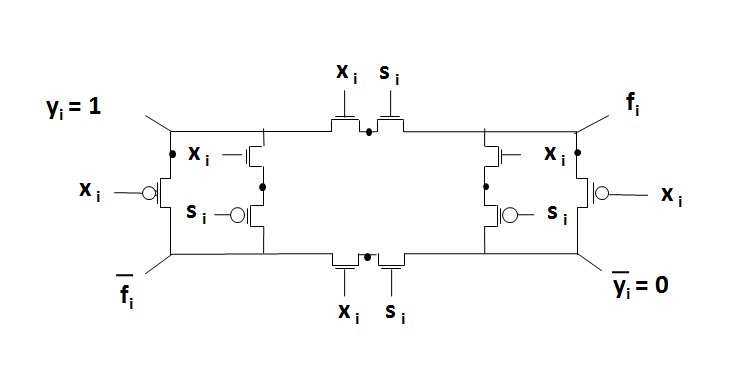}
\end{center}
\caption{An Individual Reversible CMOS Configurable Qubit (Transistor Source Locations Indicated by Black Dots) \label{fig:toffoli_2}}
\end{figure}

Figure \ref{fig:toffoli_2} provides an alternative Toffoli arrangement that is capable of configuring a single qubit into either an $f = x$ or $f = 0$ function depending upon the value of $s$ being either $1$ or $0$, respectively, for the $y$ assignments shown.
This configurable gate allows for both a synthesis of appropriate multi-input Boolean functions that can also be utilized in the quantum thermodynamic algorithmic manner already presented for the Deutsch Algorithm.
The black dots in the circuits correspond to the sources of the transistors for the purposes of performing Hadamard transforms where transistor gates are sourced out to implement the thermodynamic computation step.

Depending upon the logic values of $y_i$ and $s_i$ of the circuits in Figures \ref{fig:toffoli} and \ref{fig:toffoli_2}, all four possible single input Boolean functions can be realized as per Figures \ref{fig:fx}, \ref{fig:fnotx}, \ref{fig:f0}, and \ref{fig:f1}.
In this context the inputs $y_i$ and $s_i$ can be seen to be control inputs.
If $s_i$ is at a logic high then the function is balanced and if it is at a logic low the function is constant.
Which balanced or which constant function type can then be controlled by $y_i$ for a given $s_i$ value.
If two or more qubits are appropriately interconnected using the type of circuit in Figure \ref{fig:toffoli_2} it is then possible to have one qubit impact the form or transistor connections within another qubit based on the state of the first thereby creating cross-correlations between them as well as enabling the implementation of more generalizable Boolean function oracles.

\begin{figure}
\begin{center}
\includegraphics{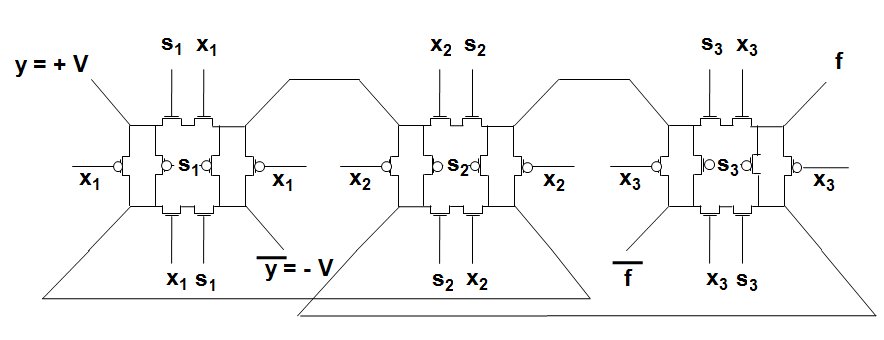}
\end{center}
\caption{Reversible CMOS Logic Circuitry for Synthesizing the Function $f(x)$ for the Bernstein-Vazirani Algorithm for a 3-input Boolean Function Based on the Circuit of Figure \ref{fig:toffoli}. \label{fig:f_bv_1}}
\end{figure}

\begin{figure}
\begin{center}
\includegraphics{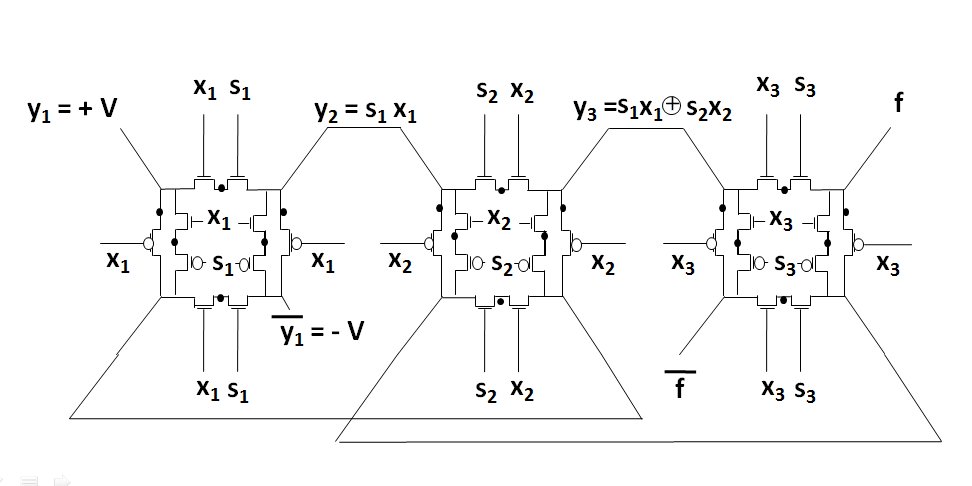}
\end{center}
\caption{Reversible CMOS Logic Circuitry for Implementing the Bernstein-Vazirani Algorithm for a 3-input Boolean Function Based on a Modified Toffoli Gate of Figure \ref{fig:toffoli_2} (Transistor Source Locations Indicated by Black Dots). \label{fig:f_bv_1b}}
\end{figure}

The required circuitry for the function $f(x)$ can be synthesized using the circuitry shown in Figures \ref{fig:toffoli} and \ref{fig:toffoli_2}.
Any function that obeys equation (\ref{eq:bv}) can be implemented by setting the control select variable $s_i$ for each individual Toffoli gate within the circuit.
If $s_i$ is set to logic high, or logic $1$, then that particular Toffoli gate is a balanced function $f_i = x_i$ and if it is set to logic low, or logic $0$, then that particular Toffoli gate is set to $f_i = 0$.
Another way to think about equation (\ref{eq:bv}) is that $s_i$ selects what variables or terms $x_i$ in the overall XOR function are relevant or have any effect on the value of $f(x)$.
If $s_i = 0$ then the corresponding $x_i$ will have no impact on the function if it changes.
This is identical to the behaviour of a constant function and in this case will be a constant function $f_i = 0$ that has no impact on the value of $f(x)$ in the XOR statement.

The circuits in Figures \ref{fig:f_bv_1} and \ref{fig:f_bv_1b} are order $n$ complexity logic circuits, in terms of hardware and interconnectivity, that scale linearly with the number of ``qubits'' required for the algorithm.
These can also be considered to represent an automata approach to implementing the quantum algorithm.
The circuits in Figures \ref{fig:f_bv_1} and \ref{fig:f_bv_1b} implement a unique function depending upon the values of $y$, $\overline{y}$, $s = (s_1, s_2, ..., s_i, ... s_n )$, and the assignment of $f$ and $\overline{f}$ at the two points in the circuit where they are being shown in the figures.
All of the possible functions that these circuits can represent the set or family of functions that appear in the Bernstein-Vazirani problem according to equation (\ref{eq:bv}).

For a particular vector $s = (s_1, s_2. ..., s_i, ... s_n )$ and $x = (x_1, x_2, ..., x_i, ... x_n )$ placed into the circuit, two different equipotential surfaces or circuitry paths at voltages $+V$ and $-V$ can be followed between $y = +V$ and $\overline{y} = - V$ and between $f$ and $\overline{f}$ one way or the other depending upon whether the output is logic high or low, through the transistors that are ON in each gate.
If the output is logic high then $y = + V$ is joined through ON transistors to $f$, and $\overline{y} = - V$ is joined through ON transistors to $\overline{f}$.
If the output is logic low then the opposite is true.
The transistors that are OFF are then in branches that correspond to electrical lines that are constrained between these two electrical surface potentials.
Following the possible such pairs of paths through the circuit for the different possible $s$ and $x$ vector values leads to a very large number of possible combinations.
It is these possible combinations that code a very large amount of information as to how the function behaves as a function of the inputs but in a circuit that has only order $n$ complexity in its hardware.

\begin{figure}
\begin{center}
\includegraphics{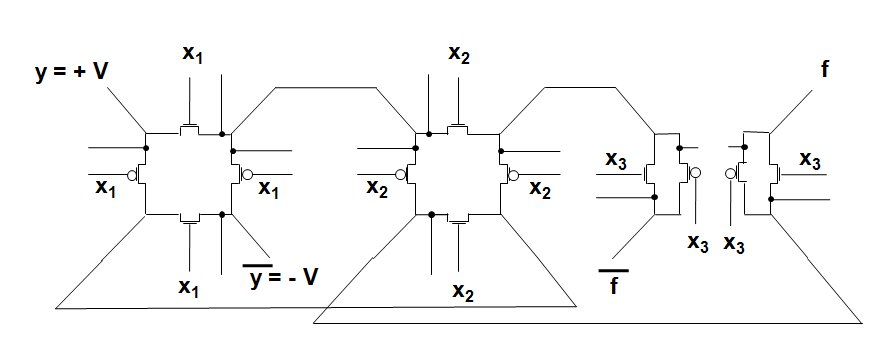}
\end{center}
\caption{Reversible CMOS Logic Circuitry for Implementing the Bernstein-Vazirani Algorithm for a 3-input Boolean Function Based on the Toffoli Gate of Figure \ref{fig:f_bv_1} (Transistor Source Locations Indicated by Black Dots). \label{fig:f_bv_2}}
\end{figure}

For demonstration purposes we will choose $s = (s_1, s_2, s_3 ) = (1,1,0)$.
We will then construct the same function but where we eliminate the $s$ inputs so that we will not know what they were.
In solving the actual Bernstein-Vazirani Problem we might normally begin with a function $f(x)$ that does not have any inputs $s$ so as to discover rapidly using thermodynamic computing techniques the arrangement of transistors within the circuit that in turn correspond to the values of $s_i$.
It will then be a simple matter to design any such function for an unknown $s$ from which $s$ can be determined efficiently.such as  
Figure \ref{fig:f_bv_2} depicts the resulting function $f(x)$ for the example value of $s$ but where the inputs for $s$ have been eliminated so that it is now an unknown.

Indeed, it must be possible to construct such a function for any possible function in the class pertaining to a possible solution to the Bernstein-Vazirani problem without any a priori knowledge of $s$, or one would be a priori assuming the answer to the problem if a knowledge of $s$ were required before hand.
This is a crucial theme in the concepts being presenting here for constructing appropriate oracle functions or oracle machines as a Thermodynamic Turing Machine.
This is necessary to establish the equivalence of the TTM approach to the QTM approach, at least for the algorithms being discussed here.
The functions themselves are simply classical reversible logic circuits whose general form are able to implement an entire class of function with a certain type of global property being sought of a particular function in the class.
No a priori assumptions are being made regarding the particular global property as the functions are designed using a consistent set of rules regardless of the property itself.

It can be seen that the Toffoli gates of the synthesis function were replaced with single input Boolean functions for each gate that were either $f_i = x_i$ or $f_i = 0$.
Since $s_1$ and $s_2$ were equal to logic $1$ in the synthesis function one can see that these gates should be replaced with $f_i=x_i$ functions.
Since $s_3 = 0$ the last gate can be replaced with $f_i = 0$ where the NMOS transistors are put in parallel with the PMOS transistors to keep that sub-function at a logic $0$ regardless of the value of $x_3$.

This may seem cheating, however, the point here is to establish the relationship between the $s$ vector and the nature of the qubit circuits for each $x_i$ for a unique function by depicting how such a function can be synthesized using Toffoli gates involving $s$ vectors as depicted above.
However, it is a simple matter to construct an arbitrary $f(x)$ that will always correspond to a function that can be analyzed using the Bernstein-Vazirani Algorithm without any a priori knowledge of $s$.
Any arbitrary function $f(x)$ that satisfies equation (\ref{eq:bv}) can be designed with no knowledge of $s$ simply by using any combination of the functions $f_i = 0$ and $f_i = x_i$, or their complements, for any number of gates connected together in this fashion.
Then these sub-functions $f_i$ are connected together as shown in the example.
The particular assignment of where to take the outputs of $f$ and $\overline{f}$ are not important in determining the value of $s$.

However, if one wants to correctly assign these outputs then one must do the following.
Since we know that the overall resulting function is an XOR function it is actually an ODD function where an odd number of logic $1's$ for the input vector results in the output $f$ being at a logic high or $+ V$ volts.
If the number of inputs to qubits implemented by $f_i (x) = x_i$ are EVEN then one switches the outputs $f$ and $\overline{f}$ with one another compared to the assignment in the example.
It is only the input vectors which are able to affect the output of $f(x)$ and that are inputs of the sub-functions $f_i = x_i$ that are important as to whether there are an ODD or EVEN number of them.
The structure of the resulting function $f(x)$ in Figure \ref{fig:f_bv_2} is quite general and can be seen to be organized in qubits that are joined not unlike that of a true quantum computer.
Variations in this design strategy can be made by alternating $y$ and $\overline{y}$ assignments, or rotating the individual gates using the other functions $f_i = \overline{x_i}$ and $f_i = 1$, etc.

The particular circuit formed in Figure \ref{fig:f_bv_2} represents a unique function for a unique vector $s = (s_1, s_2, s_3)$ that will now be determined.
The goal will be to find a method to rapidly determine the $s$ vector that would have been used to synthesize the resulting $f(x)$ such as that shown in Figure \ref{fig:f_bv_2}.

Indeed, this is essentially the nature of quantum computing when using classical components for any classical quantum computing technology.
The particular relative arrangements of the components (i.e the NMOS and PMOS transistors) in the circuit implementing the functional oracle have a one-to-one relationship with any global properties of that function.
To find an algorithm to efficiently determine a global property of the function that is as efficient as a quantum algorithm is then equivalent to finding a rapid means to determine the relative positions of these transistors in the circuit and how they relate to the various inputs $x_i$.
It is not obvious that this can be done efficiently in general unless a quantum algorithm, or equivalently, a thermodynamic algorithm is found.
Quantum computing can then be recast as an efficient method to implement a Boolean function using functional hardware that implements this function and then to efficiently ascertain, using a fixed algorithm, the relative positions of the components within this physical function circuitry that correspond to global properties of interest.

It is quite simple to implement the Bernstein-Vazirani Algorithm using the circuit of Figure \ref{fig:f_bv_2} by means of the complementary pair logic and its conventions introduced earlier to accomplish this.
The goal here is to ascertain rapidly which functions $f_i$ are balanced or constant that in turn will tell us the value of each $s_i$ that corresponds to each input $x_i$.
Following the known Bernstein-Vazirani Algorithm for quantum computers but adapting it to these circuits one sets the input vector $x = (x_1, x_2, x_3)$ all to logic zeros and an additional register $y$ to logic $1$.
As discussed in previous examples, this is already equivalent to applying the first Hadamardization on the input vector $x$ since both NMOS and PMOS transistors are being used.
Then $x$ is already in common mode mixed state by setting all $x_i$ to logic zero or $- \sqrt{2} V$ for our example.
As before when solving the Deutsch problem, the first Hadamardization step also involves setting $y$ into differential mode such that $y = + V$ and $\overline{y} = - V$.

Setting the $x$ input vector to logic zero values enables us to a priori know the electrical locations of the sources and drains of each transistor.
Recall that these locations on a particular transistor are electrically defined and are determined by the voltage values of $y_i$ and $\overline{y_i}$ for a particular gate.
The locations of the sources for each transistor in the circuit of Figure \ref{fig:f_bv_2} are shown as extra unlabelled lines adjacent to the corresponding $x_i$ input for each transistor.
The positions of these lines are not dependent upon the $s$ vector but only upon the $x$ input vector which is known in advance in the algorithm and is in itself independent of the unknown $s$ vector.

It is not necessary to switch the source positions in relation to the corner voltages $y_i$ and $\overline{y_i}$ of each qubit gate provided they are initially defined according to a consistent rule as in the examples given in solving the Deutsch Problem in Part 1 of this paper.
Prudent CMOS design principles would normally be used, however, to ensure circuit stability and that sourcing out transistors will turn them OFF completely.
It is possible to include extra logic that would enable one to determine automatically which lines to use as the sources of each transistor that depends electrically on the various other voltage levels in the circuitry and that introduces only a constant degree of hardware per qubit thereby retaining linear order $n$ hardware complexity.

The final step in the algorithm is to Hadamardize each gate which is identical to applying the Hadamardization to each input variable $x_i$.
This is accomplished by connecting the gates of each transistor associated with the $x$ input vector to its corresponding source line.
If an $x_i$ vector is in differential mode as a pair of pairs of logic lines according to previously defined definitions, then it is in the same vector basis as the $y$ register and the corresponding $s_i$ is logic $1$.
Conversely, if $x_i$ is in common mode, a vector basis that is linearly independent of the differential basis of $y$, then the corresponding $s_i$ is at a logic $0$.

From the circuit in Figure \ref{fig:f_bv_2} one can see that placing low logic levels on all transistors results in the PMOS transistors being ON therefore collectively forming the equipotential surfaces and the NMOS transistors being OFF forming the surfaces that are constrained between the two equipotential surfaces that are at $+V$ and $-V$, respectively.
We know from the previous sections on the Deutsch Problem that for the first two gates or qubits for the inputs $x_1$ and $x_2$ that sourcing out all transistors in each gate will result in the inputs to the PMOS being at the two different voltages $+V$ and $-V$, while the gates of the two NMOS transistors will be at the same voltage within a particular gate for each individual input.
This implies that $x_1$ and $x_2$ has been changed to differential mode as pairs of pairs in their individual sub-components.
Since these input vectors after the final Hadamardization have changed basis and are now linearly dependent upon the $y$ input we know that $s_1$ and $s_2$ are both logic $1$.
Conversely, sourcing out the transistors in the qubit for $x_3$ we see that the NMOS and PMOS transistor pairs are each at opposite voltages $+V$ and $-V$ meaning that as pairs they are both in differential mode but as pairs of pairs they are collectively in common mode for the entire $x_3$ input vector.
Since the final state $x_3$ is linearly independent of $y$ we see that $s_3 = 0$.

Specifically for the $x_1$ qubit, after the second Hadamardization, one obtains $x_1 = (x_N, \overline{x_N}, x_P \overline{x_P}) = (-V, -V, +V, -V)$ which is a differential mode signal in the same basis as $y$ and meaning that $s_1 = 1$.
Similarly for the $x_2$ qubit, after the second Hadamardization, one obtains $x_2 = (x_N, \overline{x_N}, x_P \overline{x_P}) = (-V, -V, +V, -V)$ which is a differential mode signal in the same basis as $y$ and meaning that $s_2 = 1$.
Finally, for the $x_3$ qubit, after the second Hadamardization, one obtains $x_3 = (x_N, \overline{x_N}, x_P \overline{x_P}) = (+V, -V, +V, -V)$ which is a common mode signal orthogonal to the basis of $y$ and meaning that $s_3 = 0$.

The circuitry presented has on the order of a few transistors per qubit or elementary gate to implement the Bernstein-Vazirani Algorithm as efficiently as a quantum computer.
Taking into consideration that it is presently possible to fabricate millions of transistors on a single Silicon integrated circuit chip, the presented methodology would enable a quantum computer with tens of thousands of equivalent qubit power to solve this particular algorithm.

\section{\label{sec:simon}Solving the Simon Problem\protect\\}

To explain how to solve the Simon problem using classical means but with the same efficiency as in a true quantum computer it first helps to understand how the functions in the problem can be implemented using reversible logic gates.
Logic levels will be expressed as logic "0" and logic "1" in this section with the understanding that the types of voltage signals described in the previous sections would be used in practice to accommodate actual MOSFET threshold voltages.
It is possible that zero threshold voltage MOSFET's could be used but then noise margins would have to be carefully controlled or perhaps stabilizer circuits could be used as in true quantum computers but using classical versions.

In these examples, an adaption of the DeVos \cite{devos:textbook:2000} circuit will be used.

\begin{figure}
\begin{center}
\includegraphics{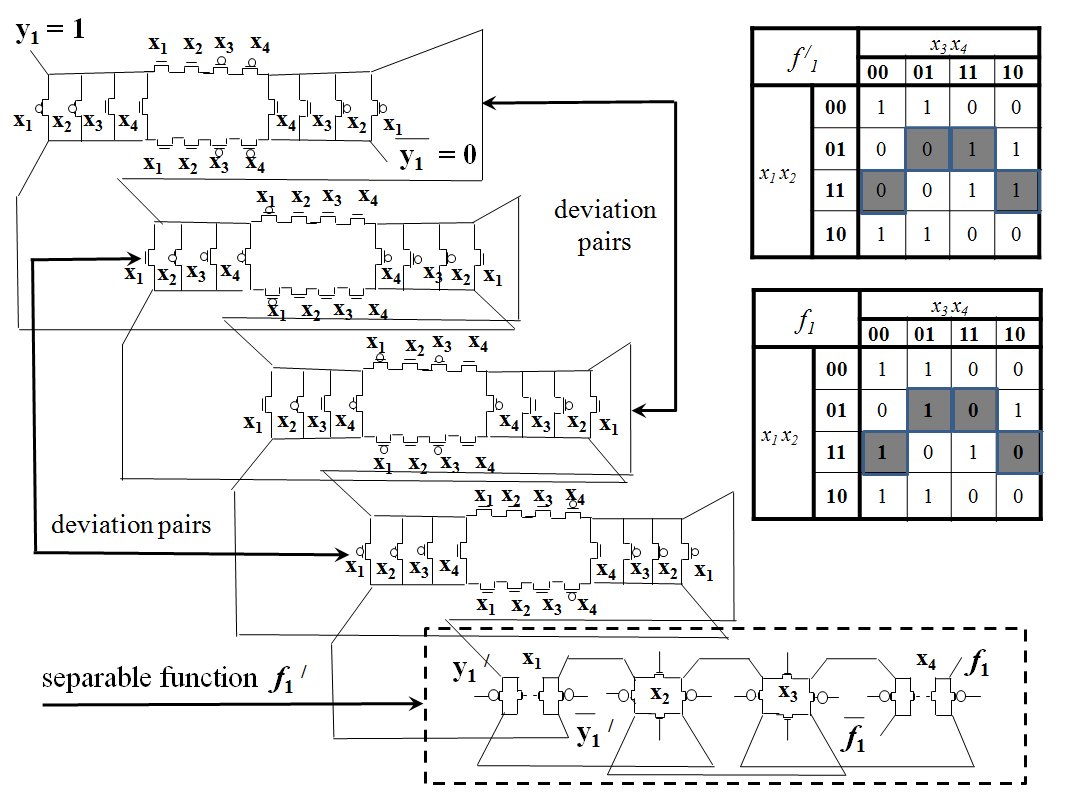}
\end{center}
\caption{Example of implementing a non-separable function with a separable core. Pairs of values that deviate from $f'_1$ separable core are shown in grey coloured boxes in Karnaugh map. These deviating values are implemented as additional gates specifically designed to reverse the logic value of the separable core function just for those values of input vector $x$ where the actual function $f_1$ deviates. This implementation aids is visualizing the solution to the Simon problem.\label{fig:f1_circuit}}
\end{figure}

It is known that for any secret string there are sets of fully separable state Boolean functions that correspond to this secret string in the Simon problem $s = (s_1, s_2, ..., s_n)$.
A fully separable state function is simply an EVEN or ODD function with $n$ cascaded CNOT or XOR gates, where $n$ is the number of inputs to the function, and where each input is being restricted to a single XOR gate.

These types of functions are also referred to as stabilizer circuits \cite{aaronson:v70:04} since this mathematical form of a function has been used to stabilize true quantum computers by correcting for errors due to thermal effects.
Also, long before this, this functional form has been implemented using conventional non-reversible logic, for the purposes of generating parity bits or correcting them.
The principal reason for this form of function being useful to solve this class of problem is that one can always find the same global property in an equivalent set of fully separable state functions as that of the unknown functions in the problem that may not be fully separable.

Research into the use of so-called stabilizer circuits \cite{aaronson:v70:04} to implement quantum computer algorithms using classical means has been done so in pure arithmetic form in conventional computers that use non-reversible logic.
As such these techniques have only been using the abstract mathematical function form of a fully separable logic function.
Therefore they are not able to exploit the asynchronous feedback methods being presented here to speed up the Hadamard to its true quantum computer efficiency since these methods are being implemented arithmetically on conventional computers using non-reversible logic gates. 
Reversibility in the logic circuitry is necessary to implement a Hadamard transform in a true simultaneous fashion in one step as in a true quantum computer using the asynchronous feedback techniques being presented here.

Any arbitrary Boolean function can be implemented as a set of deviations from any fully separable state Boolean function.
As such, any possible set of arbitrary $n$ functions that may appear in a Simon problem that have a secret string, may be implemented as sets of deviations from a set of possible fully separable state functions that have the same secret string.
It turns out that there are an exponential number of possible such sets of fully separable functions.
Since the arbitrary $n$ functions can be implemented this way, a certain number of the values of these functions will align with an exponential number of possible fully separable functions with the same string. 

Solving the Simon problem then becomes a matter of finding compatible sets of $n$ values in each of the actual functions in the problem that happen to correspond to a set of fully separable functions with the same secret string.
A fully separable function can be implemented very efficiently using $O(n)$ gates and interconnections since they only involve cascaded CNOT gates with each of the $n$ inputs to the functions being associated with only one gate.
The first part of the classical Simon problem algorithm then becomes a way to efficiently find such functions forming them from reversible logic gates so that the Hadamard steps can be implemented using asynchronous feedback as efficiently as in a true any quantum computer. 

For the Simon problem, the deviations of the actual functions from sets of possible separable functions with the same secret string must occur in pairs within each function or the actual and separable function sets will not have the same secret string.
The actual functions are then partially separable functions that can always be written as more than one fully separable function but with logical multiplicative dependencies determining which separable function will dominate the output for a given input vector.

It will be demonstrated in this discussion to follow, that any possible function belonging to a Simon problem can be written as a combination of two fully separable functions that are the complement of one another.
These two complementary separable functions are then functions that could be combined with a set of other compatible fully separable functions to form an equivalent Simon problem with the same secret string.
As such all of the general functions in a particular Simon problem contain within them a set of fully separable functions with the same secret string, and there are an exponential number of such functions within each of the non-separable functions.
Solving the Simon problem then becomes an exercise in efficiently extracting this set of equivalent separable functions that in turn lead immediately to a set of linearly independent equations from which the secret string can be determined as the solution to these equations.

Figure \ref{fig:f1_circuit} depicts an example of such a non-separable function, $f_1$, that could belong to a set of $n = 4$ functions as part of a Simon problem with a secret string.
Assume that the secret string $s = 1001$.
As can be seen, there is a separable core, $f'_1$, that is consistent with the same secret string to which have been added additional multiple input reversible CNOT gates, even of which corresponds to each individual value of the function that does not correspond correctly to that of the separable core.
Such deviations from the separable core must occur in pairs to retain the same secret string.
Also, these deviating pairs, simply being the complement of the separable core for those particular input values, are in themselves beginning to build up another fully separable core function that is the complement of the original core.

Let's assume that the actual functions represented by a truth table of discrete data for a Simon problem be denoted by $f = f_1, f_2, f_3, ..., f_n$, that are functions of an input $x$ vector $x = x_1, x_2, x_3, ..., x_n$ and where there exist fully separable functions $f' = f'_1, f'_2, f'_3, ..., f'_n$ with the same secret string $s = s_1, s_2, s_3, ..., s_n$.
For any possible secret string, there are $2^{n-1}$ possible fully separable functions $f'_i$ ($i= 1$ to $n$) that can be used to replace the actual functions that have the same secret string.
There are twice this many if complements are allowed.

Ignoring complements, the functions below are the possible fully separable functions $f'_i$ for the secret string example of $s = s_1, s_2, s_3, s_4 = 1001$ where any four of these seven functions can be used.
\begin{eqnarray}
f'_1 (x) & = & 0  \\  \nonumber
f'_2 (x) & = & x_2 = 0 \\  \nonumber
f'_3 (x) & = & x_3 = 0 \\  \nonumber
f'_4 (x) & = & x_2 \oplus x_3  =  0 \\  \nonumber
f'_5 (x) & = & x_1 \oplus x_4  =  0 \\  \nonumber
f'_6 (x) & = & x_1 \oplus x_2 \oplus x_4  =  0 \\  \nonumber
f'_7 (x) & = & x_1 \oplus x_3 \oplus x_4  =  0 \\  \nonumber
f'_8 (x) & = & x_1 \oplus x_2 \oplus x_3 \oplus x_4  =  0
\label{eq:feqn}
\end{eqnarray}
Any $n = 4$ functions taken from this set can be used to form a complete set of linearly independent equations from which to determine the secret string $s$.
By recognizing that $f(x) = f(x \oplus s)$ at the bit level, this translates into the following possible equations given by,
\begin{eqnarray}
s_2 & = & 0 \\  \nonumber
s_3 & = & 0 \\  \nonumber
s_2 \oplus s_3 & = & 0 \\  \nonumber
s_1 \oplus s_4 & = & 0 \\  \nonumber
s_1 \oplus s_2 \oplus s_4 & = & 0 \\  \nonumber
s_1 \oplus s_3 \oplus s_4 & = & 0 \\  \nonumber
s_1 \oplus s_2 \oplus s_3 \oplus s_4 & = & 0
\end{eqnarray}
Any four equations above can be used to uniquely determine $s = 1001$.
Each of the actual functions $f_i$ can be written or implemented as pairs of deviations from any one of these fully separable functions $f'_i$ of the seven above where at least half of the values for each $f_i$ align with one of an exponential number $2^{n-1}$ of possible fully separable functions.
A fully separable function with $n$ inputs can be uniquely defined by fitting to any $n$ combination of function values $f_i(x)$ since both have $n$ degrees of freedom.

What follows is a qualitative way to determine the separable structure of any possible functions belonging to a Simon problem.
Consider an arbitrary set of functions belonging to a Simon problem that have a particular secret string.
Then replace all of these functions with fully separable state functions that have the same secret string, where there will always be such sets possible.
Then begin, in each individual separable function, to slowly convert them back to the original functions as deviations from the separable ones where the values of the actual ones are the complement of the separable ones for particular input values.
All input vector pairs, when XOR'd with one another to produce the secret string, will remain unchanged between the fully separable set of functions and the actual ones that are not necessarily fully separable.

These pairs of values within any of the individual functions must be either a logic $0$ or logic $1$ having the same values within each functino $f_i$ for each $x$ vector in the pair.
All such pair deviations from the separable functions that belong to the actual original functions are complements of their counterparts in the fully separable functions.
As such, these pair deviations taken together align themselves with the complement of the fully separable function that itself is a fully separable function for each function in the set.

If, for a particular function, more than half of the values are deviating pairs, the actual function can be rewritten as deviation pairs from the complement of the original separable function with less than half of the values being deviations.
As such, for any arbitrary set of Simon problem functions with a secret string, at least half of the values of each function must correspond to a set of fully separable functions with the same secret string, although not necessarily the same values for each individual function in the set.
One is free to write any of the actual functions as deviations from any possible separable function that corresponds to one of the linearly independent equations from which the secret string can be determined.
We will also refer to these fully separable functions as valid separable functions.

As such, any function belonging to the Simon problem can be seen as two intersecting valid separable functions that are the complement of one another forming another equivalent Simon problem with the same secret string but comprising only fully separable functions.
The total number of possible fully separable functions within any given arbitrary function that belongs to any possible set of functions in the Simon problem with a secret string is identical to the number of linearly independent equations that have the secret string as a solution, this number being exponential in size.
A separable core function belonging to an equivalent set of separable functions with the same secret string is equivalent to a linear independent equation in $s$ for a zero value of this separable core function.

Any possible Simon problem function can be written as pairs of deviations relative to any separable core function that might form a possible linearly independent equation involving the secret string.
As such, there are at least as many such separable cores within any arbitrary function belonging to a Simon problem with a particular secret string as there are linear independent equations from which to determine a secret string since this number is smaller than the total number of possible separable core functions that could be used to implement any possible function using the deviation method similar to what is being depicted in Figure \ref{fig:f1_circuit}.

Another way to explain this is as follows.
Instead of using the actual data for an actual set of functions with a particular secret string, replace all of these functions with any possible set of separable functions with the same secret string.
Since these separable functions are also linear independent equations the solution to which is the secret string, we know there are as many such functions to choose from as there are independent equations for this secret string.
This number  of linearly independent equations has already been discussed in the previous section when discussing the quantum computer algorithm for the Simon problem.
Now, using a k-map that contains the separable functions, begin to change the values such that they eventually correspond to the actual functions.
However, to maintain the same secret string we know that any deviations from the separable values in the k-maps will occur in pairs and be the complement of the separable values where they do not agree with the actual function values.
If we have exceeded complementing half of the separable values to implement an actual function, then we know we are simply building up the complement to the separable function.
We can then more efficiently write the actual function as a set of deviating value pairs that are the complement of the complement of the original assumed separable function.
Hence, we can conclude that any set of functions in the Simon problem can be written, an exponential number of ways, as intersections between two complementary fully separable functions with the same secret string where at least half of the values in an individual function correspond to a separable function.

We already know that there are an exponential number of such separable core functions that correspond to linear independent equations embedded within any possible valid function in any possible Simon problem.
We also know any possible function can be written in k-maps or truth tables as deviations from such a separable core function such that as least half of its values align with one of an exponential number of such separable functions.
As such, the method shown below to find such separable functions will do so in $O(n)$ execution steps since all operations are in reality manipulating an exponential amount of data within the functions themselves.

It is these facts that allow the method presented in the following sections to provide the necessary exponential speedup over the conventional classical approach to solving the Simon problem where the conventional classical approach has $O(2^{n-1})$ executions steps.
What is also important to note is that it was only necessary to resort to simple Boolean algebraic mathematical principles to arrive at a possible means to solve the Simon problem as efficiently as in a true quantum computer.
It will also be seen in the following sections, that the solution is found resorting only to purely asynchronous switching methods in classical reversible logic circuits.
As such the approach presented here involves the essential classical aspects of what must be occurring in any true quantum computer to solve this class of problems.
The equi-potential wires within the classical circuits discussed are analogies to the interlocking and configurable equi-electro-chemical potential surfaces within and between quantum systems such as atoms in a true quantum computer. 
These surfaces or paths are further influenced by random thermodynamic fluctuations at the quantum level in the quantum systems in an analogous manner as in the wires and transistors being used in the classical circuits.
Another feature of these equi-electro-chemical potential logical paths are that both the logic low and logic high paths are interlaced within one another weaving through separate yet interlocking paths through the gates or qubits forming the logical functions.
This combined with the fact that they are bathed in random thermodynamic fluctuations at a particular temperature allows these systems to exist simultaneously and non-locally in more than one state at the same time.
The quantum qubits within true quantum computers, and the classical circuits in the classical quantum computer, are both being influenced by electro-chemical potential thermal Fermi gradients that drive the systems under local and global feedback to a self-consistent solution that can be interpreted as Hadamard transforms in both the quantum and classical systems.

\begin{figure}
\begin{center}
\includegraphics{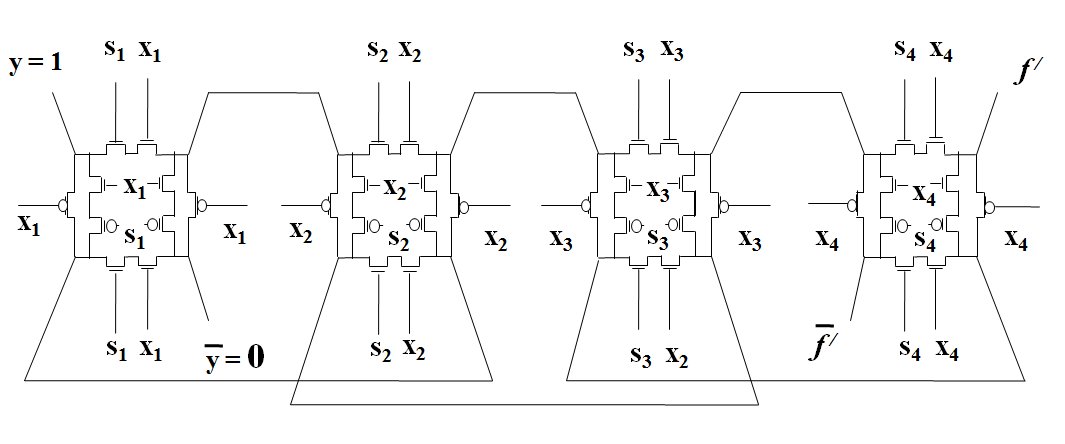}
\end{center}
\caption{Configurable or programmable fully separable core function $f'_i$ comprised of cascaded Toffoli gates that can be seen a controlled CNOT gates. \label{fig:fp_cct}}
\end{figure}

Figure \ref{fig:fp_cct} depicts a programmable fully separable function.
The Boolean inputs to the function are $x_i$ and the controls are $s_i$, respectively.
Each sub-gate is a controlled-not (CNOT) which is also known as an exclusive-or (XOR) gate implemented using the approach of Vos \cite{devos:textbook:2000} that were originally designed as a means to implement adiabatic reversible CMOS logic gates for classical cryptography purposes.

It should be emphasized that Vos did not design these circuits for use to implement quantum computer algorithms, but to perform classical operations only.
It is one of the results of this paper that circuits such as these can be used to implement what was originally believed to be a non-classical gate or transform, the Hadamard transform, with the same efficiency as in a true quantum computer.

Each of these individual CNOT gates can be thought of a synthetic qubits, and will be referred to here simply as qubits on occasion.
Each of these individual logic gates representing one qubit is also a Toffoli gate implemented using the Vos approach.

Placing the separate Toffoli circuits together in the manner shown in Figure \ref{fig:fp_cct} these circuits can implement separable functions that are also known as EVEN or ODD Boolean functions.
Here $f_i$ will be used to represent an actual function in the Simon problem that will only ever be represented partially by discrete data being entered into the machine. 
The functions $f'_i$ will represent fully separable functions as programmed through iterations of the classical version of the algorithm being shown here, into a reversible logic circuit of the general form in Figure \ref{fig:fp_cct} that will be part of the actual unknown function $f_i$ when the algorithm finds the proper form of $f'_i$. 

This type of function has the form:
\begin{eqnarray}
\label{eq:fp}
f'_i(x) & = &  ( s_1 x_1 ) \oplus  ( s_2 x_2 )  \oplus  ( s_3 x_3 )  \oplus ... ( s_n x_n ) \\ \nonumber
& = &  q_1 \oplus q_2 \oplus q_3 \oplus ... q_n 
\end{eqnarray}
If $s_i = 1$ within an individual qubit, then the qubit implements the basic function $q_i = x_i$ which is a balanced function.
If $s_i = 0$ then the function is the constant function $q_i = 0$.
For particular values of $s = (s_1, s_2, s_3 ..., s_n)$ a particular function also forms an equation in $s_i$, the solution of which can be the secret string when combined with $n-2$ other such linearly independent equations that come from the other equivalent separable functions in the problem.

The goal of solving the Simon problem will be to find the equivalent set of $n$ separable functions $f' = (f'_1, f'_2, f'_3, ... f'_n)$ of the form in equation (\ref{eq:fp}) that have the same secret string as the actual functions formed by a discrete set of data.  
Once these functions are found, they also become a set a linearly independent equations in $s_i$ the solution to which is the secret string $s = (s_1, s_2, s_3 ..., s_n)$.

It is now possible to more formally estimate the order of execution using more quantitative arguments.
It is a matter of fitting $n$ data points corresponding to $n$ pairs of $x$ and $f(x)$ values from the Simon problem being solved to $n$ fully separable functions that also happen to have the same secret string. 
When the proper fully separable functions $f'_i$ have been found they also correspond to a set of linearly independent equations in $s$ from which the secret can be found as the solution setting each function output value to a logic $0$ or logic $1$.
In general complements need to be considered where the final value of $y_i$ for the particular separable function circuit can be used.
Whether or not to use the complement equation arises naturally from the final values of $y_i$ that occur after the final iteration. 
If the final $y_i$ value has changed from its original setting at the beginning of the iteration procedure, then one must use the complement of the final function or use the uncomplemented variable version but setting its output to logic $1$ instead of logic $0$ to conduct the Gaussian elimination procedure.

Hence, the probability of obtaining a set of suitable fully separable functions $f'_i$ with the same secret string as the actual ones in the Simon problem to be solved is identical to the probability of obtaining a set of linearly independent equations in $s$ in the true quantum computer algorithm already described in section \ref{sec:simon_oracle}.

The probability of obtaining a suitable set of fully separable functions can be estimated as follows.
The circuits in Figure \ref{fig:fp_cct} represent any possible fully separable function with $n$ inputs, where $n=4$ is used in these examples.
Any unique set of $n$ $x$-$f_i(x)$ values used to determine the settings of the control lines $s_i$ including the settings of $y_i$ and $\overline{y_i}$, will result in a unique fully separable function $f'_i$.
We also know that any arbitrary function $f_i$ can be written in terms of any arbitrary fully separable functions where at least half of its values correspond or align with actual function.
As such we are free to choose any fully separable function to express an arbitrary function this way.

Say for the first fully separable function $f'_1$ we will nearly always have a suitable fully separable function to use where we will have a probability of $(2^n - 2^0)/ 2^n = (2^n - 1)/2^n =$ nearly unity probability of obtaining it.
There are $2^n$ possible settings for $s_i$ and therefore $2^n$ possible functions we could encounter.
There is one function that we cannot use it being $f = 0$ or $f = 1$ which has the same settings for $s_i$ since a unique value of the secret string, if it had only one bit, could not determined from a setting of $s_i = 0$.

However, whatever this function might be, there is a second one that we can no longer use for the other functions beyond the first one found, since it will be linearly dependent on whatever function we obtain.
This is because having only one function or equation in $s$ will uniquely determine only one bit within the secret string if all of the other functions happen not to be linearly independent with respect to all of the input variables except one.
There is therefore a second separable function we might encounter we cannot use that would also be linearly dependent upon at least the one input variable, since each bit can only take on a logic $0$ or $1$.

Given a first function found after $n$ data elements have been encountered, we now have two less functions to choose from to obtain a second suitable function to maintain linearly independence of at least one input variable.
As such the probability of encountering it becomes $(2^n - 2^1)/2^n$.
However, there now exist two other functions that we can no longer use that will become linearly dependent upon either the first or second functions we found with respect to at most two input variables corresponding to at most two bits in the secret string. 
This is because there are four possible combinations of logic values that these two bits in the secret string can take represented by those particular four functions, two of which we have found whose unique combinations will determine the two bits in question.

Using this pattern, we see that with each additional function $f'_{(k+1)}$ there are $2^k$ functions that we can no longer use since they will be linearly dependent upon the previous $f'_k$ suitable functions we did happen to find.
Another way to put this is that for each new additional suitable fully separable function we halve the number from which we can obtain new potential fully separable functions for a given secret string value.

As such the probability of finding a suitable function becomes $(2^n - 2^k)/2^n$ for each function $f'_{(k+1)}$ from $k=0$ to $k = n-1$.
The overall probability of encountering a set of $n$ suitable fully separable functions that will lead to a set of linaerly independent equations in $s$ then becomes the product of these individual probabilities as the iteration proceeds in parallel with all $n$ functions being found simultaneously in the two dimensional $f'_i$ circuit array.
The product then becomes,

\begin{eqnarray}
\left(\frac{2^n - 2^0}{2^n} \right) \; \left(\frac{2^n - 2^1}{2^n} \right) \; \;  \dots\; \;  \left(\frac{2^n - 2^{(n-1)}}{2^n} \right) \\ \nonumber
 =  \left(1 - \frac{1}{2^n} \right) \; \left( 1- \frac{2}{2^n} \right) \; \;  \dots\; \;  \left(1- \frac{1}{2} \right)
\label{eq:prob:2}
\end{eqnarray}
The lower bound of this product becomes as $n$ goes to infinity,
\begin{eqnarray}
\prod^{\infty}_{k=1} \left( 1 - \frac{1}{2^k} \right) \approx 0.28879
\label{eq:prob:3}
\end{eqnarray}
Hence we obtain precisely the same probability of obtaining a suitable set of fully separable functions in order $n$ execution steps, as in a true quantum computer for obtaining a set of linearly independent equations but using a purely classical means with the same hardware and execution efficiency as in a true quantum computer.

Hence, the probability of obtaining a suitable equivalent set of linearly independent equations is about 30\% for each $n$ sequence of data encountered.
Hence, one might expect that a suitable set would be found when order $3n$ data values have been entered into the circuitry as per the true quantum computer algorithm.

Another consideration in fitting function circuits to valid separable function sets, is to understand that it is necessary to determine the influence of each input variable $x_i$ to see how it may influence the configuration of each Toffoli sub-circuit in each separable function circuit .
As such, although in principle random data can be entered, it would be wise to consider random data but constrained in a way that ensures that each $x_i$ value changes an equal number of times to cover the truth table for each function in the Simon problem in a way to ensure that each qubit Toffoli sub-circuit is properly configured.
In the example to follow solving the Simon problem, the data is restricted to just allowing one $x_i$ to change or toggle at a time from data value to data value being entered into the circuit network to configure them to a set of valid fully separable functions with the same secret string.
Also, if a particular $x_i$ changes, a rule is used to determine if the corresponding $s_i$ control value should change the configuration of the particular $i^{th}$ Toffoli qubit gate in the circuit.

Random values of data may be selected relaxing the requirement that only one $x_i$ change at a time, if the following is done to ensure consistency between the configuration of a function and the latest data elements being entered. 
If after entering a new $x$-$f_i(x)$ pair into a particular function $f'_i$ results in a toggling of the $y_i$ line for that function, then change the parity (from ODD to EVEN or EVEN to ODD) of the $s_i$ lines being at logic $1$ that also corresponding to the $x_i$ values that changed compared to the previous $x_{i-1}$ data values. 
It can be seen that this is a generalization of the above described specific rule for having only one $x_i$ changing per new data value where the number of $s_i$ values corresponding to the $x_i$ value being changed is only one and is EVEN if a logic $0$ and ODD if a logic $1$. 
Hence changing its parity if the corresponding $y_i$ toggles amounts to toggling that specific value of $s_i$. 
It is also preferable, when randomly selecting data values from the problem being entered, that on average all $x_i$ values have changed at least once per $n$ data elements being entered and that on average all $s_i$ values have been updated per $n$ data elements being entered.  
This will ensure that any functions with long strings of one logic level will not interfere with the convergence process.
This is something that would be expected to occur naturally if true randomness (or a good pseudo-random method) were being used in the selection of the data values being entered.  

In order to demonstrate how to solve the Simon problem using asynchronous methods in reversible transistor based logic gates, it is best to use an example.
The example will involve functions with four inputs, $x = (x_1, x_2, x_3, x_4)$ and four functions $f = (f_1, f_2, f_3, f_4)$ that are 2:1 functions known to have a secret string $s = (s_1, s_2, s_3, s_4)$.

The first procedure in the example will be to demonstrate how an equivalent set of separable functions of the form in equation (\ref{eq:fp}) can be found, which is the focus of this section.
The second procedure will show how the secret string can be extracted efficiently without using Gaussian elimination by extending the same asynchronous techniques used throughout.
It will be understood at this time that this second procedure would normally be performed after each iteration of the first procedure to ascertain when to stop iterating in the first procedure.
It will be seen that both procedures have $O(n)$ execution steps.
Combining them one would execute the second procedure that takes $O(n)$ steps with each step in the first procedure for a total of $O(n^2)$ execution steps.

This happens to be identical in order to the best known previous solution in classical electronic circuits using Gottesman-Knill theory \cite{aaronson:v70:04} to simulate the Simon problem in a conventional computer with one important distinction.
Since the Hadamard transform portions of the algorithm cannot implemented in a true simultaneous fashion using Gottesman-Knill theory, using this method results in another embedded $O(n^2)$ slowdown for the Simon problem compared to using the approach in this work.
This degree of slowdown can be considerable if the number of qubits in the problem reaches useful ranges in the thousands to tens of thousands, where the degree of slowdown will scale as the square of these numbers.

\begin{figure}
\begin{center}
\includegraphics{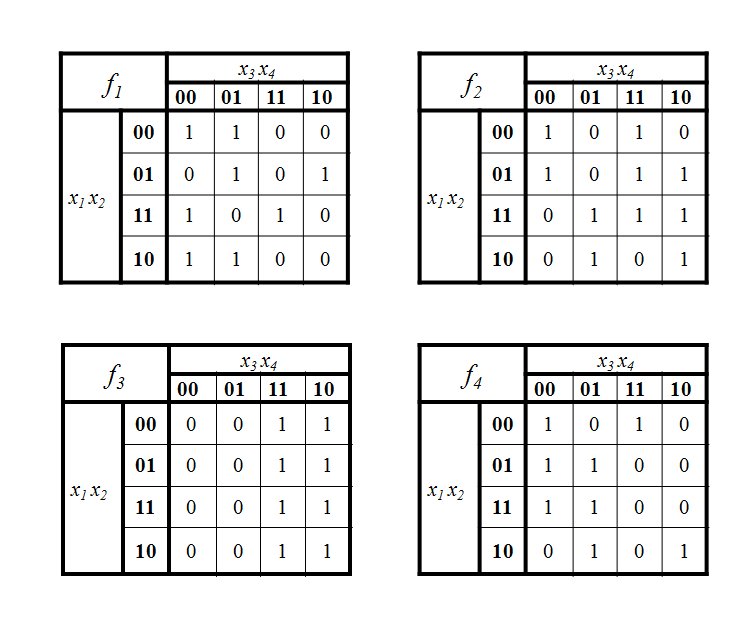}
\end{center}
\caption{K-Maps of the Simon problem functions for the example being considered. \label{fig:kmap_f_ex_1}}
\end{figure}
\begin{figure}
\begin{center}
\includegraphics{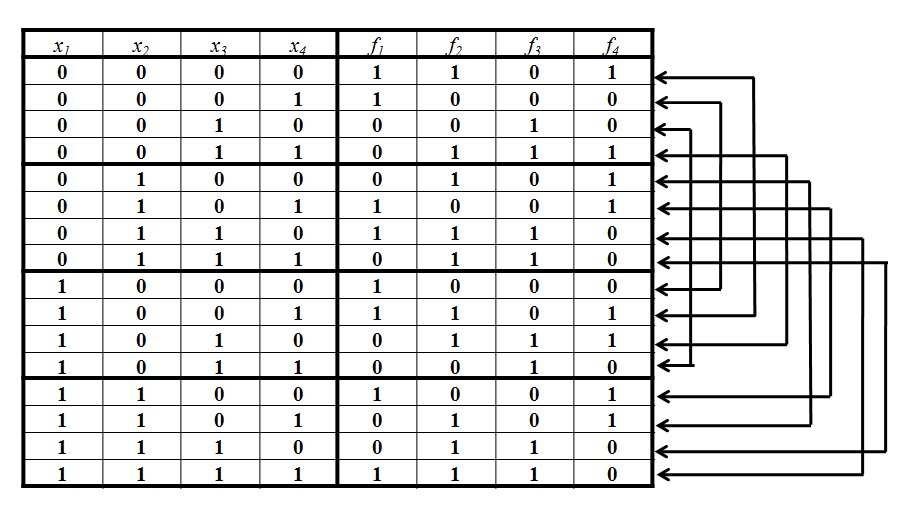}
\end{center}
\caption{Truth tables of the Simon problem functions for the example being considered. \label{fig:t_table_f_ex_1}}
\end{figure}
\begin{figure}
\begin{center}
\includegraphics{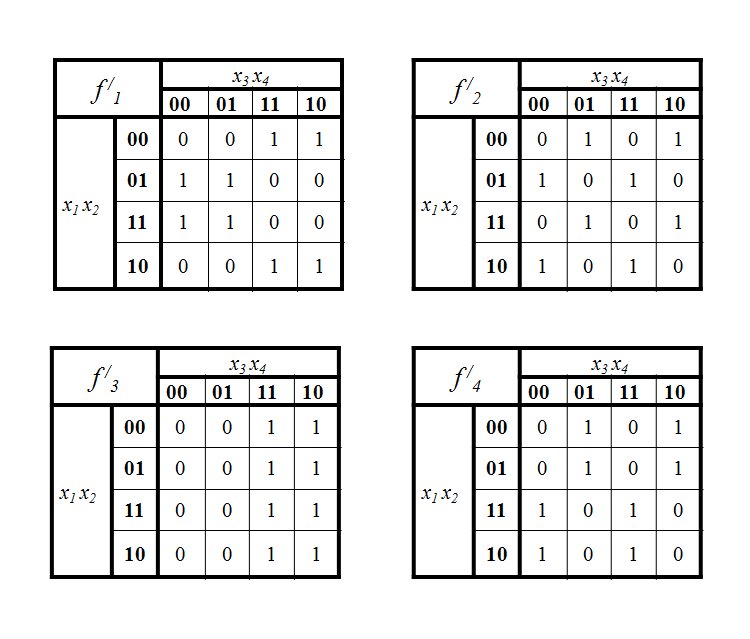}
\end{center}
\caption{K-Maps of the equivalent separable functions for the Simon problem for the example being considered.}\label{fig:kmap_fp_ex_1}
\end{figure}
\begin{figure}
\begin{center}
\includegraphics{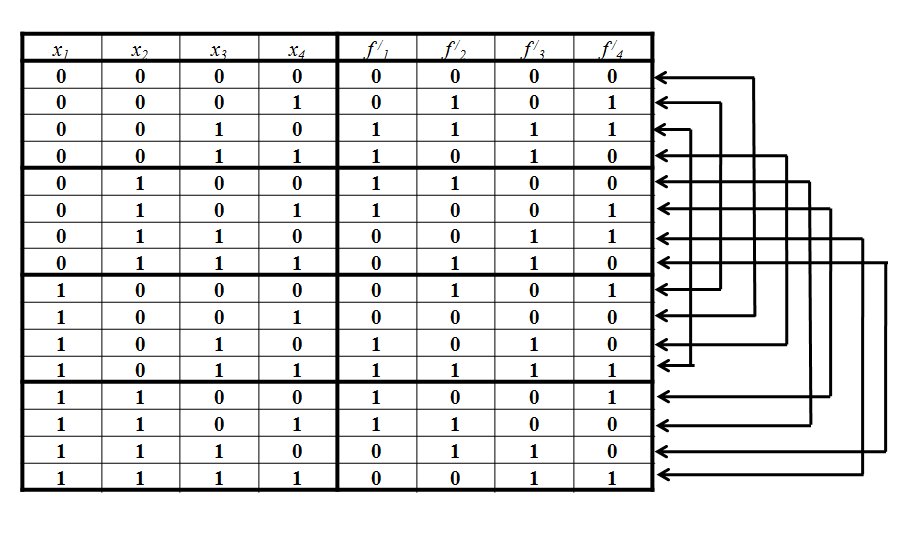}
\end{center}
\caption{Truth tables for the equivalent separable functions for the Simon problem for the example being considered.}\label{fig:t_table_fp_ex_1}
\end{figure}

Figures \ref{fig:kmap_f_ex_1} and \ref{fig:t_table_f_ex_1} show the k-maps and truth tables for four functions of a particular example of the Simon problem that will be solved for the secret string, that happens to be $s = 1001$.
Figures \ref{fig:kmap_fp_ex_1} and \ref{fig:t_table_fp_ex_1} show the k-maps and truth tables for the four separable functions that will be found using the method described below that have the same secret string. 

\begin{figure}
\begin{center}
\includegraphics{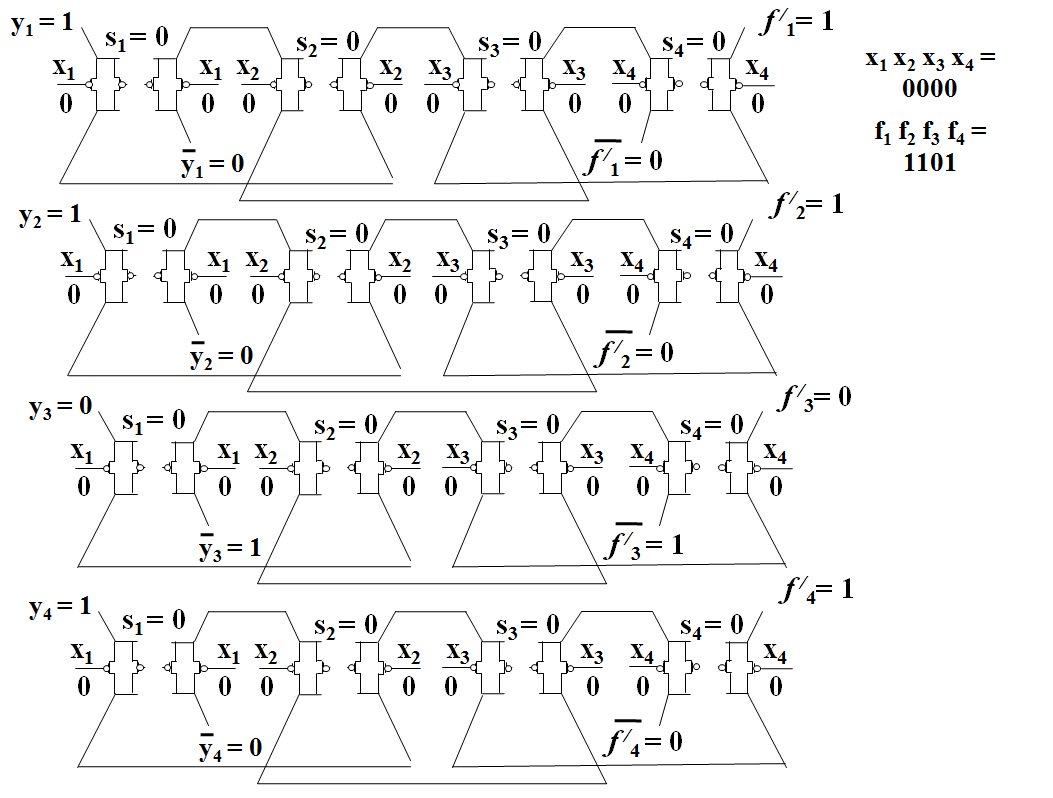}
\end{center}
\caption{Set $x$ vector to $0000$. Set $y$ values for each circuit such that outputs are the proper values of $f = 1101$ for $x = 0000$.  \label{fig:simon_sim:1}}
\end{figure}

\begin{figure}
\begin{center}
\includegraphics{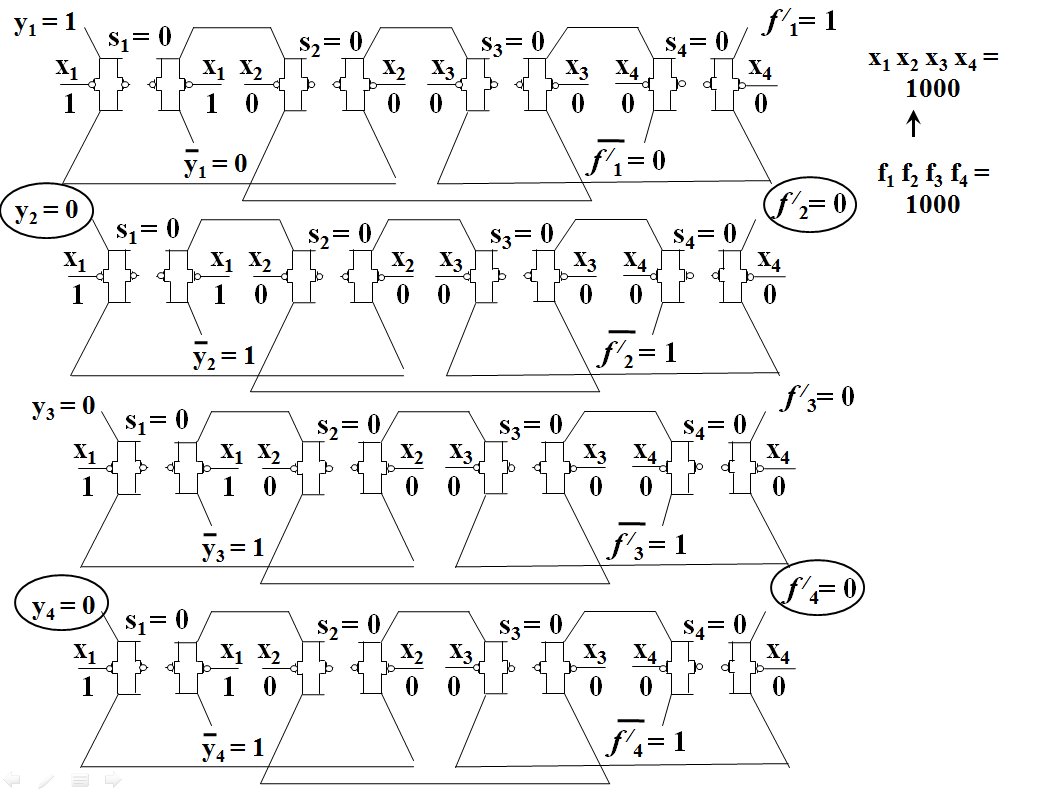}
\end{center}
\caption{$x = 1000$ and $f = 1000$. $y_2$ and $y_4$ toggle.\label{fig:simon_sim:2}}
\end{figure}

\begin{figure}
\begin{center}
\includegraphics{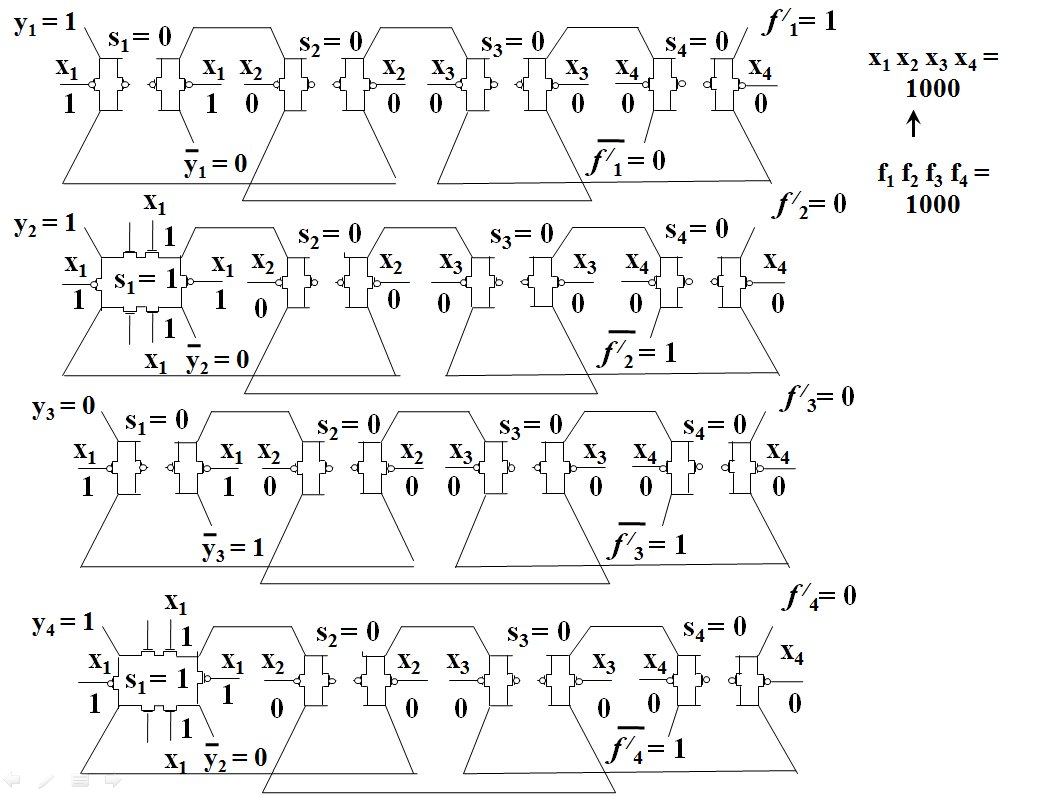}
\end{center}
\caption{Toggle $s_1$ values for $x_1$ qubits in $f'_2$ and $f'_4$ function circuits.\label{fig:simon_sim:3}}
\end{figure}

\begin{figure}
\begin{center}
\includegraphics{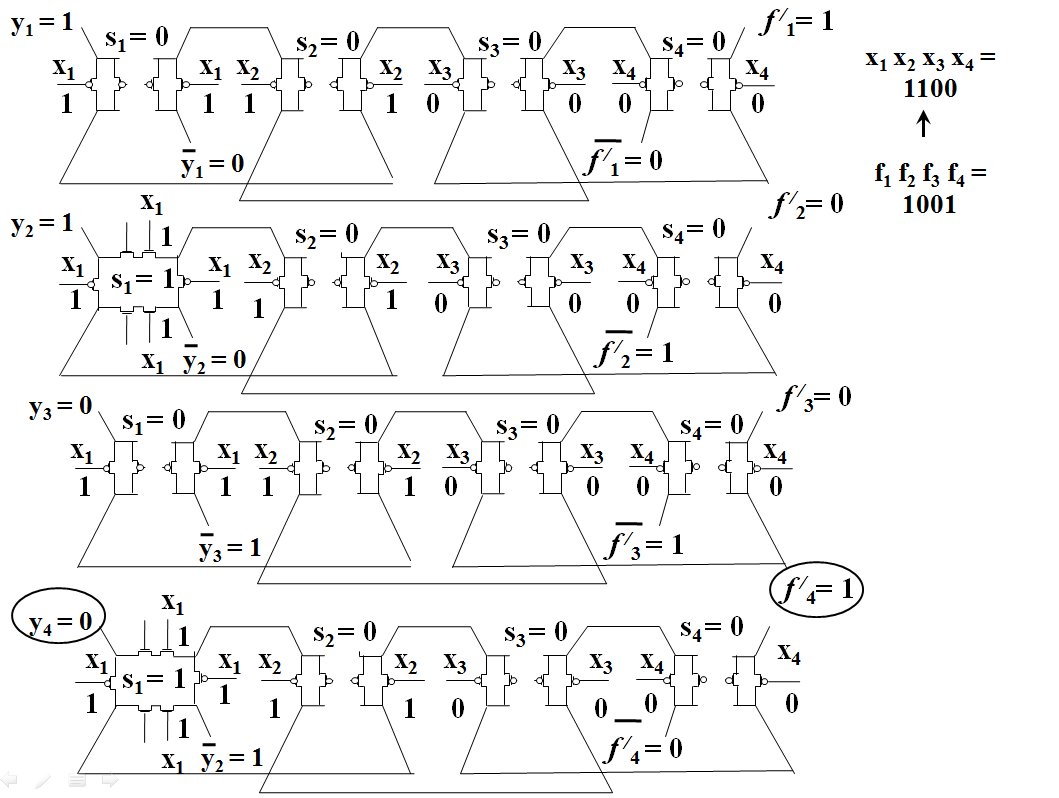}
\end{center}
\caption{$x = 1100$ and $f = 1001$. $y_4$ toggles.\label{fig:simon_sim:4}}
\end{figure}

\begin{figure}
\begin{center}
\includegraphics{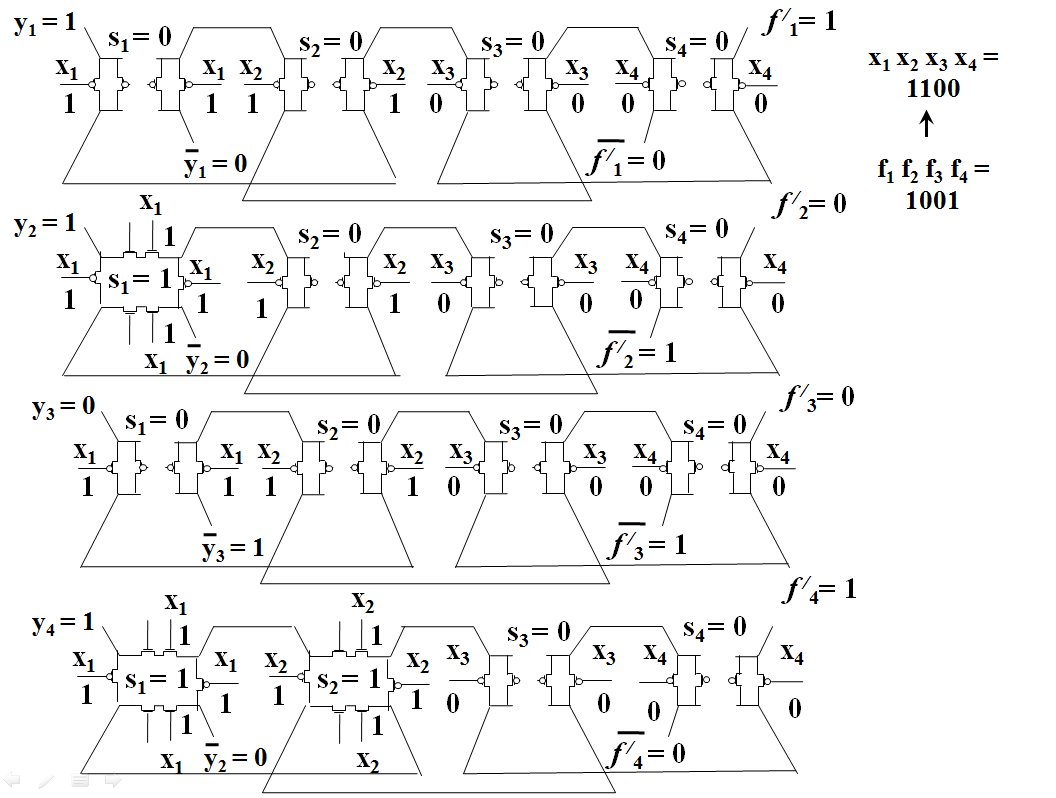}
\end{center}
\caption{Toggle $s_2$ value for $x_2$ qubit in $f'_4$ function circuit.\label{fig:simon_sim:5}}
\end{figure}

\begin{figure}
\begin{center}
\includegraphics{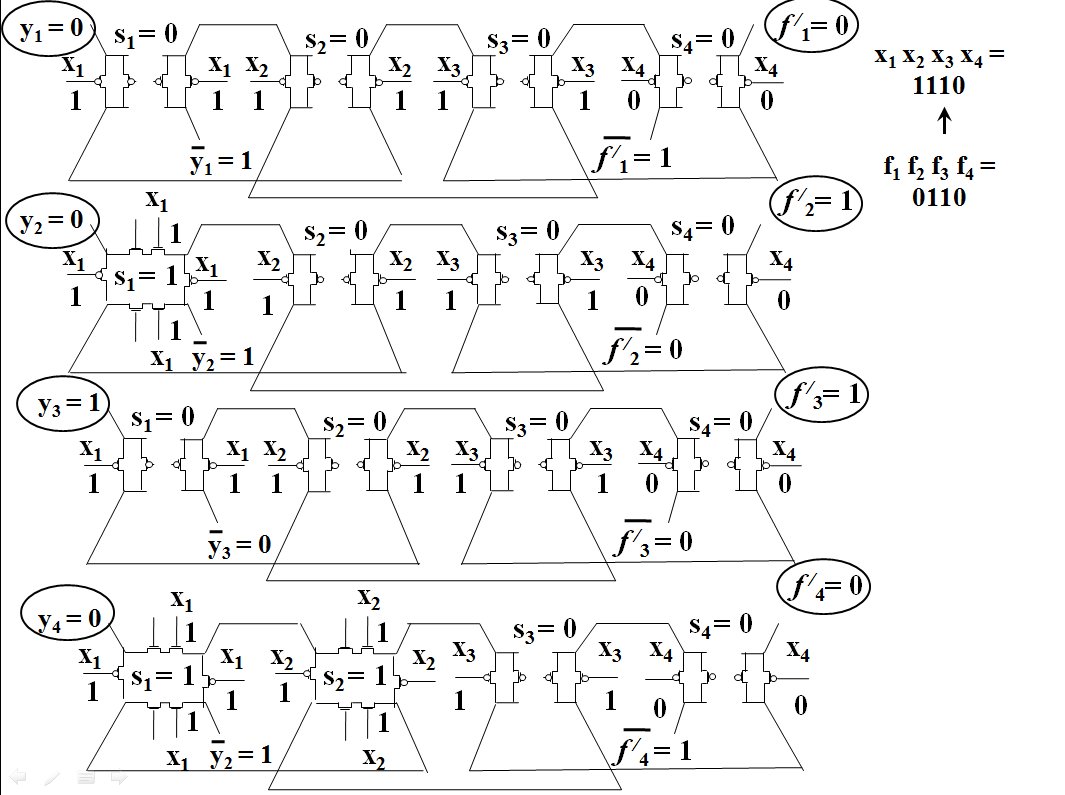}
\end{center}
\caption{$x = 1110$ and $f = 0110$. $y_1$, $y_2$, $y_3$, and $y_4$ toggle.\label{fig:simon_sim:6}}
\end{figure}

\begin{figure}
\begin{center}
\includegraphics{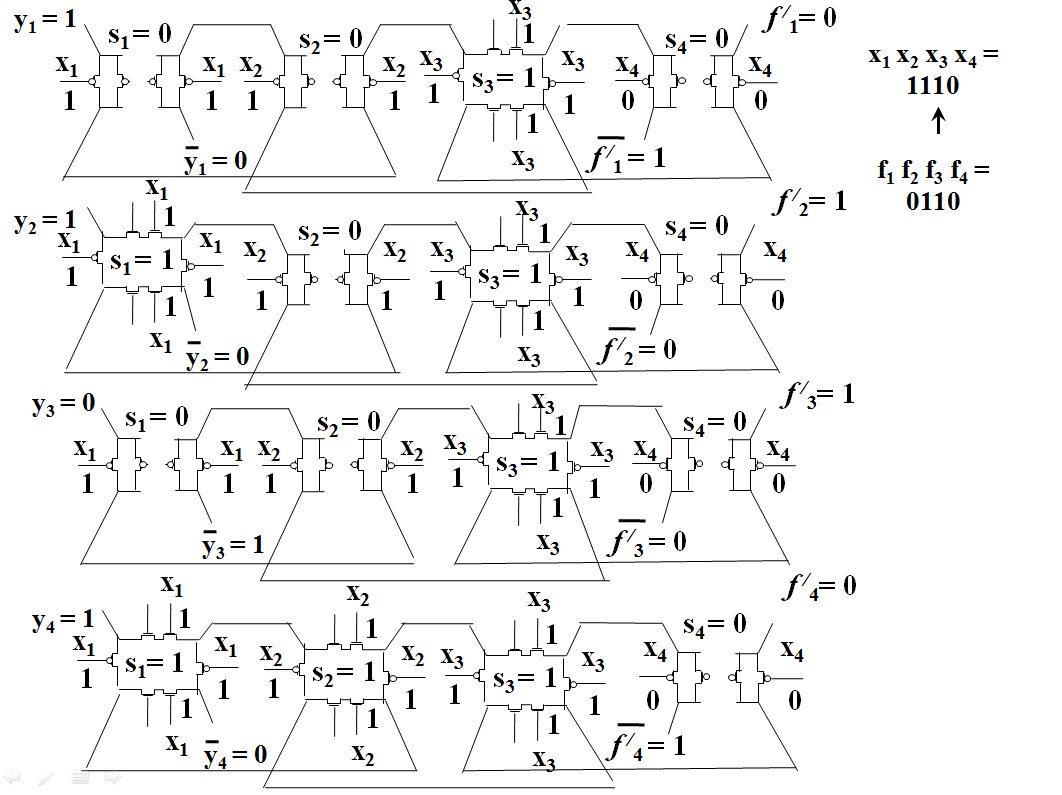}
\end{center}
\caption{Toggle $s_3$ values for $x_3$ qubits in $f'_1$, $f'_2$, $f'_3$, and $f'_4$ function circuits.\label{fig:simon_sim:7}}
\end{figure}

\begin{figure}
\begin{center}
\includegraphics{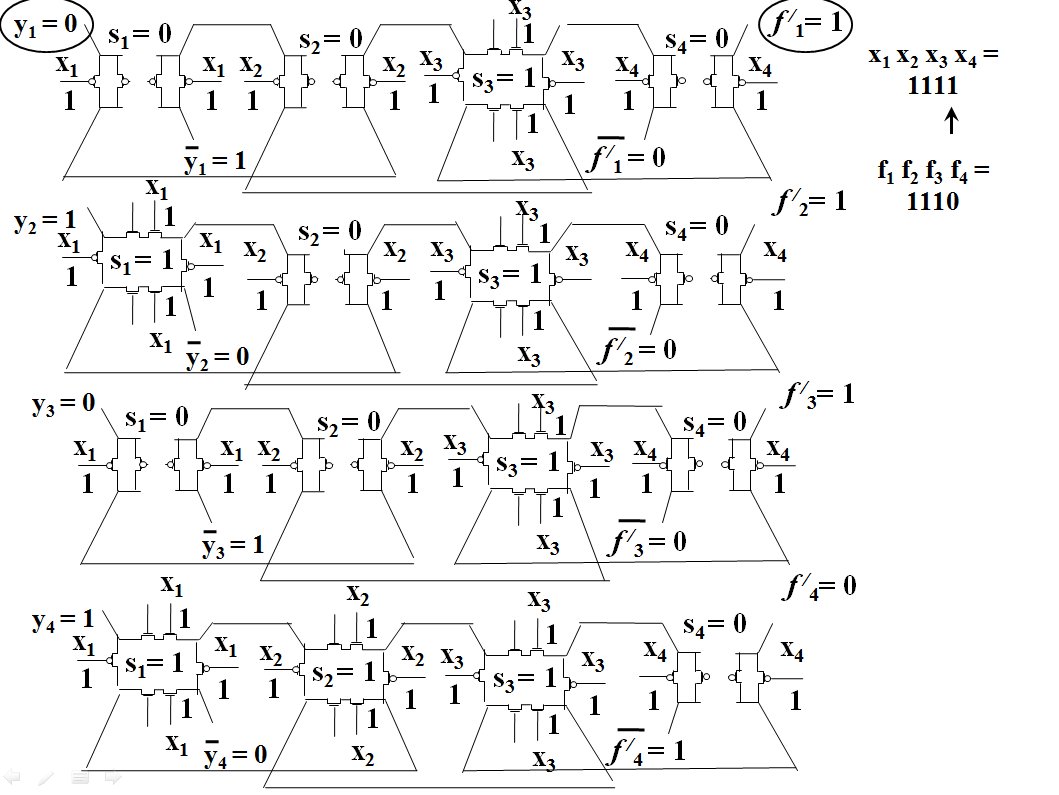}
\end{center}
\caption{$x = 1111$ and $f = 1110$. $y_1$ toggles.\label{fig:simon_sim:8}}
\end{figure}

\begin{figure}
\begin{center}
\includegraphics{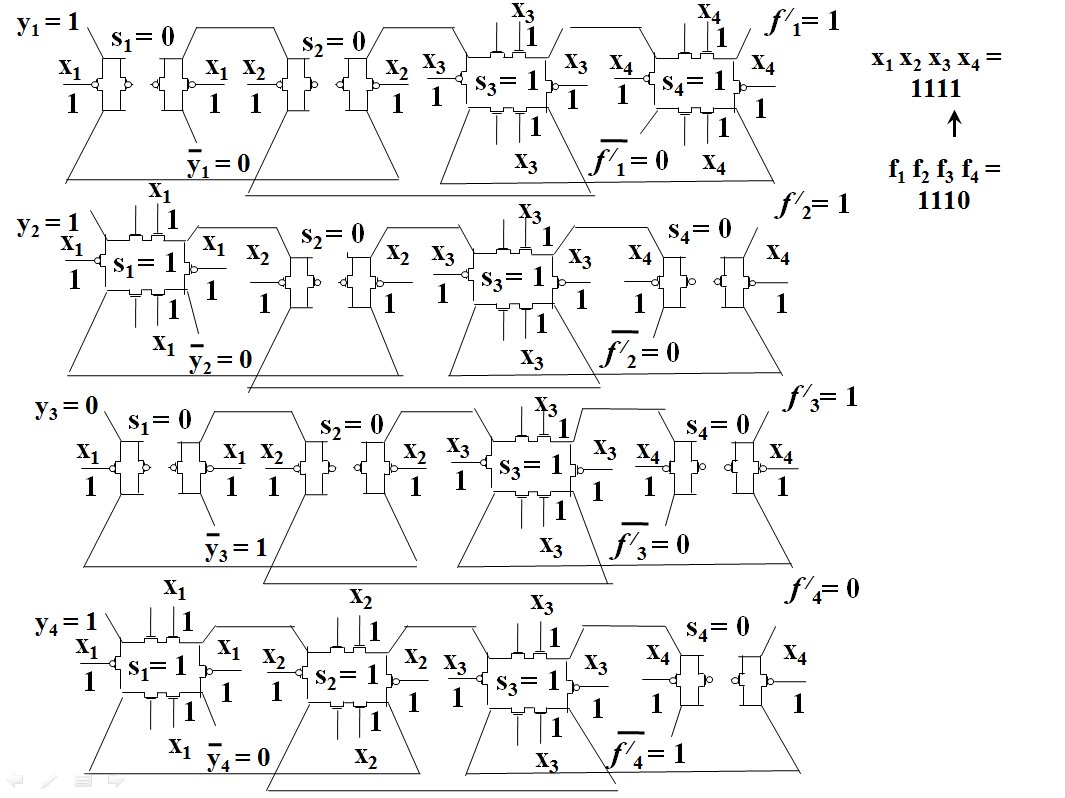}
\end{center}
\caption{Toggle $s_4$ value for $x_4$ qubit in $f'_1$ function circuit.\label{fig:simon_sim:9}}
\end{figure}

\begin{figure}
\begin{center}
\includegraphics{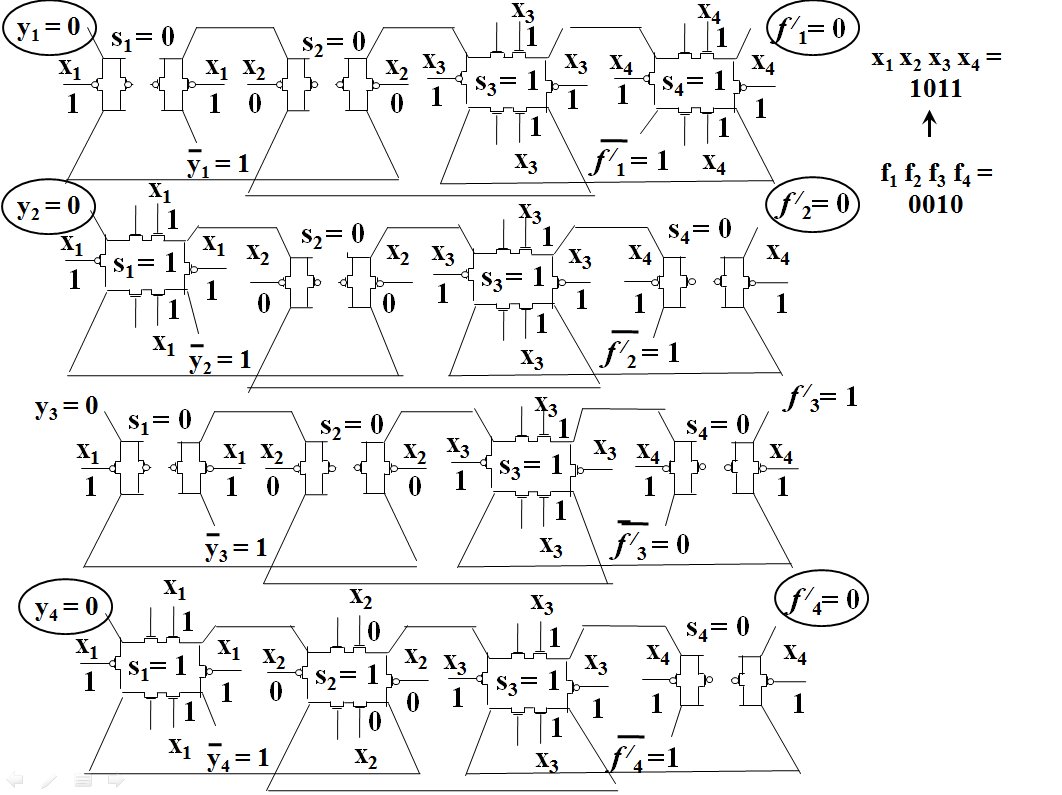}
\end{center}
\caption{$x = 1011$ and $f = 0010$. $y_1$, $y_2$, and $y_4$ toggle.\label{fig:simon_sim:10}}
\end{figure}

\begin{figure}
\begin{center}
\includegraphics{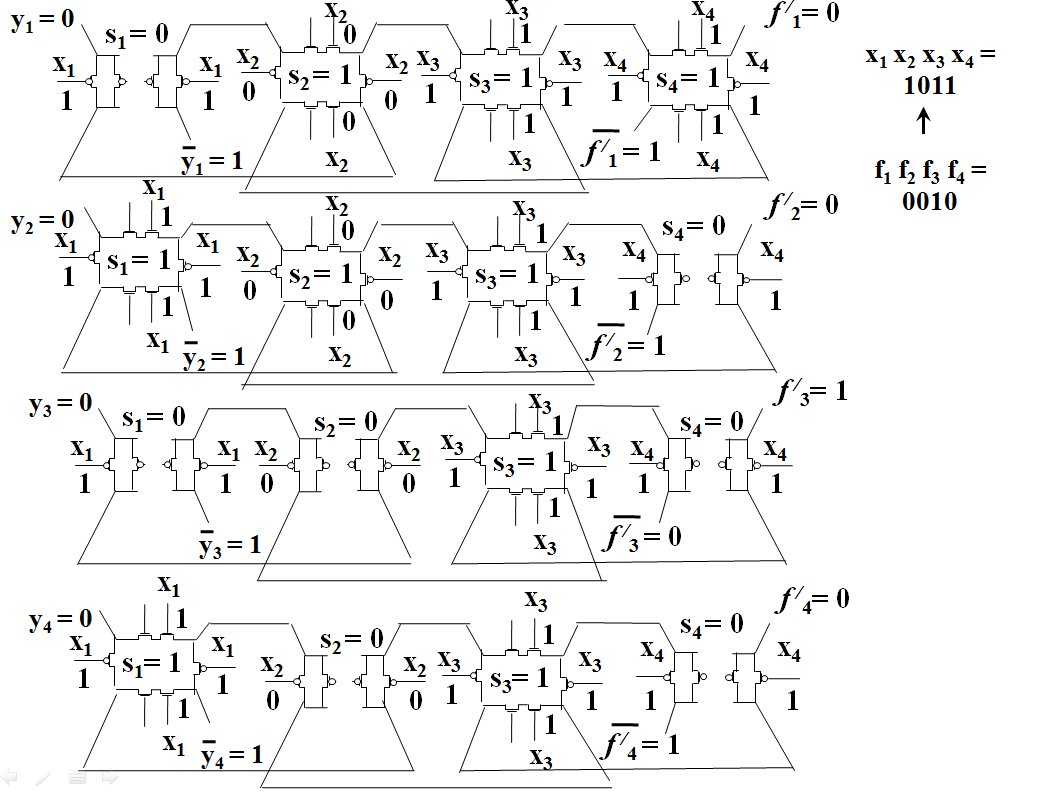}
\end{center}
\caption{Toggle $s_2$ values for $x_2$ qubits for $f'_1$, $f'_2$, and $f'_4$ function circuits.\label{fig:simon_sim:11}}
\end{figure}

\begin{figure}
\begin{center}
\includegraphics{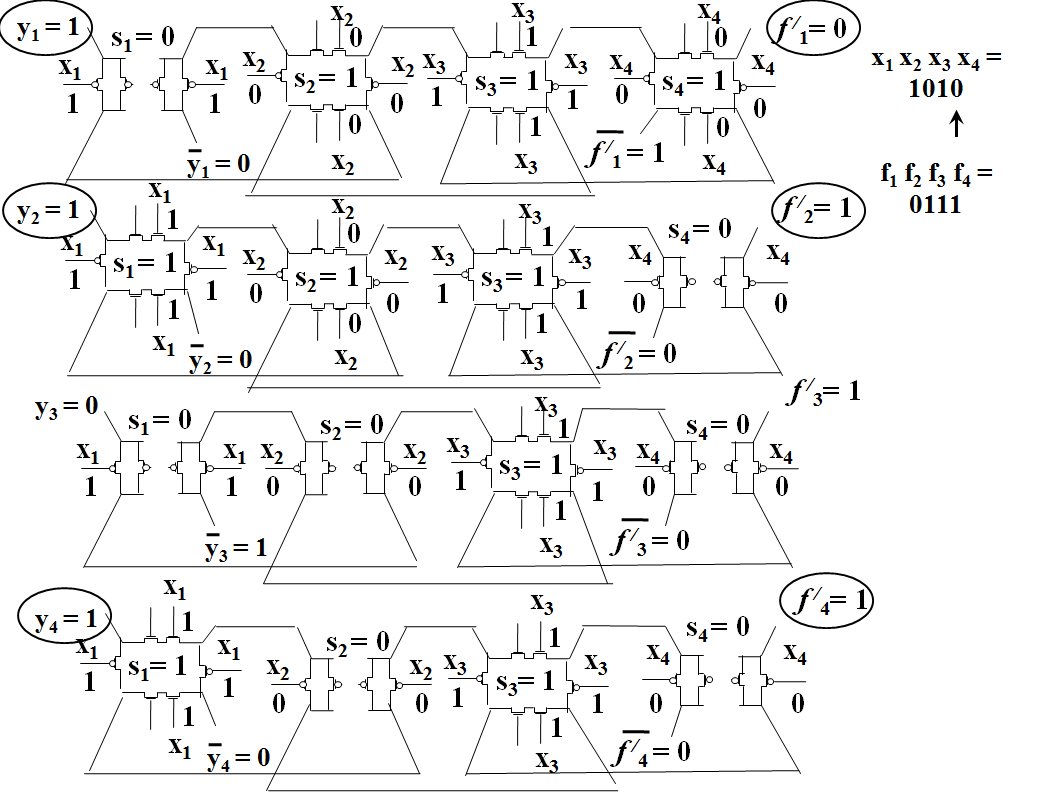}
\end{center}
\caption{$x = 1010$ and $f = 0111$. $y_1$, $y_2$, and $y_4$ toggle.\label{fig:simon_sim:12}}
\end{figure}

\begin{figure}
\begin{center}
\includegraphics{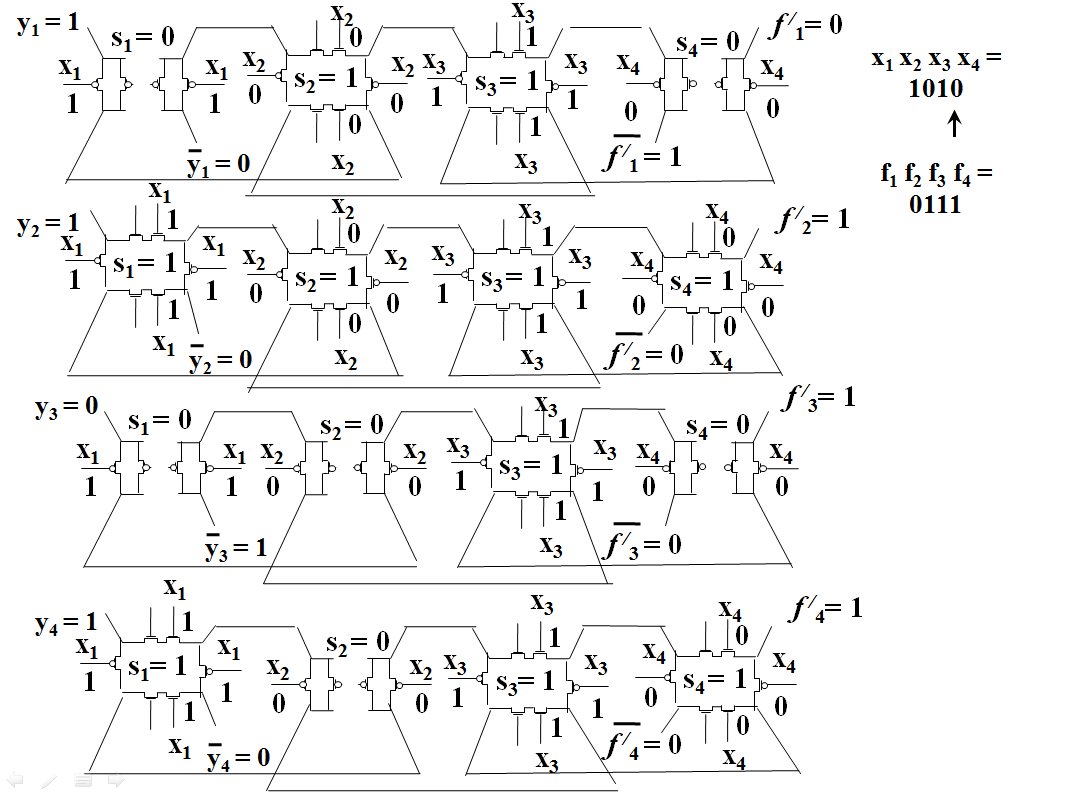}
\end{center}
\caption{Toggle $s_4$ values for $x_4$ qubits for $f'_1$, $f'_2$, and $f'_4$ function circuits. This is the last iteration required as the functions formed by the circuits are a valid set of separable functions with the same secret string as the actual ones. \label{fig:simon_sim:13}}
\end{figure}

Figures \ref{fig:simon_sim:1} through \ref{fig:simon_sim:13} depict how a machine consisting of $n$ circuits of the form in Figure \ref{fig:fp_cct} can be constructed and then used to find the equivalent separable functions for any Simon problem with four functions and four inputs or qubits.
Obviously the system can be scaled to any practical size.

First the algorithm to iterate to a correct set of separable functions $f_i$ will be described, followed by a specific example to clarify.
All input values $x_i$ and $s_i$ values are set to zero in all circuits for $f'_i$.
Then the $y_i$ (and the $\overline{y_i}$) values for each separable function circuit is set to give the correct output $f_i$ for $x = 0$ vector from the available data for the problem.
Then each iteration consists of randomly selecting new $x$ and $f$ values from the problem itself and imposing them on the $x_i$ inputs and $f'_i$ outputs of the circuits as depicted. 
Since the circuits are electrically reversible, placing data on the so-called function outputs $f'_i$ influences the $y_i$ (and the $\overline{y_i}$) values for a given set of $x_i$ and $s_i$ values placed on the particular function circuit.
If the data is consistent with the existing circuit as it already exists for the particular existing values of the control signals $s_i$, there will be no change in the $y_i$ (and the $\overline{y_i}$) values for that circuit.
If the data being imposed on the circuit is not consistent with the existing form of the circuit, then the values of $y_i$ (and the $\overline{y_i}$) will toggle or change for the particular circuit.

If $y_i$ (and the $\overline{y_i}$) changes or toggles as a result of new $x_i$ and $f_i$ values being placed onto that particular circuit, the $s_i$ value is toggled or changed for the particular qubit circuit or Toffoli gate for which an $x_i$ value also changed compared to the previous data set.
This may or may not toggle the $y_i$ (and the $\overline{y_i}$) value, but it does not matter.
One can arrange to select $x_i$ and corresponding $f_i$ data values for each function randomly such that only one $x_i$ value changes or toggles per data set.
Alternatively, one can allow any number of $x_i$ values to change when randomly selecting data changing only one of the $s_i$ values for one of the qubit circuits for which one of the $x_i$ values changed if the $y_i$ (and the $\overline{y_i}$) value changed.
Which $s_i$ value to change could also be random but constrained only to a qubit where an $x_i$ value changed.

This process is identical to fitting $n$ data elements to a fully separable function.
Regardless of what $n$ data values are encountered it will always be possible to fit them to a unique fully separable function since both the function and that many data values have $n$ degrees of freedom.
Given that there are an exponential number of ways in which at least half of the data within any possible function in a Simon problem will be a suitable fully separable function, the probability of encountering $n$ data elements in $O(n)$ iterations that will form a suitable fully separable core function that corresponds to a valid independent equation for the secret string is quite high.

Only changes are detected allowing automatically for the possibility of complements of the separable core functions to be included in the search.
For each data set encountered, any new logical functions that result represent new possible equations from which to determine the secret string.

What follows is a particular example to help visualize the process of finding a secret string in the Simon problem given a set of functional data.
It is not necessary to have the complete set of data that would require an exponential amount of storage space.
It is only necessary to have $O(n)$ data elements and it is not necessary to have function values that happen to be identical.

Figures \ref{fig:simon_sim:1} through \ref{fig:simon_sim:13} can be followed through from the original zero input state to a final state that corresponds to a valid set of separable functions that form a linearly independent set of equations in the unknown secret string $s = 1001$.
The Figure captions of each describe the specific actions that are taken. 
Input data is being taken from the truth table of Figure \ref{fig:t_table_f_ex_1} that represents the Simon problem from which to determine the secret string.
Also indicated on this figure are the pairs of inputs that, when XOR'd with one another, result in the same secret string.

It happens in this example that a correct set of separable functions is not reached until the last iteration shown in Figure \ref{fig:simon_sim:13}.
Here the final separable functions are:
\begin{eqnarray}
f'_1 (x) & = & x_2 \oplus x_3 \\ \nonumber
f'_2 (x) & = & x_1 \oplus x_2 \oplus x_3 \oplus x_4 \\ \nonumber
f'_3 (x) & = & x_3 \\  \nonumber
f'_4 (x) & = & x_1 \oplus x_3 \oplus x_4
\label{eq:fps1}
\end{eqnarray}
Using the fact that $f'_i (x) = f'_i (x \oplus s)$, where $x \oplus s = (x_1 \oplus s_1, x_2 \oplus s_2, ..., x_n \oplus s_n)$, using the above functions for each $f'_i$, one can write, 
\begin{eqnarray}
x_2 \oplus x_3 & = & (x_2 \oplus s_2) \oplus (x_3 \oplus s_3) \\ \nonumber
x_1 \oplus x_2 \oplus x_3 \oplus x_4 & = & (x_1 \oplus s_1) \oplus (x_2 \oplus s_2) \\ \nonumber
& & \oplus \; \; (x_3 \oplus s_3) \oplus (x_4 \oplus s_4) \\ \nonumber
x_3 & = & (x_3 \oplus s_3) \\  \nonumber
x_1 \oplus x_3 \oplus x_4 & = & (x_1 \oplus s_1) \oplus (x_3 \oplus s_3) \\ \nonumber
& & \oplus \; \; (x_4 \oplus s_4)   
\label{eq:fps2}
\end{eqnarray}
This can then be re-written as:
\begin{eqnarray}
x_2 \oplus x_3 & = & (x_2 \oplus x_3 ) \oplus (s_2 \oplus s_3) \\ \nonumber
x_1 \oplus x_2 \oplus x_3 \oplus x_4 & = & (x_1 \oplus x_2 \oplus x_3 \oplus x_4) \\ \nonumber
& & \oplus \; \; (s_1 \oplus s_2 \oplus s_3 \oplus s_4)  \\ \nonumber
x_3 & = & x_3 \oplus s_3 \\  \nonumber
x_1 \oplus x_3 \oplus x_4 & = & (x_1 \oplus x_3 \oplus x_4) \\ \nonumber
& & \oplus \; \; (s_1 \oplus s_3 \oplus s_4)  
\label{eq:fps3}
\end{eqnarray}
Knowing that $x_i \oplus x_i = 0$,  XOR'ing $x_i$ elements out on both sides yields the final equations in $s_i$ such that,
\begin{eqnarray}
s_2 \oplus s_3 & = & 0 \\ \nonumber
s_1 \oplus s_2 \oplus s_3 \oplus s_4 & = & 0 \\ \nonumber
s_3 & = & 0 \\  \nonumber
s_1 \oplus s_3 \oplus s_4 & = & 0
\label{eq:fps4}
\end{eqnarray}
from which the only non-zero solution is $s_1 = 1$, $s_2 = 0$, $s_3 = 0$, $s_4 = 1$.

For the example given for the unknown secret string $s = 1001$, the data is chosen at random but where only one value of $x_i$ is allowed to change or toggle at a time.
Then it is easy to know which qubit $s_i$ to change it being the same qubit for which $x_i$ changed if $y_i$ also changed from one data set to the next.
As already discussed, this is not necessary but convenient and is unlikely to have any impact on average on the convergence speed of the procedure.
However, it may be that this type of data is not available requiring the more general approach already described that enables any random data order to be used.
This requires a slightly more complicated circuit arrangement to select which qubit control signals $s_i$ to alter given which $x_i$ values changed dependent upon which $y_i$ values changed in the process.
Any equivalent procedure should enable a more or less equally efficient fitting of the functions to a sequence of $n$ random data elements encountered that will arrive at a suitable set of separable functions from which to determine the secret string.

It is not necessary to solve the equations using traditional Gaussian elimination.
The concept of implementing a Hadamard using asynchronous feedback in reversible switching networks can be extended to including the entire set of $f'_i$ function circuits where all of the inputs are connected together to form on large 2-D mesh.
This mesh always exists throughout the procedure to solve the Simon problem since all of the inputs $x_i$ for each $f'_i$ circuit will be shorted to one another anyways locally within the circuit.
However, in circuit diagrams up to this time the fact that the inputs $x_i$ to each of the function circuits $f'_i$ are locally shorted to one another has not been shown explicitly.

Using this feedback technique, if the circuits form a set of linearly independent equations then the entire mesh will have only two stable input states for zero output for all $f'_i$ functions.
The first input value will be all $x_i = 0$, and the next stable state will be $x_i = s_i$, where $s_i$ will be the nonzero secret string.
Note that the secret string being all $s_i = 0$ is not an allowed value in the problem or the functions would not be 2:1 functions as required.

With each iteration it is then possible to check to see if the functions obtained form a linearly independent equating all $f'_i = 0$.
We are using all $f'_i = 0$ where since we have no need to change the equations from their original uncomplemented form since it can be seen by comparing Figures \ref{fig:simon_sim:1} and \ref{fig:simon_sim:13} that the $y_i$ values have remained unchanged after the iterations were completed.
If they were to change then the corresponding $f'_i = 1$ would be used to obtain the complement of the equations if they were obtained using the approach demonstrated in equations (\ref{eq:fps1}) to (\ref{eq:fps4}).
For each time the alternate Gaussian elimination procedure to be described were to be used in lieu of doing a traditional Gaussian elimination, after each iteration to obtain a new set of functions $f'_i$, it would have to be determined whether or not the $y_i$ had toggled or not compared to the initial state of the circuitry.
If the $y_i$ had toggled at any time after the function fitting iteration the corresponding $f_i$ value would be forced to a logic $1$ keeping the $y_i$ the same to carry out the Gaussian elimination procedure.

For a given set of function values $f'_i$, the network will have two possible solutions for $x_i$ that allow the network to be stable.
The first solution is all $x_i = 0$, and the second solution is $x_i = s_i$, where $s_i$ are the secret string values.
The procedure will then be to first set all $x_i = 0$ keeping all $f_i = 0$ at the outputs.
Then the network will be coaxed to transition to the alternative stable state such that all of the inputs $x_i$ will settle on $s_i$ values without falling back to the zero state.

This situation is analogous to two stable energy states of an atomic system.
The zero state of the circuit network where $x_i = 0$ and $f_i = 0$, is similar to the ground state of an atom. 
The state of the circuit network $x_i = s_i$ and $f_i = 0$ is similar to the next higher energy state of an atom.

To coax the circuit network into the next higher state, one way to do this would be to place a single logic high (logic $1$), on one of the inputs $x_i$, say $x_1$.
If it turns out that $s_1 = 1$, then the remainder of the inputs will be forced to move to their proper non-zero $s_i$ values on their own to reach the next stable state.
This is because by placing a logic $1$ on one of the inputs that also happens to be the correct non-zero state of this input, extra charge is introduced to the system that is not allowed to leave due to the quasi-static nature of the complementary transistors.
As such the system has no choice but to set the other $x_i$ input values to the value of the secret string since falling back to the zero state would required removal of the extra charge.

This is similar to what occurs within atomic systems.
Once a quanta of energy is added to the atom, say via a photon that strikes an electron increasing its energy, provided the energy is large enough to overcome the quantum energy gap between energy states, the atom will become excited to the next energy level according to the amount of energy added.
Unless there is a mechanism for the energy to leave the atom, the atom must accommodate the new energy by reconfiguring itself in a way that this energy is allowed due to the constraints of its atomic quantum structure.

A mathematical way to describe what is occurring within the circuits is that there are only two possible solutions to the set of linearly independent equations, $x_i = 0, f_i = 0$, and $x_i = s_i, f_i = 0$.
One simply needs to guess at which of any of the $x_i = 1$ is the proper solution to the higher state so the remainder will follow to move to the correct values corresponding to the secret string.
As such, in general, it is necessary to try each $x_i$ by placing a logic $1$ on them until the network arrives at the proper state.
This takes only $O(n)$ execution steps.
Whether or not the proper secret string is found by looking at the resulting input values $x_i$ after the network settles can be determined quickly by going to the original function data.
The input value $x = 0$ for the Simon problem is always paired with the input $x = s$ since $s_i \oplus 0 = s_i$.
Therefore, to check to see if a particular string obtained is the secret string, one merely needs to check the actual $f$ values of the Simon problem data for both the input value $x = 0$ and the input value corresponding with the secret string guess to see if they are indeed the same.
If they are the same then the proper secret string has been found.

For the network to behave in this manner moving to another stable state as described above, it is first necessary to alter the design of the basic qubit cell, which in its programmable form is essentially a Toffoli gate.
This design alteration would then be used for all aspects of the algorithm representing the final design form for the entire machine as it would not interfere with other operations already discussed.

\begin{figure}
\begin{center}
\includegraphics{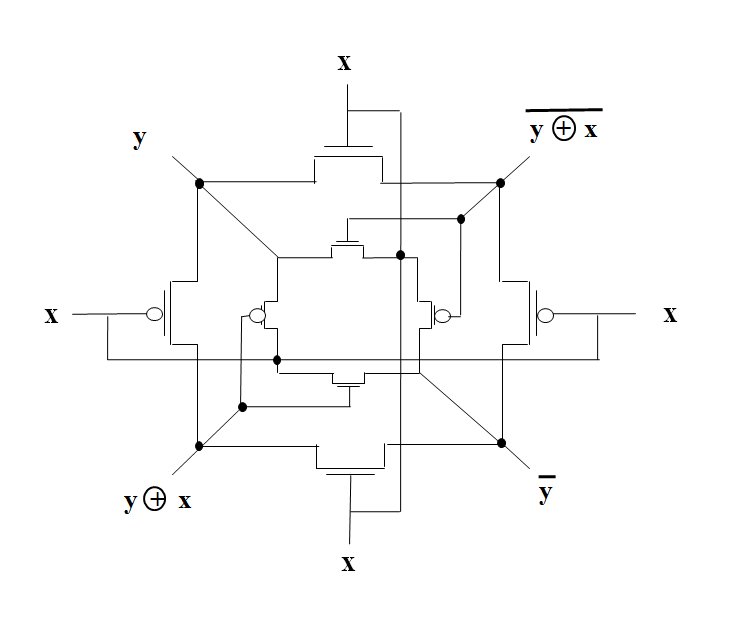}
\end{center}
\caption{Qubit with internal feedback sub-circuit.  \label{fig:qubit_1}}
\end{figure}

The present individual qubit CNOT circuit design establishes voltages at the corners of the circuit dependent upon the voltages input to the $x_i$ variable at the gates of the transistors.
In order to enable the laws of thermodynamics to govern the operation of the circuits, it is necessary to provide reversibility to the basic qubit design on a larger scale.
If an additional CNOT gate is placed within the first CNOT gate as depicted in Figure \ref{fig:qubit_1}, then the inputs to the gates can be influenced by the voltages at the corners and vice versa.

We want to arrange things so that if the top and bottom lines of the outer CNOT circuit in Figure \ref{fig:qubit_1} are each equipotential, or each shorted through the NMOS transistors in these lines, such that there are either two logic highs (1) at the top and two logic lows (0) at the bottom, or vice versa, then there will be a logic high on all gates of the transistors such that the qubit state will be logic high. 
If the right and left lines are equipotential, or shorted through the PMOS transistors in these lines, then we want a logic low to be applied to all transistor gates where the qubit is in a logic low state.

\begin{figure}
\begin{center}
\includegraphics{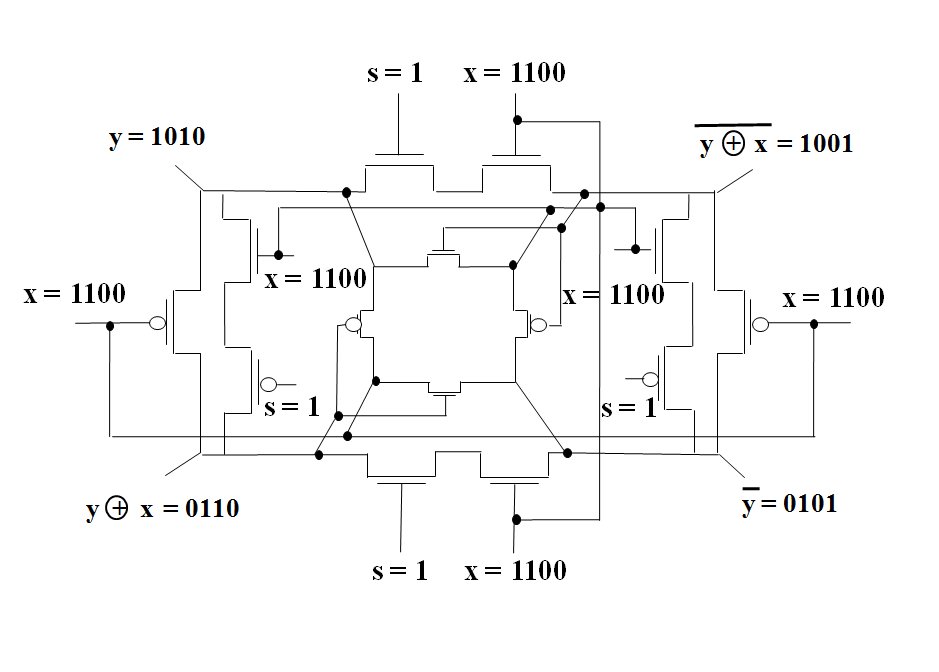}
\end{center}
\caption{Programmable qubit (Toffoli gate) with internal feedback circuit.  \label{fig:qubit_2}}
\end{figure}

Figure \ref{fig:qubit_2} shows the programmable version of this more advanced qubit where the various logic low and logic high combinations at the corners versus at the gate inputs are shown in all possible combinations allowed.
In this figure we see the origonal programmable qubit being a Toffoli gate as in Figure \ref{fig:toffoli} but with the additional center circuitry that resembles another CNOT qubit gate. 
This internal gate are actually dual connected CMOS inverters, each inverter containing a PMOS and NMOS transistor, to perform the operation already described through internal feedback.
This circuit is able to perform all of the original functions of the simple one described in Figure \ref{fig:toffoli} with the additional feature of allowing the outputs of the circuits, ($y, \overline{y}, y \oplus x, \overline{y \oplus x}$) to influence the four inputs $x$ or vice versa. 
This programmable version also enables the overall qubit circuit to represent the Boolean function $x$ or $0$ depending on whether $s = 1$ or $s = 0$, respectively.
The combinations of logic values that the corner variables $y, \overline{y}, y \oplus x, \overline{y \oplus x}$ in the programmable qubit circuit with internal feedback can take are shown in Figure \ref{fig:qubit_2}.

These more advanced programmable qubits of Figure \ref{fig:qubit_2} can now be placed into the overall separate function circuits replacing the simpler Toffoli gates in Figure \ref{fig:fp_cct} being used for all aspects of the machine to solve the Simon problem.
When $s_i = 0$ for a particular qubit, the internal feedback CNOT element can either remain connected where it will have no logical impact on the system, or some extra circuitry could be used to have it disconnected when this occurs.

\begin{figure}
\begin{center}
\includegraphics{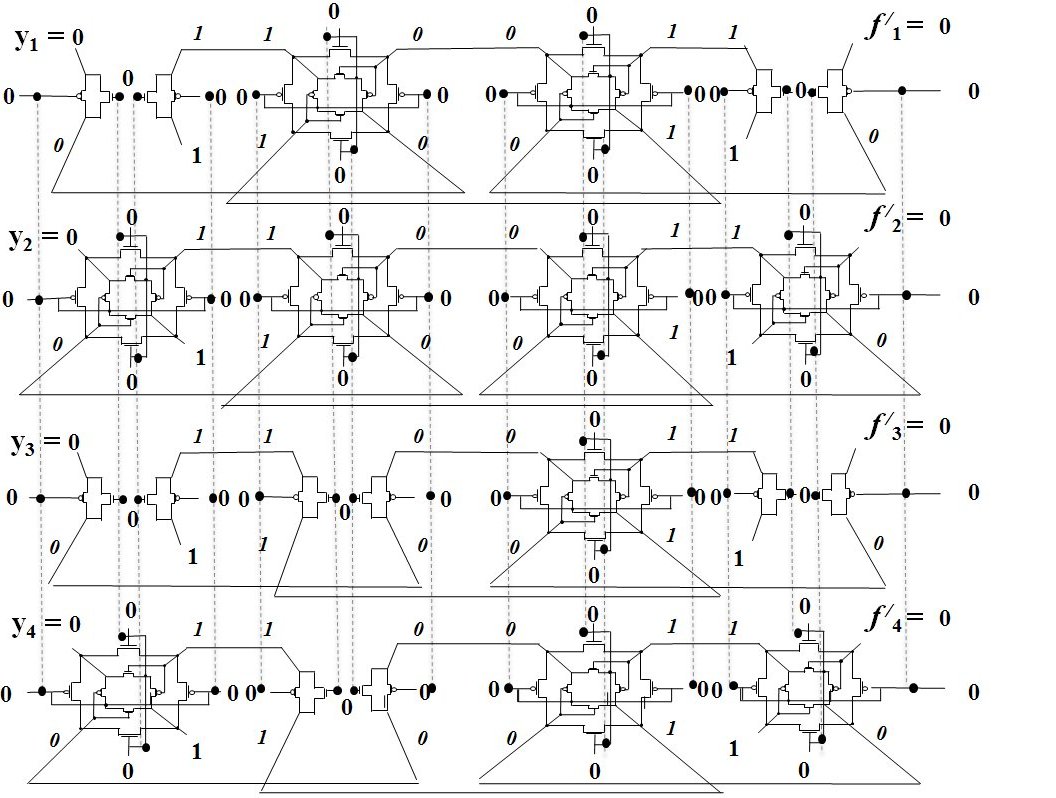}
\end{center}
\caption{Function network depicting Gaussian elimination in $O(n)$ steps with input and output vectors initialized to zero. \label{fig:elim_zero}}
\end{figure}

\begin{figure}
\begin{center}
\includegraphics{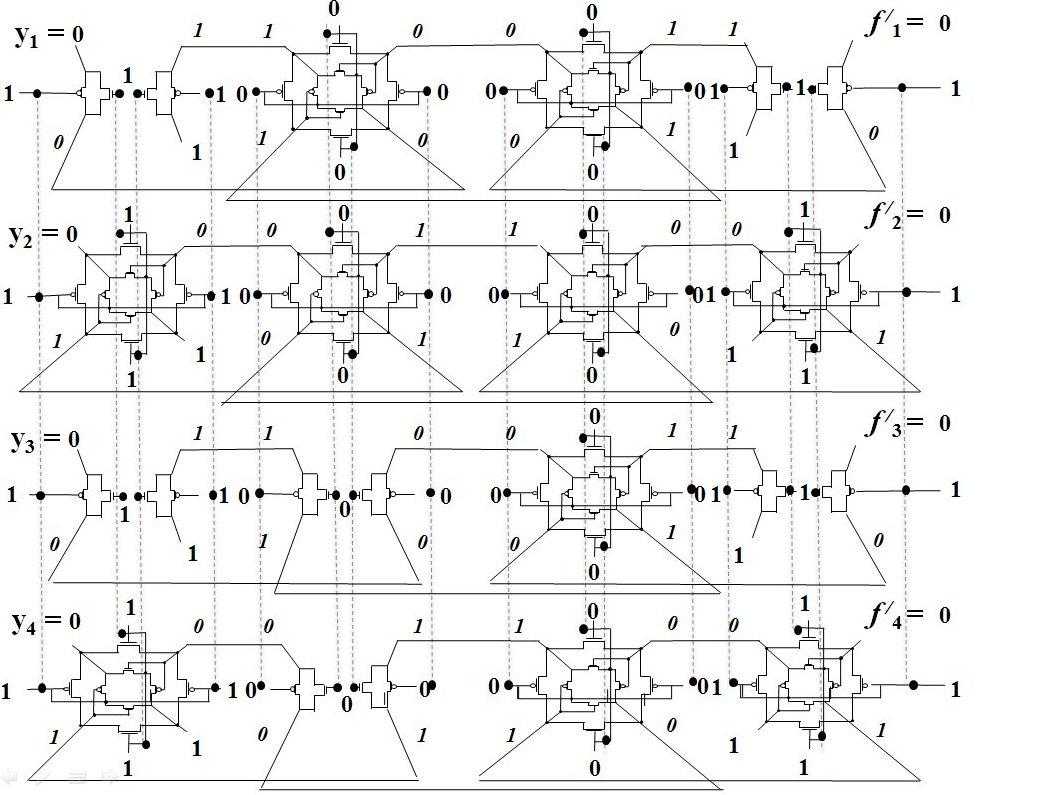}
\end{center}
\caption{Function network depicting Gaussian elimination in $O(n)$ steps in next available state for zero output with secret string at inputs $x = 1001$. \label{fig:elim_one}}
\end{figure}

Figures \ref{fig:elim_zero} and \ref{fig:elim_one} show the final configuration of the separable functions arrived at in Figure \ref{fig:simon_sim:13} but where the internal feedback CNOT circuits are included for $s_i = 1$ cases only to avoid clutter in the circuit diagram and where they would have an impact on the system operation.
This mesh with the internal feedback mechanism in the center of all of the qubits will enable thermodynamics to govern their operation in determining the secret string.
In principle the operation of allowing the circuits to settle on the $x$ inputs becoming the secret string $s$ thereby solving the system of linearly independent equations represented by the two-dimensional circuit mesh, would be conducted after each iteration in Figures \ref{fig:simon_sim:1} through to \ref{fig:simon_sim:13} to ascertain whether or not it was possible to determine the secret string from the equations that had been obtained up to that iteration number.

For a given set of separable logic functions, if they form a linearly independent set of equations, there will be only two possible stable states for these functions as depicted in Figures \ref{fig:elim_zero} and \ref{fig:elim_one} if their $f'_i$ outputs are set to zero.
This is equivalent to writing $n$ functions in the known string $s_i$ but where the string appears in the $x_i$ lines instead since the $s_i$ are controlling the circuit configurations arrived at through the first iteration procedure.

To emphasize the interconnectivity, Figures \ref{fig:elim_zero} and \ref{fig:elim_one} show how all of the input lines $x_1$ through $x_4$ are also connected as they would be at all times even during the first iteration steps.
It can then be seen that all of the functions with their qubits are connected together in one larger two-dimensional mesh.
The behaviour of this mesh is not unlike a giant flip-flop with two stable states, one being the zero state and one being where the inputs $x_i$ are at the secret string $s_i$ logic values where for both states the function outputs $f'_i$ are forced to logic $0$.

Finally, the required circuitry is not shown to go between toggling $y_i$ values and toggling $s_i$ controls throughout Figures \ref{fig:simon_sim:1} to \ref{fig:simon_sim:13}.
Any number of approaches could be used that use known technology, in particular edge triggered flip-flops but that are best designed to be reversible.
One approach would be to use dynamic reversible flip-flop logic that only sensed toggling values transferring the toggling of a $y_i$ line to the required $s_i$ line dependent on which $x_i$ also changed from one data set to the next.
This would require only a linear order amount of extra transistor overhead per qubit.

\section{\label{sec:hadamard}Methods to Implement Hadamard Transforms in One Step Using Feedback Circuitry\protect\\}

Thus far how to implement Hadamard transforms for various examples has been described verbally.
In this section example circuitry is given how to achieve this for the various cases discussed, including solving the Deustch problem, which can be extended trivially to solve the Bernstein-Vazirani, as well as circuitry for the Simon problem.
These methods are not unique but share the common property of involving feedback circuitry around the reversible logic gates.
To do this one must ensure stability is retained as it is easy to come up with schemes that are inherently unstable.

This section outlines possible means to implement the Hadamard transform on both individual $x$ qubits sub-circuits, as well as the $y$ registers in the Simon problem that feedback into the $s$ control lines of the separable function circuits.
The Hadamard circuits used for the $x$ qubit sub-circuit can resolve, using the same number of steps as in a true quantum computer, the global property of a single input function as being either varying or constant according to the Deutsch problem.
It is possible to use this technique for the one and two input Deutsch-Jozsa problems and the $n$-input Bernstein-Vazirani problem.
It is expected that the approach used here, or equivalent, can be used in any quantum computer algorithm where a Hadamard transform must be applied to one or more individual qubit sub-circuits.
The Hadamard circuits used for the $y$ registers enable the configuration of the logic functions by influencing the $s$ control lines that control the functional form of each qubit sub-circuit as being either varying or constant.
As such both of these Hadamard tranform examples involve either the determination of this global property of an individual qubit sub-circuit or it determines it through the setting of the control line $s$ in the Toffoli gate (controlled-CNOT) form of the qubit sub-circuit.
Application of the Hadamard as shown in this addendum or using a similar technique, such that it controls or configures the basic functional form of a qubit sub-circuit, is useful for any probabilistic quantum computer algorithm, such as the Simon problem or the Grover Search algorithm.
Both of these examples of how to implement the Hadamard transform involve asynchronous feedback using external circuitry.
It is also possible to use synchronous techniques that are synchronized to a running system clock.
These are examples of using external feedback to implement Hadamard transforms.
Internal feedback examples already exist within the original patent paper.
This includes the very form of the DeVos CNOT circuits themselves where opposite logic levels are applied to opposite corners of the circuits as explained in the paper.
Another example of using internal feedback is demonstrated in the original patent paper in solving the Simon problem.
Here the rippling of the signal from applying the function values $f_i$ from the truth table of the Simon problem to the $f'_i$ lines of the separable function circuitry through each qubit sub-circuit to the $y_i$ register lines dependent upon the settings of the control lines $s_i$ and the input vector values $x_i$ on each qubit within each separable function, is also part of the overall Hadamard transform on the $y$ register as explained in more detail below.
It should be understood that these Hadamard transform circuits are enabling the circuits themselves, through the feedback, to find a stable thermodynamic equilibrium state that corresponds to a particular vector $x_i$ of the qubit or qubits involved.
In this way the individual qubit sub-circuits, being implemented by either CNOT or Toffoli gates, are in effect classical models of artificial atoms.
The feedback methods used to implement the Hadamard transform on these qubit circuits enable the artificial atom to attain a thermodynamic equilibrium state naturally allowing a computation to take place, in one case determining the global property of the qubit function as being either varying or constant, and in the second case, enabling external data entered from the Simon problem to configure the cascaded qubit circuits representing a separable function.
In the case of the cascade qubit circuits, they form what could be said to be an artificial classical molecule that has many stable thermodynamic equilibrium states that correspond to values of the actual functions in the Simon problem.
Each value in a logic function can be thought of as a stable thermodynamic equilibrium energy state of a classical artificial molecule being modelled using a classical reversible switching network. 
The implementation of the Hadamard steps using feedback on these reversible networks enables solutions to problems to be found rapidly by enabling the overall network to find stable thermodynamic equilibrium states that are enumerated by the values of the input vectors $x_i$ of each qubit within the overall cascade of qubit sub-circuits that make up the molecule.
The circuit techniques shown are intended as practical examples of how to implement these Hadamard transforms using explicit external feedback.
It is meant to be read in conjunction with the relevant sections of patent paper as an addendum.

\section{\label{sec:h1}Hadamard Pulse Circuit for Deutsch Problem\protect\\}

\begin{figure}
\begin{center}
\includegraphics{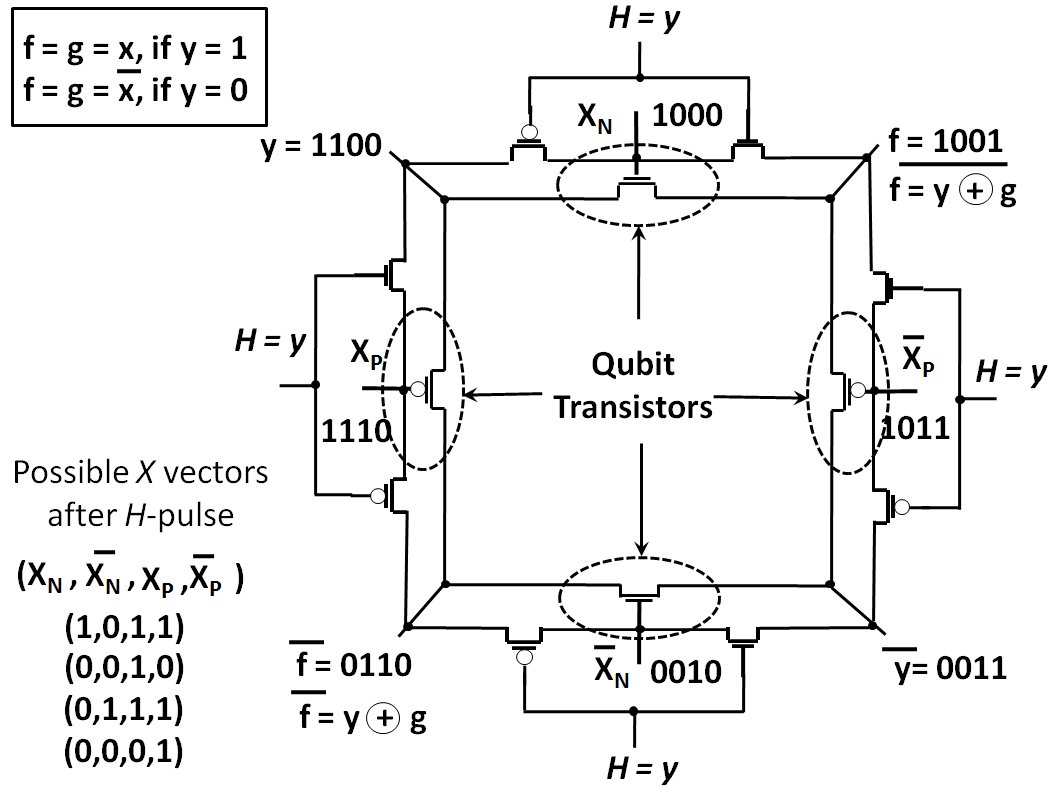}
\end{center}
\caption{Hadamard pulse (H-Pulse) circuitry for providing feedback within a single qubit sub-circuit to determine the global property of the qubit being a varying or constant logic function, where the function in this example is varying, $f = x$ or $f = \overline{x}$.  \label{fig:hx}}
\end{figure}

\begin{figure}
\begin{center}
\includegraphics{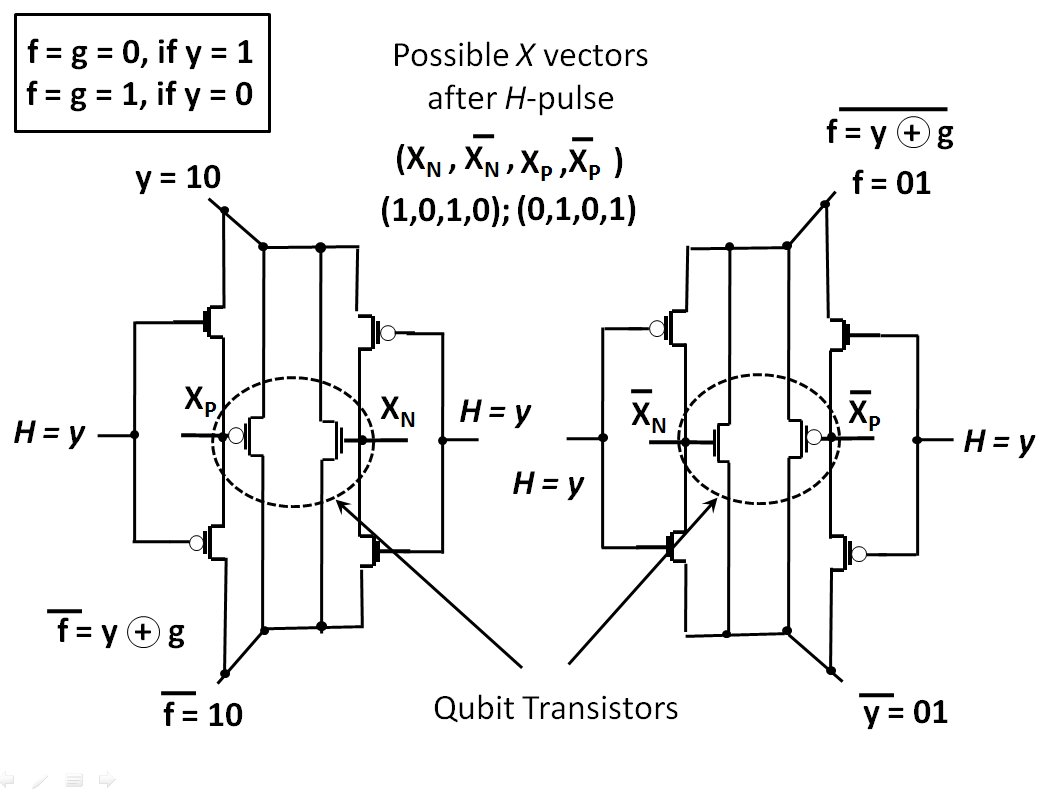}
\end{center}
\caption{Hadamard pulse (H-Pulse) circuitry for providing feedback within a single qubit sub-circuit to determine the global property of the qubit being a varying or constant logic function, where the function in this example is constant, $f = 0$ or $f=1$. \label{fig:h0}}
\end{figure}

Figures \ref{fig:hx} and \ref{fig:h0} depict how the Hadamard transform can be implemented in the Deutsch problem on a single input qubit sub-circuit for both varying and constant function examples, respectively.
The feedback circuity is identical regardless of the type of logic function being represented by the qubit sub-circuit.
The examples shown use CNOT circuits to implement the qubit logic function, but a Toffoli gate being a controlled CNOT circuit, can also be used with no modifications to the feedback circuitry.
The feedback circuitry implements the procedures being described in the patent paper using the orthogonal vector basis formed using common and differential mode logic signals as explained in detail there.
No a priori assumptions are being made using this particular feedback circuitry about the type of function being represented by the qubit sub-circuit.
The feedback circuits consist of PMOS and NMOS transistors that are able to short the gates of each of the four input transistors of the qubit sub-circuit, these gates corresponding to the overall qubit input vector $x = (x_N, \overline{x_N}, x_P, \overline{x_P})$ with its four sub-components that forms a two dimensional vector space for the logic signals.
Each of the individual sub-components can take on two logic values, that could be represented by positive and negative voltages.
In the examples shown here only abstract logic $0$ and logic $1$ values are shown.
It is assumes that the proper logic voltage levels are used on the transistor gates themselves to overcome any non-zero transistor MOSFET threshold voltage $V_t$ that is positive for NMOS and negative for PMOS transistors as explained in detail in the patent paper.

Only the final Hadamard step is discussed here for these circuits where the gates of the input transistors must be shorted out to their respective sources to establish thermodynamic equilibrium within the circuit with zero static current flow in all circuit transistors and branches.
This ``sourcing out'' procedure drains charge from the gates of the four qubit input transistors changing the logic levels that were placed there in the first part of the Deutsch algorithm procedure to that of the sources of each transistor.

The trick is to know where the source is for every possible set of logic levels at the inputs and output to the qubit sub-circuit and for every possible physical form of the qubit sub-circuit.
For a given qubit sub-circuit design, its logic function can be complemented simply by changing the logic value of $y$ and $\overline{y}$, respectively.
As such, in Figure \ref{fig:hx}, the logic function is $f = x$ for $y = 1$ and $\overline{y} = 0$, and $f = \overline{x}$ for $y = 0$ and $\overline{y} = 1$.
In Figure \ref{fig:h0}, the logic function is $f = 1$ for $y = 1$ and $\overline{y} = 0$, and $f = 0$ for $y = 0$ and $\overline{y} = 1$.
Simply by inspection for each of the possible four functions and any possible starting logic state for the qubit sub-circuits, simply by connecting the $H$ line to the $y$ line ($H$-pulse) the gates of all four transistors are automatically and properly connected to their electrical sources such that zero current will flow establishing logic levels at each of the four gate inputs together forming the overall vector $x$ representing the vector state of the qubit.
In each figure the possible $x$ vectors that could occur confirm that for the varying function cases in Figure \ref{fig:hx}, the $x$ vector is always an ODD function being a differential mode signal comprised of one common mode and one differential mode sub-signal as defined in the patent paper, and for the constant function cases in Figure \ref{fig:h0}, the $x$ vector is always an EVEN logic function being a common mode signal comprised of two differential mode sub-signals as defined in the patent paper.
What is important here is that the global properties of these qubit functions were determine through simultaneous feedback driving the system to a stable thermodynamic equilibrium state in a similar way as in a true molecular quantum computing paradigm.

Obviously this same feedback circuitry can be used in the two input Deutsch-Jozsa problem and in the $n$ input Bernstein-Vazirani problem modifying the CNOT or Toffoli gate qubit sub-circuits shown in the patent paper for these problems with the same Hadamard feedback circuitry shown in Figures \ref{fig:hx} and \ref{fig:h0} but placed into each qubit in the multiple input circuit functions.
It should be noted that to solve the Deustch-Jozsa problem, a Hadamard of this form is only applied to a single answer qubit sub-circuit.
This being the case it has been found necessary to ensure that there are no redundant input variables left in the function to be implemented that do not impact the logic output of the function in the Deutsch-Jozsa problem for this method to work properly.
This is a reasonable assumption as it is standard practice to minimize the logic associated with the implementation of any logic function using circuitry where such redundant variables are always eliminated.

\section{\label{sec:h2}Hadamard Pulse Circuit for Simon Problem\protect\\}

\begin{figure}
\begin{center}
\includegraphics{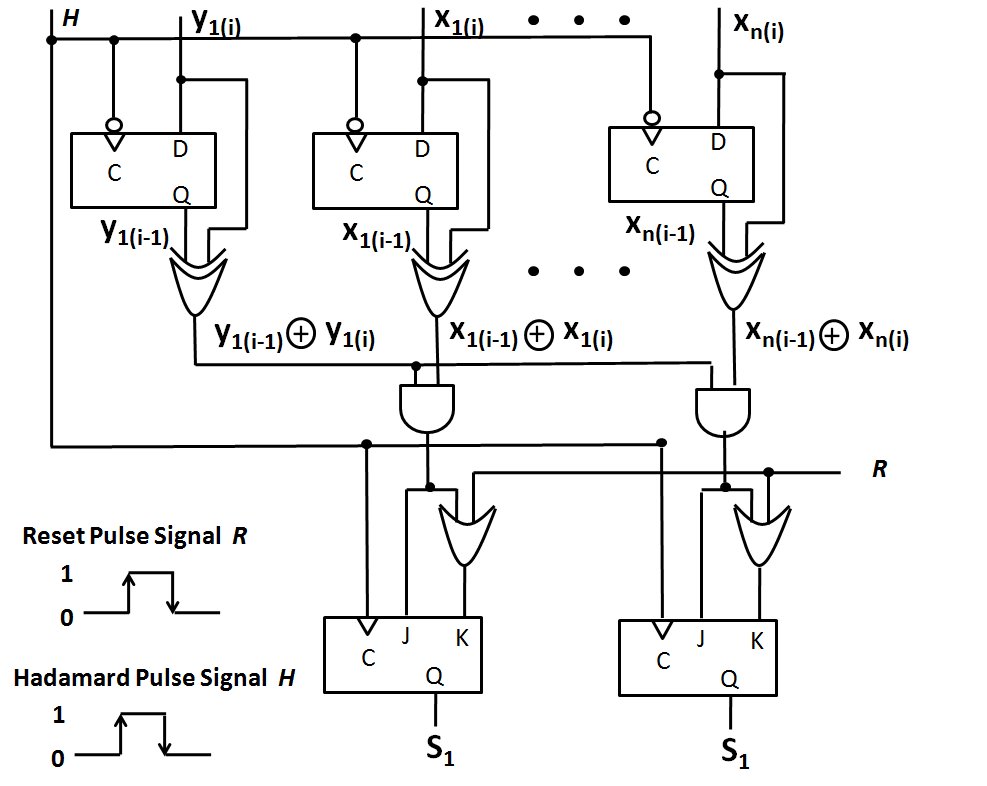}
\end{center}
\caption{Hadamard pulse (H-Pulse) circuitry for providing feedback between $y$ control registers and $s$ control lines of individual qubits circuits within separable function circuits.\label{fig:h1}}
\end{figure}

\begin{figure}
\begin{center}
\includegraphics{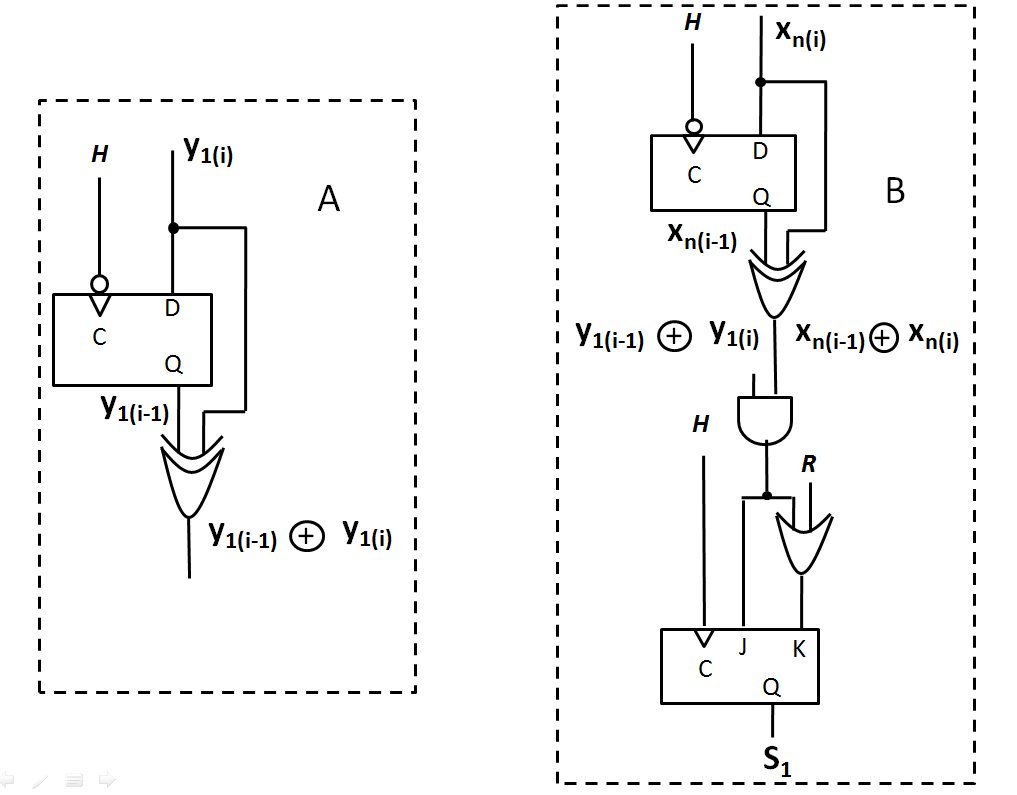}
\end{center}
\caption{Sub-division of H-pulse circuitry shown in Figure \ref{fig:h1} into $A$ and $B$ circuits for identification in overall system shown in Figure \ref{fig:h3}. \label{fig:h2}}
\end{figure}

\begin{figure}
\begin{center}
\includegraphics{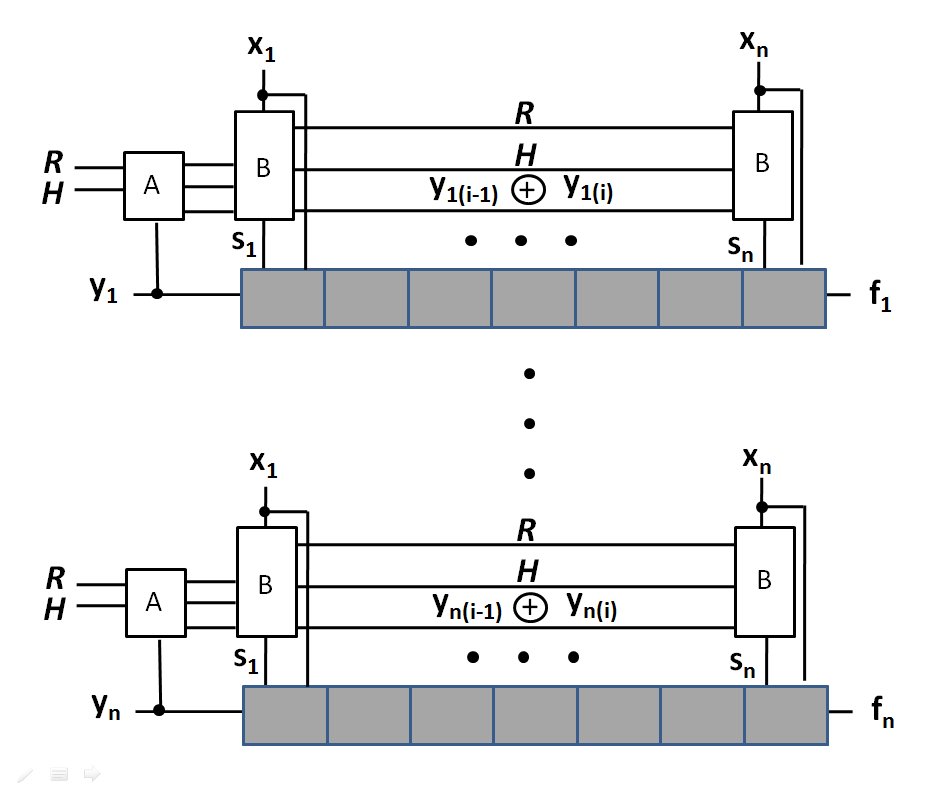}
\end{center}
\caption{Incorporation of H-pulse circuitry of Figures \ref{fig:h1} and \ref{fig:h2} into overall system of $n$ separable function circuits\label{fig:h3}}
\end{figure}

Figures \ref{fig:h1}, \ref{fig:h2}, \ref{fig:h3} depict circuitry that co-ordinates the feedback between the $y$ registers and the $s$ control signals lines that configure separable function circuits $f'_i$. 
Figure \ref{fig:h1} shows the detailed operation of system that is broken down into circuits $A$ and $B$ as shown in Figure \ref{fig:h2} where how these circuits fit into the entire system is shown in Figure \ref{fig:h3}. 

Referring to Figures \ref{fig:h1} and \ref{fig:h3}, the entire system in Figure \ref{fig:h3}, including the H-pulse circuitry, is initialized first by placing the input vector $x = x_1, x_2, x_3, ..., x_n$ = $0,0,0, ...., 0$ and the correct values of the $y = y_1, y_2, y_3, ..., y_n$ to obtain the proper $f = f_1, f_2, f_3, ..., f_n$ values according to the functions given in the Simon problem for $x = 0$ vector.
The $H$ pulse signal, that will be referred to as the `Hadamard Pulse' or H-pulse, is pulsed once to establish values of $x_{i-1}$, $x_i$ and $y_{i-1}$, $y_i$ on either side of the D-latches such that $x_i = x_{i-1}$ and $y_i = y_{i-1}$.
A reset pulse $R$ is then used to set the JK-flip flop outputs $s_i$ to zero as part of the intialization process.

The active clock signals for the D-latches and the JK-flip flops are arranged to be negative and positive edge triggered, respectively.
During a single H-pulse $H$, the positive rising edge of the pulse causes both the JK inputs to the JK-flip flops to take on the value of the output of the result of XORing the previous $x_{i-1}$, $y_{i-1}$ and present $x_i$, $y_i$ states of both the $y$ and $x$ variables to toggle the $s_i$ values if there are changes between the previous and present states of these variables as required for the Hadamard to change the $s_i$ control lines of individual qubit sub-circuits.
The following negative falling edge of the H-pulse $H$ signal then passes the present $x_i$ and $y_i$ states into the outputs of the D-latches to prepare for the next iteration of the Hadamard pulse.

Following the one initialization procedure at the beginning of the iteration to solve the Simon problem, the following iteration procedure involves placing the new values of inputs $x_i$ and function outputs $f_i$ onto the qubit circuitry, one value at a time, where the $x_i$ values are also being placed into the H-pulse circuitry.
Between each new such values there will be an H-pulse.
Time must be allowed for the logic values placed on the $f_i$ outputs to ripple through to the $y_i$ lines on each array of qubits.
This is an asynchronous action within the circuitry which represents the first part of the quantum computer Hadamard transform.
If the new $x$ input vector is not compatible with the $f$ vector through the existing form of the circuitry, then the $y$ lines for these incorrectly configured circuits will toggle or change value.

This action of allowing the $y$ lines to change will require time dependent upon how many qubits are in the system.
For thousands of qubits this could represent an appreciable amount of time compared to any clock periods being used in the system to co-ordinate other actions, such as the placement of new $x$ and $f$ values into the system.
The circuitry that might be used to co-ordinate the entire system is not shown, but can be designed using any number of well known techniques in CMOS IC design.

There are many ways that could be used to estimate this time for the $y$ lines to settle.
It is important that this delay time be allowed before pulsing the $H$ of the H-pulse circuitry since it is possible that glitches may occur in the $y$ lines due to internal races within the circuitry giving a false indication.
One way that this delay time could be estimated is to design a dummy version of a separable circuit function with the same number of Toffoli or controlled-CNOT gates as being used to implement the $f'_i$ circuit functions.
A signal could also be propagated through this circuit from one end to the other such that the other end will definitely toggle being triggered by the same signals in the larger system that would co-ordinate placement of new data on the function circuits.
The output of this signal that travels through the rippling dummy Toffoli array, when it reached the other end of the Toffoli array, could then be used to drive the $H$ pulse line of the H-pulse circuitry with precisely the correct delay it would take for all similar signals to go between the $f'_i$ and $y_i$ lines of each function circuit.
Such an array could be designed using identical circuitry as the actual $f'_i$ functions with precisely the same number of Toffoli qubit gates, or perhaps a few extra to provide for a margin or error.
This dummy function could be designed such that if it is pulsed at one end it will pulse at the other then driving the $H$ line pulse with the proper delay.
This pulsing of the input to dummy array to trigger the H-pulse with the proper delay for the $y_i$ lines to settle on their new values, could be the same pulse that places new data onto the function circuits from the Simon problem being solved.
This would then synchronize the rippling of the influence of the $f'_i$ signals through the function circuits to the $y_i$ lines with the $H$ pulse.
Such a dummy function would only require $n$ more circuits of the same complexity as the original qubits, but without all of the co-ordinating circuity per qubit sub-circuit required in the real function circuitry being used to solve the problem.
As such the hardware overhead for this array would not be high compared to the entire system as only one would be required for all of the function circuits together, where all such functions would be expected to have a similar delay through them from $f'_i$ to $y_i$.

The delay can be estimated as follows.
Assuming that the transistors have a $10$ GHz maximum oscillation frequency, which is a very conservative estimate for today's minimum geometry VLSI CMOS transistors, it would take approximately $0.1$ $nsec$ per Toffoli gate to change states.
In reality typical modern such figures of merit are closer to $100$ GHz, but we will assume that the reversible nature of the circuitry will introduce a ten times increase in switching delay due to their known non-optimal switching speed compared to non-reversible logic that uses more power.
As such these delays are quite conservative.
Assuming a thousand qubits, the delay might be on the order of $100$ billionths of a second ($100$ $nsec$) or $0.1$ microseconds ($0.1$ $\mu sec$) per gate switching another gate in the ripple.
It is expected that convergence to the correct functions when solving the Simon problem would take on the order of $2n$ to $3n$ iterations step using number of data elements being entered, where $n$ is the number of qubits or inputs to the functions.
If this Hadamard step dominated the computation due to the rippling delay per iteration step or per data element being entered, it would then require on the order of $3000 \times 0.1$ $\mu sec$ or $0.3$ one thousandths of a second ($0.3$ $msec$) to obtain convergence for thousand qubit problem because of the asynchronous delays of this nature.
This is a very small computation time compared to the hours it now takes to perform such operations using conventional computers that simulate quantum computer gates using Gottesman-Knill theory.
It is this rapid asynchronous delay that enables the significant speedup of this approach over using conventional computers that arithmetically simulate the actions of quantum gates using theories such as Gottesman-Knill theory. 

Referring to Figure \ref{fig:h1} it can be seen that if the new present values of $y_i$ and $x_i$ are different than their previous past values $y_{i-1}$ and $x_{i-1}$, respectively, then the XOR gates in the H-pulse circuitry will output a logic $1$ that will then be ANDed to each pair of JK inputs to the JK-flip flops.
If both JK inputs are logic $1$ then the JK-flip flop acts as a toggle or T-flip flop where the output state toggles between logic levels $0$ and $1$ always changing its logic value with every active clock signal, this signal being the positive rising clock edge of the H-pulse.
Each the past and present values of each the $x$ and $y$ values at the bit level are being XOR'ed.
If they have changed relative to one another from one iteration to the next placing data values from the Simon problem functions onto the system, then these XOR functions will both produce logic $1$ values that will be fed into both the J and K inputs of the JK-flip flop associated with the relevant $s_i$ control line for the relevant qubit sub-circuit.
If the inputs to these JK-flip flops are both logic $1$ then the outputs of those flip flops will toggle those $s_i$ control lines as required to reconfigure the sub-function being implemented by those particular qubit Toffoli or controlled-CNOT circuits, toggling them between a sub-logic function of $0$ or $x$.
The positive edge of the $H$ pulse will then allow the toggle flip flops to change the value of any $s_i$ control line in any function circuit where this condition exists. 
Then the negative edge of the H-pulse signal $H$ will move the present values of $x_i$ and $y_i$ into the D-latches that will allow them to become the new past values $x_{i-1}$ and $y_{i-1}$, respectively.
Only one $H$ pulse will be executed per $x$-$f(x)$ value entered into the system to provide the required $s_i$ toggle on its positive rising edge, then preparing the H-pulse circuitry for the next time it will be used on the falling edge of the pulse.
It is assumed that the $R$ reset line will remain logic $0$ after being used only once at the initialization procedure.

It is worth mentioning that a portion of this Hadamard is asynchronous and a portion is synchronous using this solution.
Both of these stages involve feedback.
The first part of this feedback is from the $f'_i$ lines through the entire array of qubits to the $y_i$ lines for a given $x_i$ input into each separable function circuit.
This is internal feedback that is possible because of the use of reversible logic circuitry that involves feedback between the $x_i$ and $y_i$ lines, since whether or not a particular $y_i$ line will toggle will depend partly upon the values of $x_i$ placed into the circuit.
The second part of the feedback that completes the Hadamard step is indeed being co-ordinated in a synchronous fashion, but it involves only a single $H$ pulse cycle that in itself does not constitute a true synchronous signal.
This is because this $H$ pulse is not part of an overall system clock that is co-ordinating the entire iteration procedure placing one data element from the outside Simon problem after another onto the circuitry as it is configured.
The timing of when to perform the $H$ pulse is determined using asynchronous delays as explained above dependent upon the settling time of all of the data including the longest settling time of the $y_i$ lines.
One function of the H-pulse circuity is simply to hold previous values of $x_{i-1}$ and $y_{i-1}$ with which to compare the next present values $x_i$ and $y_i$ to ascertain which $s_i$ lines need to toggled.

It is also worthwhile to discuss the added hardware complexity required to implement the H pulse circuitry.
It is anticipated that many extra transistor per qubit are required to co-ordinate the critical four transistors that form the basic functionality of an individual qubit.
It is entirely possible that as many of 50 to 100 such transistors would be required.
However, the use of the H-pulse circuitry is consistent with the fact that only a constant number of transistors and interconnections between them are being used per qubit meeting the requirement for a linear scaling of hardware complexity per separable function and a quadratic hardware complexity for something like the Simon problem that requires $n$ such functions.

Consider the realization of such a machine capable of implementing this class of quantum computer algorithm with an equivalent thousands of qubits of power.
This would be decades a head of what is considered to be possible using true quantum computer technology that involves exploiting true quantum systems.
Even if as many as one hundred transistors and interconnections were required per qubit, to solve something like the Simon problem, one of the most complex and useful algorithms of the Gottesman-Knill class, it would require only $100 \times n^2$ = $100$ million transistors in a CMOS IC.
Today's CMOS IC CPU's and memory chips routinely have over one billion transistors, yet using the means being presented in this paper it is possible to realize a machine capable of implementing important quantum computing algorithms with a thousand of qubits of computing power on a CMOS IC one tenth this size.
This is the significant advantage of being able to implement at least certain classes of quantum computing algorithms that are known to have efficient classical solution, using conventional CMOS transistor IC technology.
CMOS transistor IC technology is the most sophisticated technology in existence today with a well established mass production and affordable manufacturing technology where billions of components can now be reliably placed on a signal chip.
It is estimated that tens of billions of transistors will be integrated in the near future within the next few years enabling even larger quantum computing machines to be realized using these techniques.
It will be some time before molecular electronics that exploit quantum phenomenon will be able to compete with these scales of integration, where we can perhaps harness the true power of important and useful classes of quantum computers using our already existing mature transistor technologies earlier than expected until these others future true quantum technologies mature over the decades.

\section{\label{sec:simon_compare}Comparison Between Quantum and Classical Simon Algorithm\protect\\}

It is interesting to note the similarities between the quantum computer version of the Simon Oracle or Algorithm, discussed in Section \ref{sec:simon_oracle}, and the classical solution using asynchronous feedback techniques in reversible switching networks in Section \ref{sec:simon}.
Both use two registers, an input $x$ vector register and a $y$ control register.
The $y$ control register in the classical solution are the $y_i$ and $\overline{y_i}$ inputs to the cascaded Toffoli gate function circuit of Figure \ref{fig:fp_cct} used to represent each fully separable core function $f'_i$. 
The initial Hadamard transform being performed on this register, as is required in the quantum computer Oracle depicted in Figure \ref{fig:simon_algorithm}, is the placement of the two complementary voltage signals representing a logic $0$ and logic $1$ in the circuit of Figure \ref{fig:fp_cct}.
Simply by applying these two voltages to the same circuit in a CNOT gate is a superposition of the two possible states of this circuit.
This can be compared to how an initial Hadamard transform was applied to the $x$ inputs in the circuits in previous sections which was necessary to solve problems such as the Deutsch problem.

The CNOT qubit circuits themselves, the way they are formed by interleaving two equi-potential lines, the logic $0$ and $1$ voltages, into the same circuit represents a fundamental superposition of two logic states of the system as in a true quantum computer.
When more then one such qubit circuit is connected together to form more complex functions, all of the multiple states of that function are being superimposed as in a true quantum computer in situations where quantum entanglement is not required or exists.
The random quantum fluctuations of the electrons and hole in the semiconductor transistors, as well as the electrons in the connecting conductors, are joining these multiple states together in a random statistical thermodynamic Boltzmann gas.
The simultaneity of these random quantum fluctuations occurring throughout the circuit is what causes these circuits to exist dynamically in several states at the same time.
This is the only essential feature required in quantum systems to attain the same type of superposition of states within a true quantum computer.
This is a new concept being introduced in this paper and is central to the understanding of how a classical Thermodynamic Turing Machine has the same properties as a true quantum computer with regards to superposition and simultaneity of operation.
They both draw upon the same fundamental principles of thermodynamics to achieve these properties essential to quantum computing.
Just how these quantum computing properties can be extracted from a classical switching network to solve practical problems previously thought to belong only to the quantum computing realm is the subject of this paper.

It does not matter, statistically, if single electrons form statistical positional clouds within single atoms in a true quantum computer or huge numbers of electrons form a classical Boltzmann gas within macroscopic systems.
In other words, the size of the system is not relevant where both the tiny quantum systems and the macroscopic circuits are providing precisely the same physics necessary for the superposition of a Hadamard transform in quantum computing.
From this perspective, both are indeed true quantum computers with regards to superposition, since both are depending upon the same essential quantum phenomena, that of discreteness and randomness in matter within their respective physical systems to implement the algorithm.

An analogy to this situation is understanding that it does not matter if one flips a single coin randomly a large number of times, which is what is happening within microscopic atomic level quantum physical systems as one or a few electrons move randomly according to a predetermined spatial probability distribution, or if one flips a larger number of coins randomly at the same time, which is the situation corresponding to the macroscopic transistor circuits.
These ideas are fundamental to the concept of a Thermodynamic Turing Machine (TTM) described in more detail in a following section.

The next step in both the quantum and classical Simon oracles is to apply a zero state to the input vector $x$ and then to obtain an output in the second register $f(x)$ representing the $n$ functions, which also correspond to the $f'_i(x)$ outputs for each separable circuit function in the classical machine.
In the classical circuits, both the $y_i$ (including $\overline{y_i}$) and the $f'_i$ are actually the same register since the $y_i$ (and $\overline{y_i}$) control register directly changes the value of the so-called function output $f'_i$.
Hence, any operation on one impacts the other directly for a given $x$ input.
The simultaneous existence of the classical $y$ registers in both $y_i$ and $\overline{y_i}$ states represents the original Hadamard of this register that was the first part of both the quantum and classical versions of the Simon oracle
After applying a particular $x$ value to obtain an $f(x)$ value stored in the second register, this register is then allowed to remain as it is until the next iteration where a new $x$ value will be chosen.

The next step is to apply another Hadamard to the $y_i$ register in both the quantum and classical oracles.
With every iteration besides the original initialization of $x = 0$, this is interpreted in the classical oracle as imposing the correct values of $f_i$ from the external data on the $f'_i$ outputs of the circuits while at the same time imposing the new value of $x$ at the inputs to each function circuit.
The classical interpretation of this second Hadamard of the second $y$ register then is to watch to see what happens to the $y$ control registers for each individual function circuit that corresponds to both $y_i$ and $\overline{y_i}$ values. 
If the $y_i$ and $\overline{y_i}$ values of a particular function circuit change or toggle, then this toggling is allowed to influence the controls $s_i$ for the qubit of the changing values of $x_i$ as described in the classical oracle treatment given in Section \ref{sec:simon}.
These actions of allowing the changing $y_i$ values to influence the programming of the qubit with the changing $x_i$ value is the classical interpretation of this Hadamard step in the quantum version of the Simon algorithm or oracle.
This is accomplished through asynchronous feedback between various parts of the classical reversible logic circuits that can be identified as the two registers $x$ and $y$ in the quantum version of the algorithm.
This is identical to imposing a quantum superposition between the $x$ and $y$ registers in the quantum oracle of Figure \ref{fig:simon_algorithm}.

A Hadamard in true quantum computers performs a superposition where this superposition may take place within a single qubit, or with a single register of qubits, or between registers of qubits, etc.
An example of performing a Hadamard using asynchronous feedback in classical reversible logic gates within a single qubit was shown in a previous example to solve the single input Deutsch problem.
An example of performing a Hadamard on an entire register of qubits using these classical techniques was shown by solving the Bernstein-Vazirani problem in a previous section.
Finally, an example of showing how to implement a Hadamard that results in superposition between two or more registers using asynchronous feedback within reversible classical logic circuits is now being demonstrated here for the Simon problem for the second Hadamard step in the Simon algorithm.

Any Hadamard can influence the states of individual qubits, within or between qubits, or within and between entire registers of qubits that experience the superposition associated with the Hadamard.
This is why the toggling of the classical $y$ control lines in the classical function circuits in the classical Simon problem example are changing the $s_i$ lines of the qubits.
The $s_i$ and $x_i$ lines within the classical circuitry (e.g. see Figure \ref{fig:fp_cct}) can be seen to be interchangeable and indistinguishable from one another.
This is fundamental to the fact that changing the $s_i$ values is no different than changing the $x_i$ values in practice in the circuitry.
As such allowing feedback between the $y$ and $s$ lines in the classical circuitry can be interpreted as allowing a Hadamard superposition between the $y$ and $x$ registers in the quantum version of the algorithm.

The above description of the classical version of the second Hadamard on the second register is probably the most difficult aspect of the interpretation of the classical versus the quantum oracle.
Allowing feedback between the changes in the control registers $y_i$ in the classical circuits and the controls to the qubits that experience a change in the $x_i$ value involves random quantum fluctuations of sub-atomic carriers within the electronics of the circuit to influence one another through electro-chemical potential gradients following quasi-Fermi level energy distributions within the electronics no different then what occurs within quantum systems in a quantum computer.

The simple explanation is to realize that what is happening within both the quantum computer and the classical circuit implementation is that both are simply fitting a set of fully separable functions to sequences of $n$ data elements being fed into both kinds of systems with the goal of obtaining linearly independent sets of equations in the unknown secret string $s$.
The quantum computer algorithm discussed here stops at the obtaining of the necessary linearly independent equations in $s_i$ from which the secret string can be obtained.
Other more recent quantum algorithms for the Simon problem exist that attempt to improve upon the finding of the solution to these equations, once found, that are more efficient than using a Gaussian elimination as well as some that provide deterministic solutions.
It is likely that classical analogies can be made to these as well unless they depend upon quantum entanglement which is not being considered in this work.

\section{\label{sec:self_learning}Probabilistic Thermodynamic Turing Machine as a Self Learning Computer\protect\\}

It is interesting to note that the solution presented in Section \ref{sec:simon} is essentially a self configuring or self learning machine.
The machine actually designs itself at the interconnection level between transistors.
Configuration at the interconnection level is the most fundamental way for a machine to learn.

What can be concluded from Section \ref{sec:simon} is that the use of asynchronous feedback in a reversible logic network to implement or to interpret the quantum computer Hadamard leads to a self learning paradigm where implementation of the Hadamard itself using feedback is what determines how the machine is to learn or to be configured at the interconnection level.
In other words, one property or consequence of a Hadamard being implemented using asynchronous feedback in classical reversible switching networks is that is induces or endows self learning ability on the network.
The Hadamard itself appears to provide the necessary instructions behind endowing such self learning properties that may also be inherent in living organisms.

What appears to be happening in the solution presented to solve the Simon problem using a classical switching network seems to be essentially what takes place in an organic brain.
Both have reversible switching networks capable of dynamically configuring logic paths within these networks composed of equi-electro-chemical pathways.
Both configure these pathways simply by being exposed to external data.
Configuration of these pathways enables the forming of discrete logical Boolean functions that approximate or represent the data in some fashion.
Then both appear to be able to rapidly extract global properties to these dynamically  configured logic functions that correspond to a answers or knowledge otherwise hidden or embedded within the external data.
Both also appear to be able to do so by being exposed to only very small fraction of the total amount of data that would be required to completely define the actual functions in the outside world if all data were available or observed by the thinking machine.

The use of the neural network patterns within brains may merely be a more general and convenience form of approximating logic functions using reversible classical networks than using the simpler fully separable functions as in this paper.
It should also be possible, through a slight redesign of the networks presented in this paper, to access quantum entanglement functionality.
If the output of a individual qubit sub-circuit is allowed to be connected to the $s_i$ control lines of other qubit sub-circuits within the larger array of qubits forming the multiple input logic functions, then the logic state of qubits can be used to control the functionality of other qubits.
Without providing the proof in this paper, it is a simple matter to show that such an arrangement can lead to classical circuits that violate Bell Inequalities, the mathematical definition of quantum entanglement.
This being the case, it should be possible to use the approaches given in this paper to solve an algorithm, such as the Simon problem, that required quantum entanglement, such as the Shor algorithm by a simple redesign of the sub-circuits.
Quantum entanglement does not involve any new physics beyond the laws of thermodynamics.
As such, quantum entanglement should be efficiently accessible to a TTM scaling identically with a quantum computer with regards to hardware and execution complexity.

Indeed, Sir Roger Penrose in his book \cite{penrose:book:89} suggested that the human mind must have access to quantum computing ability.
However, it has not been yet decided by the research community that the brain has access to quantum mechanical effects to use for information processing as in a true quantum computer.
The methods presented in this paper prove that it is possible for the brain, or the human mind, to access at least the class of quantum computer algorithms being discussed here that do not involve quantum entanglement.
It is not necessary for quantum entanglement to be accessible to an organic brain to implement many classes of quantum computer algorithms as is evident from the work presented in this paper.

In summary, all of the concepts and methods presented here could also be present in an organic brain where the equivalent of Hadamard transforms are being implemented using thermodynamic feedback effects co-ordinated by large numbers of neurons through electro-chemical interactions and interconnected via reversible neural networks dynamically configurable through switchable equi-electro-chemical pathways. 
It is entirely possible that millions or billions of neurons are ``charged up'' being preset with energy for firing before being allowed to relax firing simultaneously to evolve the equi-electro-chemical network of logic paths from one thermodynamic equilibrium state to another to solve a problem.

Consciousness itself could simply be a continual evolution of large numbers of reversible neural network pathways from one thermodynamic energy state to another as if the brain were a kind of giant artificial molecule mimicking real molecules in Nature.
Such continual evolutions of the neural networks would enable the organism to continually poll its own internal logic providing a sense of itself through feedback.
This contrasts with a conventional classical computer paradigm where the logic gates are only activated when its inputs are externally stimulated for the purposes of performing a logical operation where answers appear at the outputs.

This discussions then raised the age old question as to what may be the difference between us, as living animate entities and non-living inanimate objects.
The answer may be nothing other than speed and memory.

Nature, in the form of non-living inanimate objects, where all such objects are made from molecules that have access to the same thermodynamic properties as an organic brain, is known to exhibit significant intellectual properties being allowed to evolve over long periods of time on the order of millions of years.
Given enough time any thermodynamic system might exhibit true intelligence, but if it is extremely slow it cannot interact on the same time scales as a living thinking organism.
As such the only essential difference between inanimate Natural objects and organic brains may be speed not substance, where both have access to thermodynamic intelligence afforded to classical reversible logic networks that evolve from one thermodynamic equilibrium state to another.
Furthermore, an organic brain has the ability to continually poll its owns internal logic configurations massively in parallel using the same thermodynamic feedback methods presented in this paper to solve a problem like the Simon problem.
It also has a memory that allows it to store the ``settings'' of previous neural network configurations that can be used to reconfigure the network as required as a means to retain knowledge gained through earlier probabilistic learning.
Nature does not necessarily have such knowledge retention capability in the form of inanimate objects.
It does, however, have a continual internal logic polling ability on a rather rapid time scale.

The classical solution shown here to solve the Simon problem for instance, enables meaning itself to be encoded in a machine.
Each secret string can be used to encode the meaning of a concept where an exponential number of such concepts could be accessible to a quite small machine in terms of qubit power.
There are an exponential number of sets of possible functions with the same secret string.
This enables a vast number of concepts to be encoded and then each mapped to a vast number of possible ways in which a single concept may appear.
This opens up the possibility of a machine being able to recognize a concept even though it has encountered new data never before seen by the machine.

It is possible to introduce a fuzzy logic concept that allows the same concept to have a range of secret strings that are numerically adjacent as in a kind of generalized analog-to-digital converter.
Then the possible function sets with this secret string will share logical properties that have a degree of similarity that can be detected by the machine through the same means as used to solve the Simon problem in this paper.
If a particular data set never encountered by the machine is beginning to look like a previous encountered data set, the machine simply needs to load the $s_i$ control line settings that were stored in memory  from a previous learning session to arrive quickly at the correct secret string range corresponding to a particular concept.

In this way, the Simon problem concept could be used to realize a more sophisticated thinking machine that could be trained to recognize any concept through an appropriate coding scheme.
Given the fact that the number of such concepts grows exponentially with the number of qubits, it would quite easy to develop a machine that was far more capable than a human in understanding concepts with rapid speed on the order of small fractions of a second.
The logical knowledge that would be accessible to such a machine and the speed with which it could access such knowledge would far exceed that of the smartest human being.

\section{\label{sec:ttm}Thermodynamic Turing Machine (TTM)\protect\\}

Although the above results can be explained using classical logic circuit concepts, it is also interesting to interpret the function of these circuits in the presence of feedback for the Hadamard transform step in terms of thermodynamic statistical concepts.

As with any electronic circuit, these reversible logic circuits are not strictly classical systems.
However, the use of the Hadamard transform through the use of feedback exploits the tendency of the circuits to reach a new thermodynamic equilibrium in performing a computation.
As with any physical system, the circuits are governed by the laws of statistical thermodynamics that in turn are governed by random sub-atomic quantum processes, exhibited by electrons and holes within the transistors and by electrons in the conducting lines that join them.
The circuits can be thought of as networks of interlocking paths that can be configured by way of switches that are in turn implemented using transistors.
For each circuit there are $2^n$ ways in which one can configure the paths such that two equi-potential paths or surfaces at different voltage levels are formed through the network between its corners to the function outputs $f$ and $\overline{f}$.
These networks also constitute a Boolean function whose inputs are the controls to the switches or to the transistor gates.
For the single input Boolean functions represented by the reversible logic networks, if the function is balanced there exist paths between the two equi-potential paths that constrain the system.
If the function is constant then these constraining paths do not exist.

The fast portion of the algorithm that is akin to quantum computing occurs during the Hadamard step where feedback is formed between the inputs to the switches controlling the network path configurations and the potentials existing in the network itself.
The logic values taken on by the inputs as a consequence have properties that are independent of the particular configuration of the network paths that enable a global property to be rapidly ascertained.
When voltages are applied directly to the transistors before this step the circuits could be thought of as being in a state of thermodynamic equilibrium where no current flows in any of the paths, but at a higher than ground state where charges exits on the gates of the transistors.
To effect the Hadamard step feedback is made between the different circuit paths and the inputs to the switches or transistors that control their configurations.
This feedback is done in a way that demands that the entire system reach thermodynamic equilibrium to enter a kind of ground state, or in more general, any particular stable state including higher ones, such that there are certain charge configurations on the gates of the transistors and where still no current is allowed to flow through any of the circuit paths.
For ground state this may mean that there are no net charges on the transistor gates.
In this case thermodynamic equilibrium is being defined as no current flowing in the circuit paths, and the particular state, either ground state or above depends upon whether or not there are charges on the transistor gates.
With charge on their gates the transistors and hence the entire circuit can be thought of as being in a higher energy storage state then with no charge on the gates.

For Hadamard operations involving the qubits themselves, when feedback is applied to drain charge from the gates through the ``sourcing out'' procedure thermodynamic quantum fluctuations in the charges within the transistors and connecting wires are responsible for the system reaching a different thermodynamic state that in turn alters the input voltages to function inputs which are the inputs to the transistors themselves.

The thermodynamic quantum fluctuations are present everywhere simultaneously in the circuit paths and as such are able to ``explore'' the entire state space of the Boolean function at the same time.
Through the feedback procedure in the Hadamard, their individual actions are superimposed upon one another but are at the same time constrained with regards to the structural topology of the circuit as to how they can influence the inputs.
For the single input functions, for instance, how they are constrained by the particular arrangement of the circuit paths, depending upon whether or not the function is balanced or constant, is essentially an interference effect between the possible configurations that the network paths can take.
In this way pure randomness, but constrained severely in the network of paths implementing the Boolean logic function, leads to certainty in the outcome to the algorithm.

Allowing thermodynamic quantum fluctuations throughout all of the possible paths to simultaneous influence the inputs that control the network path configurations is responsible for the efficient or fast aspect of the algorithm.
Hence, it is this phenomenon and how it is being used that can be thought as being equivalent to quantum computing for the separable or partially separable state quantum computer algorithms.
From this perspective this means of computing might be thought of as a Thermodynamic Turing Machine (TTM) that is equivalent to a Quantum Turing Machine (QTM) at least where quantum entanglement is not involved.
This is not say that the TTM approach cannot be applied to quantum entanglement situations as this has yet to be proven one way or the other.

It should be pointed out, however, that a TTM is not restricted to Boolean functions that can be expressed as separable or semi-separable states as defined in \cite{arvind:v56:01}, but can be applied to any generalized function that requires true quantum entanglement for analysis using a quantum computer.
The degree to which a TTM can compete with a true quantum computer then becomes a matter, not of execution speed, but of efficiency in logic circuit network implementation in terms of hardware requirements. 
Conversely it may be that a true quantum computer can be thought of as essentially a TTM since any physical system, be it quantum or classical, is governed by the laws of statistical thermodynamics that are in turn influenced by random quantum fluctuations of sub-atomic particles at one level or another according to the Standard Model of Physics.
A true quantum computer, however, by way of exploiting naturally occurring quantum systems, such as atoms etc., may possess a natural interconnectivity that provides exponential interdependencies between qubits without explicit interconnections as required in purely classical systems.
It remains an open question as to whether the TTM approach can be used to compete with the quantum approach in this regard since true quantum systems are essentially thermodynamic systems obeying the laws of statistical thermodynamics.

\section{\label{sec:random}Quantum Randomness\protect\\}

An essential conclusion that can be drawn from this work is that the only true quantum phenomenon that is required from Nature is randomness or noise to implement quantum computer algorithms.
True pure randomness only manifests itself from the quantum world, having no classical interpretation.
However, all other forms of quantum mechanical phenomena can be mimicked using classical means including quantum entanglement itself.

Quantum entanglement, although outside the scope of the work presented here, can be efficiently mimicked by slight modifications to the circuits presented in this paper such that they violate Bell inequalities.
It is only necessary to design a classical reversible logic circuit whose Boolean function output controls the elementary Boolean logic function of another such circuit where both circuits represent elementary qubits.
By combining such circuits it is possible to ``entangle'' arrays of fully separable state reversible functions which will result in recursive nested feedback between all classical qubits.
This in turn will enable exponential complexity to be embedded in linear or polynomial time execution speed.

It could be said, on both scientific and philosophical terms, that true randomness is the only true mystery of the Universe that cannot be understood on any terms. 
The entire Universe is essentially, at the most fundamental level, comprised of shaped randomness is space-time.
At the point of the Big Bang, if the Universe were truly a singularity, then it would have been comprised solely of pure randomness that was purely white in frequency spectrum over an infinite frequency range.
In practice it may be that it did contain some frequency-phase structure with a finite effective noise bandwidth that provided finiteness in space-time extent.
However, because of the Big Bang, this space-time energy randomness spread out in space-time requiring everything in it to adopt a structured spatial and temporal frequency spectrums with non-uniform  amplitude and phase versus frequency. 
In essence, everything is simply shaped noise or randomness in space time.
To change this spectrum or ``shape'', either in spatial structure or temporarily, requires energy, work and forces according to the laws of both Newtonian and quantum mechanics.
By reshaping matter, or displacing an object in space-time, one is attempting to reshape the order that contains and limits the extent of the randomness.

Analogies can be drawn with information processing.
Noise spectral shaping is a common practice adopted in modern non-linear analog-to-digital converters for instance.
A physical non-linearity is required to effect such noise shaping as is any attempt to shift or alter the frequency spectrum of a physical system.
Physical non-linearities are also required for information processing as is evidenced by the sets of universal gates required in any logic system.
As such, information processing itself in any form can be viewed or effected as a noise shaping exercise with analogies made to physical mechanics and thermodynamics itself.


\section{\label{sec:con}Conclusions\protect\\}

Methods to implement quantum computer algorithms by using asynchronous feedback in classical reversible switching logic circuit networks have been introduced.
The use of asynchronous feedback in reversible logic circuits enables the implementation of the Hadamard steps in quantum computing algorithms in one simultaneous fashion as is possible and expected within true quantum computers.
This is accomplished using identical hardware complexity and execution speed as in a true quantum computer.

Methods to implement the Hadamard in different situations that arise within true quantum computers have been demonstrated through various quantum computer algorithmic examples.
By implementing the single qubit Deutsch algorithm, it was shown how feedback within a single reversible CNOT qubit circuit can implement a true simultaneous Hadamard transform.
It was formally proven resorting to quantum computing mathematical principles that use of this type of feedback was indeed a Hadamard transform in every respect.
How a Hadamard can be applied to an entire register of qubits in a classical switching network using asynchronous feedback was demonstrated by solving the Bernstein-Vazirani problem.
Finally, how a Hadamard can be applied to entire registers of qubits to influence another register was also demonstrated using these asynchronous feedback techniques by solving the Simon problem.

Implementation of the Hadamard transform using asynchronous feedback in classical reversible logic circuits in a probabilistic algorithm, such as the Simon problem, let naturally to the realization of a self learning machine where the feedback guides the learning process.
This learning occurred through dynamic configuration of the interconnections within the network that came about because of the use of the asynchronous feedback that implemented the required Hadamard transform.

These methods can be applied to any quantum computer algorithm that does not require quantum entanglement and that are already known to require only Hadamard gates, CNOT gates and Pauli gates.
It has not been proven one way or the other whether or not these methods could also be applied successfully to quantum computer algorithms that require quantum entanglement outside of this class.

The methods presented here bring about the concept of a Thermodynamic Turing Machine (TTM) or thermodynamic computing.
A TTM distinguishes itself from the other paradigms of computing machines, such as Classical Turing Machines (CTM), classical Probabilistic Turing Machines (PTM), and Quantum Turing Machines (QMT).
In a TTM, portions of the computations that provide speedup identical to that of a QMT being fundamentally faster and more efficient than either a CTM or classical PTM, are conducted by allowing the laws of thermodynamics to govern the computations.
The machine itself becomes a controlled classical Boltzmann-like gas that is governed by the laws of discrete thermodynamic statistics.
Each "particle" within the gas is controlled.
Through asynchronous feedback in the classical reversible network, the laws of thermodynamics influenced by random quantum fluctuations and constrained by the internal interconnections as well as external data, are allowed to cause the system to converge to the answer of the computation.
In effect, random thermodynamics is doing the computations.
The desire of the machine to achieve thermodynamic equilibrium effects the computation while doing so in the most efficient manner.
This tendency to reach thermodynamic equilibrium is also harnessed to implement a self learning machine in a probabilistic context.
The machine structure and the external data to which the machine is exposed simply guides and constrains this process.
The interconnections within the machine are configured according to these constrained thermodynamic laws that allow the machine to learn how to solve the problem in the most efficient manner naturally.
These thermodynamic principles guide the machine how to best organize itself internally through auto-configuration and auto-parallelization to best solve the problem in the most efficient manner with no a prior knowledge of the outside user.

Analogies can be made between a TTM and how quantum systems such as atoms are used in true quantum computers.
Both exploit configurable equi-electro-potential paths influenced by random quantum fluctuations, where in the atoms these paths involve interactions between electron probabilistic clouds between different energy levels within and between atoms.



\end{document}